\documentclass[longauth]{aa}

\usepackage{graphicx}
\usepackage{natbib}
\usepackage{scalerel}

\usepackage{booktabs}
\usepackage{multirow}

\usepackage[table]{xcolor}

\usepackage{txfonts}
\usepackage[pdfencoding=auto,psdextra]{hyperref}
\hypersetup{
    colorlinks=true,
    linkcolor=blue,
    filecolor=magenta,      
    urlcolor=blue,
    citecolor=blue
}
\makeatletter
\renewcommand*\aa@pageof{, page \thepage{} of \pageref*{LastPage}}
\makeatother

\usepackage[utf8]{inputenc}

\usepackage[switch, modulo]{lineno}

\usepackage{euclid}
\usepackage{lipsum}
\usepackage{comment}
\usepackage[normalem]{ulem}

\usepackage{glossaries}
\glsdisablehyper
\newacronym{ia}{IA}{intrinsic alignment}
\newacronym{la}{LA}{linear alignment}
\newacronym{nla}{NLA}{non-linear alignment}
\newacronym{tatt}{TATT}{tidal alignment tidal torquing}
\newacronym{photo-z}{photo-$z$}{photometric redshift}
\newacronym{dv}{DV}{data vector}
\newacronym{wl}{WL}{weak lensing}
\newacronym{ggl}{GGL}{galaxy-galaxy lensing}
\newacronym{gc}{GC}{galaxy clustering}
\newacronym{fom}{FoM}{figure of merit}
\newacronym{snr}{S/N}{signal-to-noise ratio}
\newacronym{nnpz}{NNPZ}{Nearest Neighbours Photometric Redshifts}
\newacronym{hod}{HOD}{halo occupation distribution}
\newacronym{am}{AM}{abundance matching}
\newacronym{los}{LOS}{line of sight}
\newacronym{lss}{LSS}{large-scale structure}
\newacronym{fft}{FFT}{Fast Fourier Transform}
\newacronym{agn}{AGN}{active galactic nuclei}
\newacronym{cloe}{\texttt{CLOE}}{Cosmology Likelihood for Observables in \Euclid}
\newacronym{des}{DES}{Dark Energy Survey}
\newacronym{kids}{KiDS}{Kilo-Degree Survey}
\newacronym{hsc}{HSC}{Hyper Suprime-Cam}
\newacronym{lsst}{LSST}{Large Synoptic Survey Telescope}
\newacronym{desi}{DESI}{Dark Energy Spectroscopic Instrument}
\newacronym{rsd}{RSD}{redshift-space distortions}
\newacronym{pte}{PTE}{probability to exceed}

\begin{document}

%
%
   \title{\Euclid preparation} \subtitle{CIV. Impact of galaxy intrinsic alignment modelling choices on \Euclid $3\times 2$\,pt cosmology} 
   
\newcommand{\orcid}[1]{} 
\author{Euclid Collaboration: D.~Navarro-Giron\'{e}s\orcid{0000-0003-0507-372X}\thanks{\email{navarrogirones@strw.leidenuniv.nl}}\inst{\ref{aff1},\ref{aff2},\ref{aff3}}
\and I.~Tutusaus\orcid{0000-0002-3199-0399}\inst{\ref{aff2},\ref{aff3},\ref{aff4}}
\and M.~Crocce\orcid{0000-0002-9745-6228}\inst{\ref{aff2},\ref{aff3}}
\and S.~Gouyou~Beauchamps\inst{\ref{aff3},\ref{aff2}}
\and R.~Paviot\orcid{0009-0002-8108-3460}\inst{\ref{aff5}}
\and B.~Joachimi\orcid{0000-0001-7494-1303}\inst{\ref{aff6}}
\and J.~Ruiz-Zapatero\orcid{0000-0002-7951-4391}\inst{\ref{aff6}}
\and D.~Sciotti\orcid{0009-0008-4519-2620}\inst{\ref{aff7},\ref{aff8}}
\and N.~Tessore\orcid{0000-0002-9696-7931}\inst{\ref{aff9}}
\and J.~Blazek\orcid{0000-0002-4687-4657}\inst{\ref{aff10}}
\and G.~Ca\~nas-Herrera\orcid{0000-0003-2796-2149}\inst{\ref{aff11},\ref{aff1}}
\and P.~Carrilho\orcid{0000-0003-1339-0194}\inst{\ref{aff12}}
\and J.~M.~Coloma-Nadal\orcid{0009-0003-0538-4349}\inst{\ref{aff2}}
\and H.~Hoekstra\orcid{0000-0002-0641-3231}\inst{\ref{aff1}}
\and A.~Porredon\orcid{0000-0002-2762-2024}\inst{\ref{aff13}}
\and B.~Altieri\orcid{0000-0003-3936-0284}\inst{\ref{aff14}}
\and S.~Andreon\orcid{0000-0002-2041-8784}\inst{\ref{aff15}}
\and C.~Baccigalupi\orcid{0000-0002-8211-1630}\inst{\ref{aff16},\ref{aff17},\ref{aff18},\ref{aff19}}
\and M.~Baldi\orcid{0000-0003-4145-1943}\inst{\ref{aff20},\ref{aff21},\ref{aff22}}
\and S.~Bardelli\orcid{0000-0002-8900-0298}\inst{\ref{aff21}}
\and A.~Biviano\orcid{0000-0002-0857-0732}\inst{\ref{aff17},\ref{aff16}}
\and E.~Branchini\orcid{0000-0002-0808-6908}\inst{\ref{aff23},\ref{aff24},\ref{aff15}}
\and M.~Brescia\orcid{0000-0001-9506-5680}\inst{\ref{aff25},\ref{aff26}}
\and S.~Camera\orcid{0000-0003-3399-3574}\inst{\ref{aff27},\ref{aff28},\ref{aff29}}
\and V.~Capobianco\orcid{0000-0002-3309-7692}\inst{\ref{aff29}}
\and C.~Carbone\orcid{0000-0003-0125-3563}\inst{\ref{aff30}}
\and V.~F.~Cardone\inst{\ref{aff7},\ref{aff8}}
\and J.~Carretero\orcid{0000-0002-3130-0204}\inst{\ref{aff13},\ref{aff31}}
\and F.~J.~Castander\orcid{0000-0001-7316-4573}\inst{\ref{aff2},\ref{aff3}}
\and M.~Castellano\orcid{0000-0001-9875-8263}\inst{\ref{aff7}}
\and G.~Castignani\orcid{0000-0001-6831-0687}\inst{\ref{aff21}}
\and S.~Cavuoti\orcid{0000-0002-3787-4196}\inst{\ref{aff26},\ref{aff32}}
\and K.~C.~Chambers\orcid{0000-0001-6965-7789}\inst{\ref{aff33}}
\and C.~Colodro-Conde\inst{\ref{aff34}}
\and G.~Congedo\orcid{0000-0003-2508-0046}\inst{\ref{aff11}}
\and C.~J.~Conselice\orcid{0000-0003-1949-7638}\inst{\ref{aff35}}
\and L.~Conversi\orcid{0000-0002-6710-8476}\inst{\ref{aff36},\ref{aff14}}
\and Y.~Copin\orcid{0000-0002-5317-7518}\inst{\ref{aff37}}
\and F.~Courbin\orcid{0000-0003-0758-6510}\inst{\ref{aff38},\ref{aff39},\ref{aff40}}
\and H.~M.~Courtois\orcid{0000-0003-0509-1776}\inst{\ref{aff41}}
\and A.~Da~Silva\orcid{0000-0002-6385-1609}\inst{\ref{aff42},\ref{aff43}}
\and H.~Degaudenzi\orcid{0000-0002-5887-6799}\inst{\ref{aff44}}
\and G.~De~Lucia\orcid{0000-0002-6220-9104}\inst{\ref{aff17}}
\and H.~Dole\orcid{0000-0002-9767-3839}\inst{\ref{aff45}}
\and F.~Dubath\orcid{0000-0002-6533-2810}\inst{\ref{aff44}}
\and C.~A.~J.~Duncan\orcid{0009-0003-3573-0791}\inst{\ref{aff11}}
\and X.~Dupac\inst{\ref{aff14}}
\and S.~Escoffier\orcid{0000-0002-2847-7498}\inst{\ref{aff46}}
\and M.~Farina\orcid{0000-0002-3089-7846}\inst{\ref{aff47}}
\and R.~Farinelli\inst{\ref{aff21}}
\and S.~Farrens\orcid{0000-0002-9594-9387}\inst{\ref{aff5}}
\and S.~Ferriol\inst{\ref{aff37}}
\and F.~Finelli\orcid{0000-0002-6694-3269}\inst{\ref{aff21},\ref{aff48}}
\and P.~Fosalba\orcid{0000-0002-1510-5214}\inst{\ref{aff3},\ref{aff2}}
\and S.~Fotopoulou\orcid{0000-0002-9686-254X}\inst{\ref{aff49}}
\and N.~Fourmanoit\orcid{0009-0005-6816-6925}\inst{\ref{aff46}}
\and M.~Frailis\orcid{0000-0002-7400-2135}\inst{\ref{aff17}}
\and E.~Franceschi\orcid{0000-0002-0585-6591}\inst{\ref{aff21}}
\and M.~Fumana\orcid{0000-0001-6787-5950}\inst{\ref{aff30}}
\and S.~Galeotta\orcid{0000-0002-3748-5115}\inst{\ref{aff17}}
\and K.~George\orcid{0000-0002-1734-8455}\inst{\ref{aff50}}
\and B.~Gillis\orcid{0000-0002-4478-1270}\inst{\ref{aff11}}
\and C.~Giocoli\orcid{0000-0002-9590-7961}\inst{\ref{aff21},\ref{aff22}}
\and J.~Gracia-Carpio\orcid{0000-0003-4689-3134}\inst{\ref{aff51}}
\and A.~Grazian\orcid{0000-0002-5688-0663}\inst{\ref{aff52}}
\and F.~Grupp\inst{\ref{aff51},\ref{aff53}}
\and S.~V.~H.~Haugan\orcid{0000-0001-9648-7260}\inst{\ref{aff54}}
\and W.~Holmes\inst{\ref{aff55}}
\and F.~Hormuth\inst{\ref{aff56}}
\and A.~Hornstrup\orcid{0000-0002-3363-0936}\inst{\ref{aff57},\ref{aff58}}
\and K.~Jahnke\orcid{0000-0003-3804-2137}\inst{\ref{aff59}}
\and S.~Kermiche\orcid{0000-0002-0302-5735}\inst{\ref{aff46}}
\and A.~Kiessling\orcid{0000-0002-2590-1273}\inst{\ref{aff55}}
\and M.~Kilbinger\orcid{0000-0001-9513-7138}\inst{\ref{aff5}}
\and B.~Kubik\orcid{0009-0006-5823-4880}\inst{\ref{aff37}}
\and K.~Kuijken\orcid{0000-0002-3827-0175}\inst{\ref{aff1}}
\and M.~Kunz\orcid{0000-0002-3052-7394}\inst{\ref{aff60}}
\and H.~Kurki-Suonio\orcid{0000-0002-4618-3063}\inst{\ref{aff61},\ref{aff62}}
\and A.~M.~C.~Le~Brun\orcid{0000-0002-0936-4594}\inst{\ref{aff63}}
\and S.~Ligori\orcid{0000-0003-4172-4606}\inst{\ref{aff29}}
\and P.~B.~Lilje\orcid{0000-0003-4324-7794}\inst{\ref{aff54}}
\and V.~Lindholm\orcid{0000-0003-2317-5471}\inst{\ref{aff61},\ref{aff62}}
\and I.~Lloro\orcid{0000-0001-5966-1434}\inst{\ref{aff64}}
\and G.~Mainetti\orcid{0000-0003-2384-2377}\inst{\ref{aff65}}
\and O.~Mansutti\orcid{0000-0001-5758-4658}\inst{\ref{aff17}}
\and O.~Marggraf\orcid{0000-0001-7242-3852}\inst{\ref{aff66}}
\and M.~Martinelli\orcid{0000-0002-6943-7732}\inst{\ref{aff7},\ref{aff8}}
\and N.~Martinet\orcid{0000-0003-2786-7790}\inst{\ref{aff67}}
\and F.~Marulli\orcid{0000-0002-8850-0303}\inst{\ref{aff68},\ref{aff21},\ref{aff22}}
\and E.~Medinaceli\orcid{0000-0002-4040-7783}\inst{\ref{aff21}}
\and M.~Meneghetti\orcid{0000-0003-1225-7084}\inst{\ref{aff21},\ref{aff22}}
\and E.~Merlin\orcid{0000-0001-6870-8900}\inst{\ref{aff7}}
\and G.~Meylan\inst{\ref{aff69}}
\and A.~Mora\orcid{0000-0002-1922-8529}\inst{\ref{aff70}}
\and M.~Moresco\orcid{0000-0002-7616-7136}\inst{\ref{aff68},\ref{aff21}}
\and L.~Moscardini\orcid{0000-0002-3473-6716}\inst{\ref{aff68},\ref{aff21},\ref{aff22}}
\and R.~Nakajima\orcid{0009-0009-1213-7040}\inst{\ref{aff66}}
\and C.~Neissner\orcid{0000-0001-8524-4968}\inst{\ref{aff71},\ref{aff31}}
\and S.-M.~Niemi\orcid{0009-0005-0247-0086}\inst{\ref{aff72}}
\and C.~Padilla\orcid{0000-0001-7951-0166}\inst{\ref{aff71}}
\and S.~Paltani\orcid{0000-0002-8108-9179}\inst{\ref{aff44}}
\and F.~Pasian\orcid{0000-0002-4869-3227}\inst{\ref{aff17}}
\and K.~Pedersen\inst{\ref{aff73}}
\and W.~J.~Percival\orcid{0000-0002-0644-5727}\inst{\ref{aff74},\ref{aff75},\ref{aff76}}
\and V.~Pettorino\orcid{0000-0002-4203-9320}\inst{\ref{aff72}}
\and S.~Pires\orcid{0000-0002-0249-2104}\inst{\ref{aff5}}
\and G.~Polenta\orcid{0000-0003-4067-9196}\inst{\ref{aff77}}
\and M.~Poncet\inst{\ref{aff78}}
\and L.~A.~Popa\inst{\ref{aff79}}
\and F.~Raison\orcid{0000-0002-7819-6918}\inst{\ref{aff51}}
\and A.~Renzi\orcid{0000-0001-9856-1970}\inst{\ref{aff80},\ref{aff81},\ref{aff21}}
\and J.~Rhodes\orcid{0000-0002-4485-8549}\inst{\ref{aff55}}
\and G.~Riccio\inst{\ref{aff26}}
\and E.~Romelli\orcid{0000-0003-3069-9222}\inst{\ref{aff17}}
\and M.~Roncarelli\orcid{0000-0001-9587-7822}\inst{\ref{aff21}}
\and C.~Rosset\orcid{0000-0003-0286-2192}\inst{\ref{aff82}}
\and R.~Saglia\orcid{0000-0003-0378-7032}\inst{\ref{aff53},\ref{aff51}}
\and Z.~Sakr\orcid{0000-0002-4823-3757}\inst{\ref{aff83},\ref{aff4},\ref{aff84}}
\and A.~G.~S\'anchez\orcid{0000-0003-1198-831X}\inst{\ref{aff51}}
\and D.~Sapone\orcid{0000-0001-7089-4503}\inst{\ref{aff85}}
\and B.~Sartoris\orcid{0000-0003-1337-5269}\inst{\ref{aff53},\ref{aff17}}
\and P.~Schneider\orcid{0000-0001-8561-2679}\inst{\ref{aff66}}
\and T.~Schrabback\orcid{0000-0002-6987-7834}\inst{\ref{aff86}}
\and A.~Secroun\orcid{0000-0003-0505-3710}\inst{\ref{aff46}}
\and G.~Seidel\orcid{0000-0003-2907-353X}\inst{\ref{aff59}}
\and E.~Sihvola\orcid{0000-0003-1804-7715}\inst{\ref{aff87}}
\and P.~Simon\inst{\ref{aff66}}
\and C.~Sirignano\orcid{0000-0002-0995-7146}\inst{\ref{aff80},\ref{aff81}}
\and G.~Sirri\orcid{0000-0003-2626-2853}\inst{\ref{aff22}}
\and A.~Spurio~Mancini\orcid{0000-0001-5698-0990}\inst{\ref{aff88}}
\and L.~Stanco\orcid{0000-0002-9706-5104}\inst{\ref{aff81}}
\and P.~Tallada-Cresp\'{i}\orcid{0000-0002-1336-8328}\inst{\ref{aff13},\ref{aff31}}
\and I.~Tereno\orcid{0000-0002-4537-6218}\inst{\ref{aff42},\ref{aff89}}
\and S.~Toft\orcid{0000-0003-3631-7176}\inst{\ref{aff90},\ref{aff91}}
\and R.~Toledo-Moreo\orcid{0000-0002-2997-4859}\inst{\ref{aff92}}
\and F.~Torradeflot\orcid{0000-0003-1160-1517}\inst{\ref{aff31},\ref{aff13}}
\and J.~Valiviita\orcid{0000-0001-6225-3693}\inst{\ref{aff61},\ref{aff62}}
\and T.~Vassallo\orcid{0000-0001-6512-6358}\inst{\ref{aff17},\ref{aff50}}
\and G.~Verdoes~Kleijn\orcid{0000-0001-5803-2580}\inst{\ref{aff93}}
\and Y.~Wang\orcid{0000-0002-4749-2984}\inst{\ref{aff94}}
\and J.~Weller\orcid{0000-0002-8282-2010}\inst{\ref{aff53},\ref{aff51}}
\and F.~M.~Zerbi\orcid{0000-0002-9996-973X}\inst{\ref{aff15}}
\and E.~Zucca\orcid{0000-0002-5845-8132}\inst{\ref{aff21}}
\and M.~Ballardini\orcid{0000-0003-4481-3559}\inst{\ref{aff95},\ref{aff96},\ref{aff21}}
\and M.~Bolzonella\orcid{0000-0003-3278-4607}\inst{\ref{aff21}}
\and E.~Bozzo\orcid{0000-0002-8201-1525}\inst{\ref{aff44}}
\and C.~Burigana\orcid{0000-0002-3005-5796}\inst{\ref{aff97},\ref{aff48}}
\and R.~Cabanac\orcid{0000-0001-6679-2600}\inst{\ref{aff4}}
\and M.~Calabrese\orcid{0000-0002-2637-2422}\inst{\ref{aff98},\ref{aff30}}
\and A.~Cappi\inst{\ref{aff99},\ref{aff21}}
\and T.~Castro\orcid{0000-0002-6292-3228}\inst{\ref{aff17},\ref{aff18},\ref{aff16},\ref{aff100}}
\and J.~A.~Escartin~Vigo\inst{\ref{aff51}}
\and L.~Gabarra\orcid{0000-0002-8486-8856}\inst{\ref{aff101}}
\and J.~Garc\'ia-Bellido\orcid{0000-0002-9370-8360}\inst{\ref{aff83}}
\and J.~Macias-Perez\orcid{0000-0002-5385-2763}\inst{\ref{aff102}}
\and R.~Maoli\orcid{0000-0002-6065-3025}\inst{\ref{aff103},\ref{aff7}}
\and J.~Mart\'{i}n-Fleitas\orcid{0000-0002-8594-569X}\inst{\ref{aff104}}
\and N.~Mauri\orcid{0000-0001-8196-1548}\inst{\ref{aff105},\ref{aff22}}
\and R.~B.~Metcalf\orcid{0000-0003-3167-2574}\inst{\ref{aff68},\ref{aff21}}
\and P.~Monaco\orcid{0000-0003-2083-7564}\inst{\ref{aff106},\ref{aff17},\ref{aff18},\ref{aff16}}
\and A.~Pezzotta\orcid{0000-0003-0726-2268}\inst{\ref{aff15}}
\and M.~P\"ontinen\orcid{0000-0001-5442-2530}\inst{\ref{aff61}}
\and I.~Risso\orcid{0000-0003-2525-7761}\inst{\ref{aff15},\ref{aff24}}
\and V.~Scottez\orcid{0009-0008-3864-940X}\inst{\ref{aff107},\ref{aff108}}
\and M.~Sereno\orcid{0000-0003-0302-0325}\inst{\ref{aff21},\ref{aff22}}
\and M.~Tenti\orcid{0000-0002-4254-5901}\inst{\ref{aff22}}
\and M.~Tucci\inst{\ref{aff44}}
\and M.~Viel\orcid{0000-0002-2642-5707}\inst{\ref{aff16},\ref{aff17},\ref{aff19},\ref{aff18},\ref{aff100}}
\and M.~Wiesmann\orcid{0009-0000-8199-5860}\inst{\ref{aff54}}
\and Y.~Akrami\orcid{0000-0002-2407-7956}\inst{\ref{aff83},\ref{aff109}}
\and I.~T.~Andika\orcid{0000-0001-6102-9526}\inst{\ref{aff50}}
\and G.~Angora\orcid{0000-0002-0316-6562}\inst{\ref{aff26},\ref{aff95}}
\and S.~Anselmi\orcid{0000-0002-3579-9583}\inst{\ref{aff81},\ref{aff80},\ref{aff110}}
\and M.~Archidiacono\orcid{0000-0003-4952-9012}\inst{\ref{aff111},\ref{aff112}}
\and F.~Atrio-Barandela\orcid{0000-0002-2130-2513}\inst{\ref{aff113}}
\and L.~Bazzanini\orcid{0000-0003-0727-0137}\inst{\ref{aff95},\ref{aff21}}
\and J.~Bel\inst{\ref{aff114}}
\and D.~Bertacca\orcid{0000-0002-2490-7139}\inst{\ref{aff80},\ref{aff52},\ref{aff81}}
\and M.~Bethermin\orcid{0000-0002-3915-2015}\inst{\ref{aff115}}
\and F.~Beutler\orcid{0000-0003-0467-5438}\inst{\ref{aff11}}
\and A.~Blanchard\orcid{0000-0001-8555-9003}\inst{\ref{aff4}}
\and L.~Blot\orcid{0000-0002-9622-7167}\inst{\ref{aff116},\ref{aff63}}
\and M.~Bonici\orcid{0000-0002-8430-126X}\inst{\ref{aff74},\ref{aff30}}
\and M.~L.~Brown\orcid{0000-0002-0370-8077}\inst{\ref{aff35}}
\and S.~Bruton\orcid{0000-0002-6503-5218}\inst{\ref{aff117}}
\and B.~Camacho~Quevedo\orcid{0000-0002-8789-4232}\inst{\ref{aff16},\ref{aff19},\ref{aff17}}
\and F.~Caro\orcid{0009-0003-1053-0507}\inst{\ref{aff7}}
\and C.~S.~Carvalho\inst{\ref{aff89}}
\and F.~Cogato\orcid{0000-0003-4632-6113}\inst{\ref{aff68},\ref{aff21}}
\and S.~Davini\orcid{0000-0003-3269-1718}\inst{\ref{aff24}}
\and F.~De~Paolis\orcid{0000-0001-6460-7563}\inst{\ref{aff118},\ref{aff119},\ref{aff120}}
\and G.~Desprez\orcid{0000-0001-8325-1742}\inst{\ref{aff93}}
\and A.~D\'iaz-S\'anchez\orcid{0000-0003-0748-4768}\inst{\ref{aff121}}
\and S.~Di~Domizio\orcid{0000-0003-2863-5895}\inst{\ref{aff23},\ref{aff24}}
\and J.~M.~Diego\orcid{0000-0001-9065-3926}\inst{\ref{aff122}}
\and P.~Dimauro\orcid{0000-0001-7399-2854}\inst{\ref{aff123},\ref{aff7}}
\and V.~Duret\orcid{0009-0009-0383-4960}\inst{\ref{aff46}}
\and M.~Y.~Elkhashab\orcid{0000-0001-9306-2603}\inst{\ref{aff106},\ref{aff17},\ref{aff18},\ref{aff16}}
\and Y.~Fang\orcid{0000-0002-0334-6950}\inst{\ref{aff53}}
\and P.~G.~Ferreira\orcid{0000-0002-3021-2851}\inst{\ref{aff101}}
\and A.~Finoguenov\orcid{0000-0002-4606-5403}\inst{\ref{aff61}}
\and A.~Franco\orcid{0000-0002-4761-366X}\inst{\ref{aff119},\ref{aff118},\ref{aff120}}
\and K.~Ganga\orcid{0000-0001-8159-8208}\inst{\ref{aff82}}
\and T.~Gasparetto\orcid{0000-0002-7913-4866}\inst{\ref{aff7}}
\and E.~Gaztanaga\orcid{0000-0001-9632-0815}\inst{\ref{aff2},\ref{aff3},\ref{aff124}}
\and F.~Giacomini\orcid{0000-0002-3129-2814}\inst{\ref{aff22}}
\and F.~Gianotti\orcid{0000-0003-4666-119X}\inst{\ref{aff21}}
\and E.~J.~Gonzalez\orcid{0000-0002-0226-9893}\inst{\ref{aff125},\ref{aff126}}
\and G.~Gozaliasl\orcid{0000-0002-0236-919X}\inst{\ref{aff127},\ref{aff61}}
\and A.~Gruppuso\orcid{0000-0001-9272-5292}\inst{\ref{aff21},\ref{aff22}}
\and M.~Guidi\orcid{0000-0001-9408-1101}\inst{\ref{aff20},\ref{aff21}}
\and C.~M.~Gutierrez\orcid{0000-0001-7854-783X}\inst{\ref{aff34},\ref{aff128}}
\and A.~Hall\orcid{0000-0002-3139-8651}\inst{\ref{aff11}}
\and C.~Hern\'andez-Monteagudo\orcid{0000-0001-5471-9166}\inst{\ref{aff128},\ref{aff34}}
\and H.~Hildebrandt\orcid{0000-0002-9814-3338}\inst{\ref{aff129}}
\and J.~Hjorth\orcid{0000-0002-4571-2306}\inst{\ref{aff73}}
\and J.~J.~E.~Kajava\orcid{0000-0002-3010-8333}\inst{\ref{aff130},\ref{aff131},\ref{aff132}}
\and Y.~Kang\orcid{0009-0000-8588-7250}\inst{\ref{aff44}}
\and V.~Kansal\orcid{0000-0002-4008-6078}\inst{\ref{aff133},\ref{aff134}}
\and D.~Karagiannis\orcid{0000-0002-4927-0816}\inst{\ref{aff95},\ref{aff135}}
\and K.~Kiiveri\inst{\ref{aff87}}
\and J.~Kim\orcid{0000-0003-2776-2761}\inst{\ref{aff101}}
\and C.~C.~Kirkpatrick\inst{\ref{aff87}}
\and S.~Kruk\orcid{0000-0001-8010-8879}\inst{\ref{aff14}}
\and J.~Le~Graet\orcid{0000-0001-6523-7971}\inst{\ref{aff46}}
\and L.~Legrand\orcid{0000-0003-0610-5252}\inst{\ref{aff136},\ref{aff137}}
\and M.~Lembo\orcid{0000-0002-5271-5070}\inst{\ref{aff138}}
\and F.~Lepori\orcid{0009-0000-5061-7138}\inst{\ref{aff139}}
\and G.~Leroy\orcid{0009-0004-2523-4425}\inst{\ref{aff140},\ref{aff141}}
\and G.~F.~Lesci\orcid{0000-0002-4607-2830}\inst{\ref{aff68},\ref{aff21}}
\and J.~Lesgourgues\orcid{0000-0001-7627-353X}\inst{\ref{aff142}}
\and T.~I.~Liaudat\orcid{0000-0002-9104-314X}\inst{\ref{aff143}}
\and M.~Magliocchetti\orcid{0000-0001-9158-4838}\inst{\ref{aff47}}
\and F.~Mannucci\orcid{0000-0002-4803-2381}\inst{\ref{aff144}}
\and C.~J.~A.~P.~Martins\orcid{0000-0002-4886-9261}\inst{\ref{aff145},\ref{aff146}}
\and L.~Maurin\orcid{0000-0002-8406-0857}\inst{\ref{aff45}}
\and M.~Miluzio\inst{\ref{aff14},\ref{aff147}}
\and A.~Montoro\orcid{0000-0003-4730-8590}\inst{\ref{aff2},\ref{aff3}}
\and C.~Moretti\orcid{0000-0003-3314-8936}\inst{\ref{aff17},\ref{aff16},\ref{aff18}}
\and G.~Morgante\inst{\ref{aff21}}
\and S.~Nadathur\orcid{0000-0001-9070-3102}\inst{\ref{aff124}}
\and K.~Naidoo\orcid{0000-0002-9182-1802}\inst{\ref{aff124},\ref{aff59}}
\and A.~Navarro-Alsina\orcid{0000-0002-3173-2592}\inst{\ref{aff66}}
\and S.~Nesseris\orcid{0000-0002-0567-0324}\inst{\ref{aff83}}
\and L.~Pagano\orcid{0000-0003-1820-5998}\inst{\ref{aff95},\ref{aff96}}
\and D.~Paoletti\orcid{0000-0003-4761-6147}\inst{\ref{aff21},\ref{aff48}}
\and F.~Passalacqua\orcid{0000-0002-8606-4093}\inst{\ref{aff80},\ref{aff81}}
\and K.~Paterson\orcid{0000-0001-8340-3486}\inst{\ref{aff59}}
\and L.~Patrizii\inst{\ref{aff22}}
\and C.~Pattison\orcid{0000-0003-3272-2617}\inst{\ref{aff124}}
\and A.~Pisani\orcid{0000-0002-6146-4437}\inst{\ref{aff46}}
\and D.~Potter\orcid{0000-0002-0757-5195}\inst{\ref{aff148}}
\and G.~W.~Pratt\inst{\ref{aff5}}
\and S.~Quai\orcid{0000-0002-0449-8163}\inst{\ref{aff68},\ref{aff21}}
\and M.~Radovich\orcid{0000-0002-3585-866X}\inst{\ref{aff52}}
\and K.~Rojas\orcid{0000-0003-1391-6854}\inst{\ref{aff149}}
\and W.~Roster\orcid{0000-0002-9149-6528}\inst{\ref{aff51}}
\and S.~Sacquegna\orcid{0000-0002-8433-6630}\inst{\ref{aff150}}
\and M.~Sahl\'en\orcid{0000-0003-0973-4804}\inst{\ref{aff151}}
\and D.~B.~Sanders\orcid{0000-0002-1233-9998}\inst{\ref{aff33}}
\and E.~Sarpa\orcid{0000-0002-1256-655X}\inst{\ref{aff19},\ref{aff100},\ref{aff17}}
\and A.~Schneider\orcid{0000-0001-7055-8104}\inst{\ref{aff148}}
\and E.~Sellentin\inst{\ref{aff152},\ref{aff1}}
\and L.~C.~Smith\orcid{0000-0002-3259-2771}\inst{\ref{aff153}}
\and K.~Tanidis\orcid{0000-0001-9843-5130}\inst{\ref{aff154}}
\and F.~Tarsitano\orcid{0000-0002-5919-0238}\inst{\ref{aff155},\ref{aff44}}
\and R.~Teyssier\orcid{0000-0001-7689-0933}\inst{\ref{aff156}}
\and A.~Troja\orcid{0000-0003-0239-4595}\inst{\ref{aff17}}
\and D.~Vergani\orcid{0000-0003-0898-2216}\inst{\ref{aff21}}
\and F.~Vernizzi\orcid{0000-0003-3426-2802}\inst{\ref{aff157}}
\and G.~Verza\orcid{0000-0002-1886-8348}\inst{\ref{aff158},\ref{aff159}}
\and P.~Vielzeuf\orcid{0000-0003-2035-9339}\inst{\ref{aff46}}
\and S.~Vinciguerra\orcid{0009-0005-4018-3184}\inst{\ref{aff67}}
\and N.~A.~Walton\orcid{0000-0003-3983-8778}\inst{\ref{aff153}}
\and A.~H.~Wright\orcid{0000-0001-7363-7932}\inst{\ref{aff129}}
\and S.-S.~Li\orcid{0000-0001-9952-7408}\inst{\ref{aff160},\ref{aff161}}}
                                                                                   
\institute{Leiden Observatory, Leiden University, Einsteinweg 55, 2333 CC Leiden, The Netherlands\label{aff1}
\and
Institute of Space Sciences (ICE, CSIC), Campus UAB, Carrer de Can Magrans, s/n, 08193 Barcelona, Spain\label{aff2}
\and
Institut d'Estudis Espacials de Catalunya (IEEC),  Edifici RDIT, Campus UPC, 08860 Castelldefels, Barcelona, Spain\label{aff3}
\and
Institut de Recherche en Astrophysique et Plan\'etologie (IRAP), Universit\'e de Toulouse, CNRS, UPS, CNES, 14 Av. Edouard Belin, 31400 Toulouse, France\label{aff4}
\and
Universit\'e Paris-Saclay, Universit\'e Paris Cit\'e, CEA, CNRS, AIM, 91191, Gif-sur-Yvette, France\label{aff5}
\and
Department of Physics and Astronomy, University College London, Gower Street, London WC1E 6BT, UK\label{aff6}
\and
INAF-Osservatorio Astronomico di Roma, Via Frascati 33, 00078 Monteporzio Catone, Italy\label{aff7}
\and
INFN-Sezione di Roma, Piazzale Aldo Moro, 2 - c/o Dipartimento di Fisica, Edificio G. Marconi, 00185 Roma, Italy\label{aff8}
\and
Mullard Space Science Laboratory, University College London, Holmbury St Mary, Dorking, Surrey RH5 6NT, UK\label{aff9}
\and
Department of Physics, Northeastern University, Boston, MA, 02115, USA\label{aff10}
\and
Institute for Astronomy, University of Edinburgh, Royal Observatory, Blackford Hill, Edinburgh EH9 3HJ, UK\label{aff11}
\and
Department of Physics, Astronomy and Mathematics, University of Hertfordshire, College Lane, Hatfield AL10 9AB, UK\label{aff12}
\and
Centro de Investigaciones Energ\'eticas, Medioambientales y Tecnol\'ogicas (CIEMAT), Avenida Complutense 40, 28040 Madrid, Spain\label{aff13}
\and
ESAC/ESA, Camino Bajo del Castillo, s/n., Urb. Villafranca del Castillo, 28692 Villanueva de la Ca\~nada, Madrid, Spain\label{aff14}
\and
INAF-Osservatorio Astronomico di Brera, Via Brera 28, 20122 Milano, Italy\label{aff15}
\and
IFPU, Institute for Fundamental Physics of the Universe, via Beirut 2, 34151 Trieste, Italy\label{aff16}
\and
INAF-Osservatorio Astronomico di Trieste, Via G. B. Tiepolo 11, 34143 Trieste, Italy\label{aff17}
\and
INFN, Sezione di Trieste, Via Valerio 2, 34127 Trieste TS, Italy\label{aff18}
\and
SISSA, International School for Advanced Studies, Via Bonomea 265, 34136 Trieste TS, Italy\label{aff19}
\and
Dipartimento di Fisica e Astronomia, Universit\`a di Bologna, Via Gobetti 93/2, 40129 Bologna, Italy\label{aff20}
\and
INAF-Osservatorio di Astrofisica e Scienza dello Spazio di Bologna, Via Piero Gobetti 93/3, 40129 Bologna, Italy\label{aff21}
\and
INFN-Sezione di Bologna, Viale Berti Pichat 6/2, 40127 Bologna, Italy\label{aff22}
\and
Dipartimento di Fisica, Universit\`a di Genova, Via Dodecaneso 33, 16146, Genova, Italy\label{aff23}
\and
INFN-Sezione di Genova, Via Dodecaneso 33, 16146, Genova, Italy\label{aff24}
\and
Department of Physics "E. Pancini", University Federico II, Via Cinthia 6, 80126, Napoli, Italy\label{aff25}
\and
INAF-Osservatorio Astronomico di Capodimonte, Via Moiariello 16, 80131 Napoli, Italy\label{aff26}
\and
Dipartimento di Fisica, Universit\`a degli Studi di Torino, Via P. Giuria 1, 10125 Torino, Italy\label{aff27}
\and
INFN-Sezione di Torino, Via P. Giuria 1, 10125 Torino, Italy\label{aff28}
\and
INAF-Osservatorio Astrofisico di Torino, Via Osservatorio 20, 10025 Pino Torinese (TO), Italy\label{aff29}
\and
INAF-IASF Milano, Via Alfonso Corti 12, 20133 Milano, Italy\label{aff30}
\and
Port d'Informaci\'{o} Cient\'{i}fica, Campus UAB, C. Albareda s/n, 08193 Bellaterra (Barcelona), Spain\label{aff31}
\and
INFN section of Naples, Via Cinthia 6, 80126, Napoli, Italy\label{aff32}
\and
Institute for Astronomy, University of Hawaii, 2680 Woodlawn Drive, Honolulu, HI 96822, USA\label{aff33}
\and
Instituto de Astrof\'{\i}sica de Canarias, E-38205 La Laguna, Tenerife, Spain\label{aff34}
\and
Jodrell Bank Centre for Astrophysics, Department of Physics and Astronomy, University of Manchester, Oxford Road, Manchester M13 9PL, UK\label{aff35}
\and
European Space Agency/ESRIN, Largo Galileo Galilei 1, 00044 Frascati, Roma, Italy\label{aff36}
\and
Universit\'e Claude Bernard Lyon 1, CNRS/IN2P3, IP2I Lyon, UMR 5822, Villeurbanne, F-69100, France\label{aff37}
\and
Institut de Ci\`{e}ncies del Cosmos (ICCUB), Universitat de Barcelona (IEEC-UB), Mart\'{i} i Franqu\`{e}s 1, 08028 Barcelona, Spain\label{aff38}
\and
Instituci\'o Catalana de Recerca i Estudis Avan\c{c}ats (ICREA), Passeig de Llu\'{\i}s Companys 23, 08010 Barcelona, Spain\label{aff39}
\and
Institut de Ciencies de l'Espai (IEEC-CSIC), Campus UAB, Carrer de Can Magrans, s/n Cerdanyola del Vall\'es, 08193 Barcelona, Spain\label{aff40}
\and
UCB Lyon 1, CNRS/IN2P3, IUF, IP2I Lyon, 4 rue Enrico Fermi, 69622 Villeurbanne, France\label{aff41}
\and
Departamento de F\'isica, Faculdade de Ci\^encias, Universidade de Lisboa, Edif\'icio C8, Campo Grande, PT1749-016 Lisboa, Portugal\label{aff42}
\and
Instituto de Astrof\'isica e Ci\^encias do Espa\c{c}o, Faculdade de Ci\^encias, Universidade de Lisboa, Campo Grande, 1749-016 Lisboa, Portugal\label{aff43}
\and
Department of Astronomy, University of Geneva, ch. d'Ecogia 16, 1290 Versoix, Switzerland\label{aff44}
\and
Universit\'e Paris-Saclay, CNRS, Institut d'astrophysique spatiale, 91405, Orsay, France\label{aff45}
\and
Aix-Marseille Universit\'e, CNRS/IN2P3, CPPM, Marseille, France\label{aff46}
\and
INAF-Istituto di Astrofisica e Planetologia Spaziali, via del Fosso del Cavaliere, 100, 00100 Roma, Italy\label{aff47}
\and
INFN-Bologna, Via Irnerio 46, 40126 Bologna, Italy\label{aff48}
\and
School of Physics, HH Wills Physics Laboratory, University of Bristol, Tyndall Avenue, Bristol, BS8 1TL, UK\label{aff49}
\and
University Observatory, LMU Faculty of Physics, Scheinerstr.~1, 81679 Munich, Germany\label{aff50}
\and
Max Planck Institute for Extraterrestrial Physics, Giessenbachstr. 1, 85748 Garching, Germany\label{aff51}
\and
INAF-Osservatorio Astronomico di Padova, Via dell'Osservatorio 5, 35122 Padova, Italy\label{aff52}
\and
Universit\"ats-Sternwarte M\"unchen, Fakult\"at f\"ur Physik, Ludwig-Maximilians-Universit\"at M\"unchen, Scheinerstr.~1, 81679 M\"unchen, Germany\label{aff53}
\and
Institute of Theoretical Astrophysics, University of Oslo, P.O. Box 1029 Blindern, 0315 Oslo, Norway\label{aff54}
\and
Jet Propulsion Laboratory, California Institute of Technology, 4800 Oak Grove Drive, Pasadena, CA, 91109, USA\label{aff55}
\and
Felix Hormuth Engineering, Goethestr. 17, 69181 Leimen, Germany\label{aff56}
\and
Technical University of Denmark, Elektrovej 327, 2800 Kgs. Lyngby, Denmark\label{aff57}
\and
Cosmic Dawn Center (DAWN), Denmark\label{aff58}
\and
Max-Planck-Institut f\"ur Astronomie, K\"onigstuhl 17, 69117 Heidelberg, Germany\label{aff59}
\and
Universit\'e de Gen\`eve, D\'epartement de Physique Th\'eorique and Centre for Astroparticle Physics, 24 quai Ernest-Ansermet, CH-1211 Gen\`eve 4, Switzerland\label{aff60}
\and
Department of Physics, P.O. Box 64, University of Helsinki, 00014 Helsinki, Finland\label{aff61}
\and
Helsinki Institute of Physics, Gustaf H{\"a}llstr{\"o}min katu 2, University of Helsinki, 00014 Helsinki, Finland\label{aff62}
\and
Laboratoire d'etude de l'Univers et des phenomenes eXtremes, Observatoire de Paris, Universit\'e PSL, Sorbonne Universit\'e, CNRS, 92190 Meudon, France\label{aff63}
\and
SKAO, Jodrell Bank, Lower Withington, Macclesfield SK11 9FT, UK\label{aff64}
\and
Centre de Calcul de l'IN2P3/CNRS, 21 avenue Pierre de Coubertin 69627 Villeurbanne Cedex, France\label{aff65}
\and
Universit\"at Bonn, Argelander-Institut f\"ur Astronomie, Auf dem H\"ugel 71, 53121 Bonn, Germany\label{aff66}
\and
Aix-Marseille Universit\'e, CNRS, CNES, LAM, Marseille, France\label{aff67}
\and
Dipartimento di Fisica e Astronomia "Augusto Righi" - Alma Mater Studiorum Universit\`a di Bologna, via Piero Gobetti 93/2, 40129 Bologna, Italy\label{aff68}
\and
Institute of Physics, Laboratory of Astrophysics, Ecole Polytechnique F\'ed\'erale de Lausanne (EPFL), Observatoire de Sauverny, 1290 Versoix, Switzerland\label{aff69}
\and
Telespazio UK S.L. for European Space Agency (ESA), Camino bajo del Castillo, s/n, Urbanizacion Villafranca del Castillo, Villanueva de la Ca\~nada, 28692 Madrid, Spain\label{aff70}
\and
Institut de F\'{i}sica d'Altes Energies (IFAE), The Barcelona Institute of Science and Technology, Campus UAB, 08193 Bellaterra (Barcelona), Spain\label{aff71}
\and
European Space Agency/ESTEC, Keplerlaan 1, 2201 AZ Noordwijk, The Netherlands\label{aff72}
\and
DARK, Niels Bohr Institute, University of Copenhagen, Jagtvej 155, 2200 Copenhagen, Denmark\label{aff73}
\and
Waterloo Centre for Astrophysics, University of Waterloo, Waterloo, Ontario N2L 3G1, Canada\label{aff74}
\and
Department of Physics and Astronomy, University of Waterloo, Waterloo, Ontario N2L 3G1, Canada\label{aff75}
\and
Perimeter Institute for Theoretical Physics, Waterloo, Ontario N2L 2Y5, Canada\label{aff76}
\and
Space Science Data Center, Italian Space Agency, via del Politecnico snc, 00133 Roma, Italy\label{aff77}
\and
Centre National d'Etudes Spatiales -- Centre spatial de Toulouse, 18 avenue Edouard Belin, 31401 Toulouse Cedex 9, France\label{aff78}
\and
Institute of Space Science, Str. Atomistilor, nr. 409 M\u{a}gurele, Ilfov, 077125, Romania\label{aff79}
\and
Dipartimento di Fisica e Astronomia "G. Galilei", Universit\`a di Padova, Via Marzolo 8, 35131 Padova, Italy\label{aff80}
\and
INFN-Padova, Via Marzolo 8, 35131 Padova, Italy\label{aff81}
\and
Universit\'e Paris Cit\'e, CNRS, Astroparticule et Cosmologie, 75013 Paris, France\label{aff82}
\and
Instituto de F\'isica Te\'orica UAM-CSIC, Campus de Cantoblanco, 28049 Madrid, Spain\label{aff83}
\and
Universit\'e St Joseph; Faculty of Sciences, Beirut, Lebanon\label{aff84}
\and
Departamento de F\'isica, FCFM, Universidad de Chile, Blanco Encalada 2008, Santiago, Chile\label{aff85}
\and
Universit\"at Innsbruck, Institut f\"ur Astro- und Teilchenphysik, Technikerstr. 25/8, 6020 Innsbruck, Austria\label{aff86}
\and
Department of Physics and Helsinki Institute of Physics, Gustaf H\"allstr\"omin katu 2, University of Helsinki, 00014 Helsinki, Finland\label{aff87}
\and
Department of Physics, Royal Holloway, University of London, Surrey TW20 0EX, UK\label{aff88}
\and
Instituto de Astrof\'isica e Ci\^encias do Espa\c{c}o, Faculdade de Ci\^encias, Universidade de Lisboa, Tapada da Ajuda, 1349-018 Lisboa, Portugal\label{aff89}
\and
Cosmic Dawn Center (DAWN)\label{aff90}
\and
Niels Bohr Institute, University of Copenhagen, Jagtvej 128, 2200 Copenhagen, Denmark\label{aff91}
\and
Universidad Polit\'ecnica de Cartagena, Departamento de Electr\'onica y Tecnolog\'ia de Computadoras,  Plaza del Hospital 1, 30202 Cartagena, Spain\label{aff92}
\and
Kapteyn Astronomical Institute, University of Groningen, PO Box 800, 9700 AV Groningen, The Netherlands\label{aff93}
\and
Caltech/IPAC, 1200 E. California Blvd., Pasadena, CA 91125, USA\label{aff94}
\and
Dipartimento di Fisica e Scienze della Terra, Universit\`a degli Studi di Ferrara, Via Giuseppe Saragat 1, 44122 Ferrara, Italy\label{aff95}
\and
Istituto Nazionale di Fisica Nucleare, Sezione di Ferrara, Via Giuseppe Saragat 1, 44122 Ferrara, Italy\label{aff96}
\and
INAF, Istituto di Radioastronomia, Via Piero Gobetti 101, 40129 Bologna, Italy\label{aff97}
\and
Astronomical Observatory of the Autonomous Region of the Aosta Valley (OAVdA), Loc. Lignan 39, I-11020, Nus (Aosta Valley), Italy\label{aff98}
\and
Universit\'e C\^{o}te d'Azur, Observatoire de la C\^{o}te d'Azur, CNRS, Laboratoire Lagrange, Bd de l'Observatoire, CS 34229, 06304 Nice cedex 4, France\label{aff99}
\and
ICSC - Centro Nazionale di Ricerca in High Performance Computing, Big Data e Quantum Computing, Via Magnanelli 2, Bologna, Italy\label{aff100}
\and
Department of Physics, Oxford University, Keble Road, Oxford OX1 3RH, UK\label{aff101}
\and
Univ. Grenoble Alpes, CNRS, Grenoble INP, LPSC-IN2P3, 53, Avenue des Martyrs, 38000, Grenoble, France\label{aff102}
\and
Dipartimento di Fisica, Sapienza Universit\`a di Roma, Piazzale Aldo Moro 2, 00185 Roma, Italy\label{aff103}
\and
Aurora Technology for European Space Agency (ESA), Camino bajo del Castillo, s/n, Urbanizacion Villafranca del Castillo, Villanueva de la Ca\~nada, 28692 Madrid, Spain\label{aff104}
\and
Dipartimento di Fisica e Astronomia "Augusto Righi" - Alma Mater Studiorum Universit\`a di Bologna, Viale Berti Pichat 6/2, 40127 Bologna, Italy\label{aff105}
\and
Dipartimento di Fisica - Sezione di Astronomia, Universit\`a di Trieste, Via Tiepolo 11, 34131 Trieste, Italy\label{aff106}
\and
Institut d'Astrophysique de Paris, 98bis Boulevard Arago, 75014, Paris, France\label{aff107}
\and
ICL, Junia, Universit\'e Catholique de Lille, LITL, 59000 Lille, France\label{aff108}
\and
CERCA/ISO, Department of Physics, Case Western Reserve University, 10900 Euclid Avenue, Cleveland, OH 44106, USA\label{aff109}
\and
Laboratoire Univers et Th\'eorie, Observatoire de Paris, Universit\'e PSL, Universit\'e Paris Cit\'e, CNRS, 92190 Meudon, France\label{aff110}
\and
Dipartimento di Fisica "Aldo Pontremoli", Universit\`a degli Studi di Milano, Via Celoria 16, 20133 Milano, Italy\label{aff111}
\and
INFN-Sezione di Milano, Via Celoria 16, 20133 Milano, Italy\label{aff112}
\and
Departamento de F{\'\i}sica Fundamental. Universidad de Salamanca. Plaza de la Merced s/n. 37008 Salamanca, Spain\label{aff113}
\and
Aix-Marseille Universit\'e, Universit\'e de Toulon, CNRS, CPT, Marseille, France\label{aff114}
\and
Universit\'e de Strasbourg, CNRS, Observatoire astronomique de Strasbourg, UMR 7550, 67000 Strasbourg, France\label{aff115}
\and
Center for Data-Driven Discovery, Kavli IPMU (WPI), UTIAS, The University of Tokyo, Kashiwa, Chiba 277-8583, Japan\label{aff116}
\and
California Institute of Technology, 1200 E California Blvd, Pasadena, CA 91125, USA\label{aff117}
\and
Department of Mathematics and Physics E. De Giorgi, University of Salento, Via per Arnesano, CP-I93, 73100, Lecce, Italy\label{aff118}
\and
INFN, Sezione di Lecce, Via per Arnesano, CP-193, 73100, Lecce, Italy\label{aff119}
\and
INAF-Sezione di Lecce, c/o Dipartimento Matematica e Fisica, Via per Arnesano, 73100, Lecce, Italy\label{aff120}
\and
Departamento F\'isica Aplicada, Universidad Polit\'ecnica de Cartagena, Campus Muralla del Mar, 30202 Cartagena, Murcia, Spain\label{aff121}
\and
Instituto de F\'isica de Cantabria, Edificio Juan Jord\'a, Avenida de los Castros, 39005 Santander, Spain\label{aff122}
\and
Observatorio Nacional, Rua General Jose Cristino, 77-Bairro Imperial de Sao Cristovao, Rio de Janeiro, 20921-400, Brazil\label{aff123}
\and
Institute of Cosmology and Gravitation, University of Portsmouth, Portsmouth PO1 3FX, UK\label{aff124}
\and
Departament de F\'{\i}sica, Universitat Aut\`onoma de Barcelona, 08193 Bellaterra (Barcelona), Spain\label{aff125}
\and
Instituto de Astronomia Teorica y Experimental (IATE-CONICET), Laprida 854, X5000BGR, C\'ordoba, Argentina\label{aff126}
\and
Department of Computer Science, Aalto University, PO Box 15400, Espoo, FI-00 076, Finland\label{aff127}
\and
Universidad de La Laguna, Dpto. Astrof\'\i sica, E-38206 La Laguna, Tenerife, Spain\label{aff128}
\and
Ruhr University Bochum, Faculty of Physics and Astronomy, Astronomical Institute (AIRUB), German Centre for Cosmological Lensing (GCCL), 44780 Bochum, Germany\label{aff129}
\and
Department of Physics and Astronomy, Vesilinnantie 5, University of Turku, 20014 Turku, Finland\label{aff130}
\and
Finnish Centre for Astronomy with ESO (FINCA), Quantum, Vesilinnantie 5, University of Turku, 20014 Turku, Finland\label{aff131}
\and
Serco for European Space Agency (ESA), Camino bajo del Castillo, s/n, Urbanizacion Villafranca del Castillo, Villanueva de la Ca\~nada, 28692 Madrid, Spain\label{aff132}
\and
ARC Centre of Excellence for Dark Matter Particle Physics, Melbourne, Australia\label{aff133}
\and
Centre for Astrophysics \& Supercomputing, Swinburne University of Technology,  Hawthorn, Victoria 3122, Australia\label{aff134}
\and
Department of Physics and Astronomy, University of the Western Cape, Bellville, Cape Town, 7535, South Africa\label{aff135}
\and
DAMTP, Centre for Mathematical Sciences, Wilberforce Road, Cambridge CB3 0WA, UK\label{aff136}
\and
Kavli Institute for Cosmology Cambridge, Madingley Road, Cambridge, CB3 0HA, UK\label{aff137}
\and
Institut d'Astrophysique de Paris, UMR 7095, CNRS, and Sorbonne Universit\'e, 98 bis boulevard Arago, 75014 Paris, France\label{aff138}
\and
Departement of Theoretical Physics, University of Geneva, Switzerland\label{aff139}
\and
Department of Physics, Centre for Extragalactic Astronomy, Durham University, South Road, Durham, DH1 3LE, UK\label{aff140}
\and
Department of Physics, Institute for Computational Cosmology, Durham University, South Road, Durham, DH1 3LE, UK\label{aff141}
\and
Institute for Theoretical Particle Physics and Cosmology (TTK), RWTH Aachen University, 52056 Aachen, Germany\label{aff142}
\and
IRFU, CEA, Universit\'e Paris-Saclay 91191 Gif-sur-Yvette Cedex, France\label{aff143}
\and
INAF-Osservatorio Astrofisico di Arcetri, Largo E. Fermi 5, 50125, Firenze, Italy\label{aff144}
\and
Centro de Astrof\'{\i}sica da Universidade do Porto, Rua das Estrelas, 4150-762 Porto, Portugal\label{aff145}
\and
Instituto de Astrof\'isica e Ci\^encias do Espa\c{c}o, Universidade do Porto, CAUP, Rua das Estrelas, PT4150-762 Porto, Portugal\label{aff146}
\and
HE Space for European Space Agency (ESA), Camino bajo del Castillo, s/n, Urbanizacion Villafranca del Castillo, Villanueva de la Ca\~nada, 28692 Madrid, Spain\label{aff147}
\and
Department of Astrophysics, University of Zurich, Winterthurerstrasse 190, 8057 Zurich, Switzerland\label{aff148}
\and
University of Applied Sciences and Arts of Northwestern Switzerland, School of Computer Science, 5210 Windisch, Switzerland\label{aff149}
\and
INAF - Osservatorio Astronomico d'Abruzzo, Via Maggini, 64100, Teramo, Italy\label{aff150}
\and
Theoretical astrophysics, Department of Physics and Astronomy, Uppsala University, Box 516, 751 37 Uppsala, Sweden\label{aff151}
\and
Mathematical Institute, University of Leiden, Einsteinweg 55, 2333 CA Leiden, The Netherlands\label{aff152}
\and
Institute of Astronomy, University of Cambridge, Madingley Road, Cambridge CB3 0HA, UK\label{aff153}
\and
Center for Astrophysics and Cosmology, University of Nova Gorica, Nova Gorica, Slovenia\label{aff154}
\and
Institute for Particle Physics and Astrophysics, Dept. of Physics, ETH Zurich, Wolfgang-Pauli-Strasse 27, 8093 Zurich, Switzerland\label{aff155}
\and
Department of Astrophysical Sciences, Peyton Hall, Princeton University, Princeton, NJ 08544, USA\label{aff156}
\and
Institut de Physique Th\'eorique, CEA, CNRS, Universit\'e Paris-Saclay 91191 Gif-sur-Yvette Cedex, France\label{aff157}
\and
International Centre for Theoretical Physics (ICTP), Strada Costiera 11, 34151 Trieste, Italy\label{aff158}
\and
Center for Computational Astrophysics, Flatiron Institute, 162 5th Avenue, 10010, New York, NY, USA\label{aff159}
\and
Kavli Institute for Particle Astrophysics \& Cosmology (KIPAC), Stanford University, Stanford, CA 94305, USA\label{aff160}
\and
SLAC National Accelerator Laboratory, 2575 Sand Hill Road, Menlo Park, CA 94025, USA\label{aff161}}


%
%
   \abstract{The \Euclid galaxy survey will provide unprecedented constraints on cosmology, but achieving unbiased results will require an optimal characterisation and mitigation of systematic effects. The \glspl{ia} of galaxies are one of the dominant contaminants of the \gls{wl} and \gls{ggl} probes. In this work, we assess \gls{ia} modelling choices for \Euclid DR1 $3\times 2$\,pt analyses by using synthetic data vectors and comparing the performance of the two most commonly used \gls{ia} models, \gls{nla} and \gls{tatt}, along with several variations. Our analyses combine three perspectives: (i) the constraining power on the \gls{ia} and cosmological parameters for each \gls{ia} model, (ii) the bias that results when the \gls{ia} analysis model differs from the model used to generate the synthetic data vector, and (iii) the degeneracies between \glspl{ia} and \gls{photo-z} nuisance parameters. Amongst the \gls{ia} models analysed, the redshift-dependent TATT model (\tattz) provides the most flexible description of \glspl{ia}, with a constraining power similar to simpler \gls{ia} models, making it a suitable choice for \Euclid DR1 $3\times 2$\,pt analyses.}
%
%
%
%
   \titlerunning{Intrinsic alignments on \Euclid $3\times 2$\,pt cosmology}
   \authorrunning{Euclid Collaboration: D. Navarro-Gironés et al.}
   
   \maketitle
%
%
%
%
\glsresetall
\nolinenumbers
\section{Introduction}\label{sec:Introduction}

Over the past decade, large galaxy surveys have provided vast amounts of data, allowing us to map the cosmic \gls{lss} with unprecedented precision. A key probe of this structure is weak lensing (\glsentryshort{wl}\glsunset{wl}; \citealt{cosmic_shear_review_1, cosmic_shear_review_2}), which measures coherent distortions in the shapes of background galaxies caused by the intervening matter distribution. When combined with \gls{gc} and \gls{ggl}, these measurements form the so-called $3\times 2$\,pt analysis, which has been identified as one of the most powerful tools for constraining dark energy and testing cosmological models \citep{DE_task_force}.

The first $3\times 2$\,pt cosmological analyses -- performed by the Kilo-Degree Survey (\glsentryshort{kids}\glsunset{kids}; \citealt{KiDS_1000}), the Dark Energy Survey (\glsentryshort{des}\glsunset{des}; \citealt{DES_overview}), and the Hyper Suprime-Cam (\glsentryshort{hsc}\glsunset{hsc}; \citealt{HSC_overview}) -- provided competitive constraints on the $\Lambda$ cold dark matter ($\Lambda$CDM) and $w_{0}w_{a}$CDM cosmological models \citep{KiDS_3x2pt, DES_Y3_3x2pt, HSC_3x2pt}. Despite their impressive depth and area, these surveys remain largely statistically limited. However, this will no longer be the case for the new generation of galaxy surveys, such as \Euclid \citep{Euclid_overview}, the Vera C. Rubin Observatory Large Synoptic Survey Telescope (\glsentryshort{lsst}\glsunset{lsst}; \citealt{LSST_overview}), the Dark Energy Spectroscopic Instrument (\glsentryshort{desi}\glsunset{desi}; \citealt{DESI_overview}), and the \textit{Nancy Grace Roman} Space Telescope \citep{Roman_overview}, which will reach fainter magnitudes and cover even larger areas. For these surveys, statistical uncertainties will be sub-dominant, and cosmological inference will instead be limited by systematic effects if they are not accurately modelled.

Multiple systematic sources affect $3\times 2$\,pt analyses \citep{CLOE_blot}, including observational systematics, such as redshift uncertainties and shape estimation biases. While \Euclid is designed to minimise these, analyses can also be impacted by astrophysical systematics, such as magnification, baryonic effects, and the \glspl{ia} of galaxies. Here, we focus on \glspl{ia} (see \citealt{IA_review} for a review on the topic), one of the most challenging contaminants, arising from the intrinsic correlations in galaxy orientations that are caused by gravitational interactions with the surrounding \gls{lss}. These contaminants can mimic \gls{wl} and \gls{ggl} signals and thus bias cosmological constraints.

A variety of \gls{ia} models aim to describe this effect. The most popular are the non-linear alignment (\glsentryshort{nla}\glsunset{nla}; \citealt{NLA_1, NLA_2}) and the tidal alignment tidal torquing (\glsentryshort{tatt}\glsunset{tatt}; \citealt{TATT_blazek}) models, which are based on perturbative expansions of the gravitational tidal field, up to the first and second order, respectively, and can be extended into hybrid models through evaluation using non-linear matter power spectra. Other more complex models attempt to describe \glspl{ia} by performing perturbation theory in Lagrangian \citep{IA_model_lagrangian_1, IA_model_lagrangian_2}, hybrid Lagrangian \citep{IA_model_hybrid}, or Eulerian \citep{IA_model_eulerian_1, IA_model_eulerian_2, IA_model_eulerian_3} space. Extensions, such as the halo model \citep{halo_model_IA_schneider, halo_model_IA_fortuna}, can be used to describe \glspl{ia} at small scales. To date, the \gls{nla} and the \gls{tatt} models have been the ones adopted in $3\times 2$\,pt and \gls{wl}-only analyses in previous surveys due to their ability to describe \glspl{ia} based on the uncertainties of these surveys. In particular, the \gls{tatt} model was adopted by \gls{des} for the analysis of their $3\times 2$\,pt Y3 and Y6 results \citep{DES_Y3_3x2pt, DES_Y6_3x2pt}, while \gls{nla} (and variations of it) was adopted in the \gls{des} Y1 results \citep{DES_Y1_3x2pt}, the combined \gls{kids} and \gls{des} results \citep{DES_KiDS}, and in \gls{wl}-only studies by \gls{kids} \citep{NLA_kids_1, NLA_kids_2, NLA_kids_3, KiDS_Wright_cosmic_shear} and \gls{des} Y6 \citep{DES_Y6_WL}. Additionally, both the \gls{nla} and the \gls{tatt} models (and variations of them) have been employed to perform \gls{lsst} and \Euclid forecasts, assuming a $w_{0}w_{a}$CDM cosmology \citep{DES_w0wa_stage_4, TATT_blazek}.

This is the third of a series of three papers aimed at studying \glspl{ia} in \Euclid. The first paper, \citealt{Hoffmann_IA}, describes the implementation of \glspl{ia} in the Flagship simulation \citep{Flagship}. The second paper (\citealt{Paviot_IA}, hereafter P26) demonstrates that \gls{ia} models, particularly \gls{nla} and \gls{tatt}, are able to fit the \glspl{ia} from Flagship, providing similar constraints as in observations while also characterising the \gls{ia} signal for different galaxy properties. Building on these results, the main objective of this third paper is to inform the best \gls{ia} model choice for \textit{Euclid}’s first data release (DR1) analysis. 

We considered three different perspectives to accomplish this goal and employed \textit{Euclid}-like synthetic data vectors. We first studied the constraining power of \gls{nla} and \gls{tatt}. As \gls{tatt} is a more flexible model than \gls{nla}, it is expected to allow for a more accurate description of \glspl{ia} at small scales. However, if the small scales do not include significant additional information, the higher-order terms will not be constrained, reducing the constraining power due to the larger number of parameters with respect to \gls{nla}. Here, we want to quantify if this loss of constraining power is significant. Even though our fiducial setup consists of $3\times 2$\,pt analyses, we also included \gls{wl}-only and $2\times 2$\,pt (\gls{gc} and \gls{ggl}) constraining power analyses as well as some variations in the assumed cosmological model, scale cuts, and priors on the \gls{ia} parameters. Secondly, we quantified the impact of \gls{ia} mismodelling, which refers to the scenario where a data vector follows a certain \gls{ia} model but is modelled with a different one. Since we do not know the actual model that best describes \glspl{ia} in the Universe, it is necessary to quantify the potential biases that mismodelling \glspl{ia} can induce, not only in the \gls{ia} parameters but also in the cosmological ones \citep{Campos_model_selection, IA_mismodelling_samuroff, IA_mismodelling_Paopiamsap}. Finally, we evaluated the degeneracies between \gls{ia} and \gls{photo-z} estimates, since these degeneracies can mask or undermine the effect that \glspl{ia} have on cosmological analyses \citep{IA_degeneracies_shun_sheng, IA_degeneracies_fischbacher, Leonard_IA_photoz_forecast, IA_degeneracies_mcdonald}, and provided some recommendations on how to avoid them.

The paper is structured as follows. In Sect.~2, we present the modelling description used to describe the angular power spectra and the \glspl{ia} of galaxies. In Sect.~3, we introduce the forecasting pipeline, describing the likelihood, the sampled parameters, and their priors; the employed covariance matrix; and the scale cuts. In Sect.~4, we describe the measurement of the values of the \gls{ia} parameters from the Flagship simulation and how the synthetic data vectors are generated. The main results showing the constraining power, the \gls{ia} mismodelling effect, and the degeneracies of \glspl{ia} with \glspl{photo-z} are presented in Sect.~5. We finish with the conclusions in Sect.~6.

\section{Modelling}\label{sec:Modelling}

Here, we describe the theoretical setup for modelling a $3\times 2$\,pt analysis in harmonic space by defining the harmonic angular power spectra (Sect.~\ref{sec:Cls}). We also pay particular attention to different approaches to model \glspl{ia} (Sect.~\ref{sec:IA_modelling}).

\subsection{Angular power spectra}\label{sec:Cls}

We modelled the cross-correlation of different observables that are split in redshift bins. In the Limber approximation \citep{Limber_approximation}, assuming a flat universe, this can be accomplished by writing the angular power spectra in harmonic space as
\begin{equation}\label{eq:general_cl_limber}
    C_{ij}^{AB} (\ell) = \int_{0}^{\chi_{\mathrm{hor}}} {\rm d} \chi \frac{W_{i}^{A}(\chi)W_{j}^{B}(\chi)}{\chi^{2}} P_{AB} \left [\frac{\ell + 1/2}{\chi}, z(\chi) \right ]\,, 
\end{equation}
where \textit{A} and \textit{B} indicate the tracers that are correlated, and $P_{AB}(k, z)$ their corresponding power spectra, as summarised in Table~\ref{tab:power_spectra}. Here, $P_{\delta \delta}$ corresponds to the matter power spectrum, while $P_{\delta \mathrm{I}}$ and $P_{\mathrm{II}}$ are the \gls{ia} power spectra described in Sect.~\ref{sec:IA_modelling}. The conversions for $P_{\mathrm{g}\delta}$, $P_{\mathrm{gI}}$, and $P_{\mathrm{gg}}$ in Table~\ref{tab:power_spectra} assume a linear galaxy bias modelling approach \citep{Linear_galaxy_bias}, with $b_{1}$ the linear galaxy bias term\footnote{As justified in Sect.~\ref{sec:parameters_and_priors}, we leave more complex galaxy bias relations for future work.}, which relates the mean overdensities of galaxies and mass. The upper integration limit $\chi_{\rm hor}$ corresponds to the comoving distance $\chi$ at the horizon. The kernels $W_{i}^{A}(\chi)$ and $W_{j}^{B}(\chi)$ depend on the observables of interest, and the subscripts $i$ and $j$ to the redshift bins of the tracers.

The angular power spectrum in harmonic space for the observed cosmic shear (LL) consists of a purely \gls{wl} contribution ($\gamma \gamma$), a contribution from \glspl{ia} (${\rm II}$) and a combination of both (${\rm I \gamma}$ and ${\gamma \rm I}$). These contributions can be written as
\begin{equation}\label{eq:cl_wl}
    C_{ij}^{\rm LL}(\ell) = C_{ij}^{\gamma \gamma}(\ell) +
            C_{ij}^{\gamma\rm I}(\ell) + C_{ij}^{\rm I \gamma}(\ell) + C_{ij}^{\rm II}(\ell)\,,
\end{equation}
where the subscripts $i, j$ correspond to the source tomographic redshift bins. In the case of $A=\gamma$, the term $W_{i}^{\gamma}(\chi)$ corresponds to the \gls{wl} weighting function:
\begin{equation}\label{eq:weight_function_shear_Euclid}
    W_{i}^{\gamma}(\chi) =
            \frac{3H_0^{2}\Omega_{\rm m}}{2c^{2}}
            \frac{\chi}{a(\chi)}
            \int_{z}^{z_{\rm max}}{{\rm d}\chi^{\prime} n_{i}^{\rm S}[z(\chi^{\prime})]\frac{{\rm d}z}{{\rm d}\chi^{\prime}}
            \frac{\chi^{\prime} - \chi}{\chi'}}\,,
\end{equation}
with $H_0$ as the Hubble constant, $c$ as the speed of light, $a$ as the scale factor, $z_{\rm max}$ as the maximum redshift of the survey, and $n_{i}^{\rm S}$ as the source redshift distribution. In the case of $A= \rm{I}$, the kernel becomes
\begin{equation}\label{eq:weight_function_IA_Euclid}
    W_{i}^{\rm{I}}(\chi) = n_i^{\rm{S}}(z)\
            \frac{{\rm d} z}{{\rm d} \chi}\,.
\end{equation}

The observed \gls{ggl} angular power spectrum (GL) arises from contributions from the distortions of \gls{wl} by foreground galaxies (${\rm{g} \gamma}$), together with smaller contributions from \glspl{ia} (originated from source galaxies located nearby lens galaxies, denoted as gI) and magnification bias, which correlates with both \gls{wl} ($\mu \gamma$) and \glspl{ia} ($\mu \rm I$). The total observed \gls{ggl} signal can be described as
\begin{equation}\label{eq:cl_cross}
    C_{ij}^{\rm GL} = C_{ij}^{\rm{g} \gamma}(\ell)
            + C_{ij}^{\rm{g I}}(\ell) + C_{ij}^{\mu \gamma}(\ell)
            + C_{ij}^{\mu \rm I}(\ell)\,,
\end{equation}
where the subscripts $i,j$ correspond to the tomographic redshift bins from sources and lenses, respectively. Now, for $A= \rm{g}$, the kernel becomes
\begin{equation}\label{eq:weight_function_IA_Euclid}
    W_{i}^{\rm{g}}(\chi) = n_i^{\rm{L}}(z)\
            \frac{{\rm d} z}{{\rm d} \chi}\,,
\end{equation}
with $n_i^{\rm{L}}(z)$ as the lens redshift distribution. In the case of $A= \mu$, the kernel for magnification is
\begin{equation}\label{eq:kernel_magnification}
    W_i^\mu(\chi) = 2(\alpha_{i}-1)W_{i}^{\gamma}(\chi)\,,
\end{equation}
where $\alpha_{i}=5s_{i}/2$, with $s_{i}$ as the logarithmic slope of the magnitude distribution of the lenses \citep{weak_gravitational_lensing_bartelmann, Joachimi_Bridle, KiDS_1000_Fortuna}.

Finally, the observed \gls{gc} angular power spectrum (GG) originates from purely galaxy-galaxy (gg) and magnification-magnification ($\mu \mu$) contributions, together with a combination from both (g$\mu$ and $\mu$g):
\begin{equation}\label{eq:cl_gc_phot}
    C_{ij}^{\rm GG} = C_{ij}^{\rm{gg}}(\ell)
            + C_{ij}^{\rm{g}\mu}(\ell) + C_{ij}^{\mu \rm{g}}(\ell)
            + C_{ij}^{\mu \mu}(\ell)\,,
\end{equation}
where now the subscripts $i,j$ correspond to the lens redshift bins. In our study, we only considered the GG auto-correlations of lens redshift bins since the contribution from cross-correlations is expected to be sub-dominant, their inclusion might introduce additional systematics \citep{Porredon_cross_correlations}, and the computation time is decreased by ignoring them.

We note that we made some assumptions in Eqs.\,(\ref{eq:cl_wl}), (\ref{eq:cl_cross}), and (\ref{eq:cl_gc_phot}) to simplify our analyses. In particular, we neglected the contribution of magnification to the \gls{wl} $C_{\ell}$ \citep{magnification_wl}, and we did not consider redshift-space distortions (\glsentryshort{rsd}\glsunset{rsd}; \citealt{RSD}). Although all of these effects are important for precision cosmology \citep{RSD_euclid, Limber_euclid}, they are sub-dominant contributions, which we chose to neglect to focus on the effect of \glspl{ia}.

\begin{table}
\centering
\caption{Power spectra cases for the \textit{AB} combinations in Eq.~(\ref{eq:general_cl_limber}).}
\label{tab:power_spectra}
\begin{tabular}{c c}
\hline
\hline
\textit{AB} & $P_{AB}$ \\ \hline
$\gamma \gamma$, $\mu \gamma$, $\mu \mu$ & $P_{\delta \delta}$ \\ 
$\gamma$I, I$\gamma$, I$\mu$ & $P_{\delta \mathrm{I}}$ \\ 
II & $P_{\mathrm{II}}$ \\ 
g$\gamma$, g$\mu$, $\mu$g & $P_{\mathrm{g}\delta}=b_{1}P_{\delta\delta}$ \\ 
gI & $P_{\mathrm{gI}}=b_{1}P_{\delta \mathrm{I}}$ \\ 
gg & $P_{\mathrm{gg}}=b_{1}^{2}P_{\delta\delta}$ \\ 
\hline
\end{tabular}
\tablefoot{The indexes $\delta$, $\gamma$, $\mu$, I, and g refer to the density, shear, magnification, intrinsic shape and galaxy components, respectively.}
\end{table}

\subsection{Models for intrinsic alignment of galaxies}\label{sec:IA_modelling}

In this section we present the \gls{nla} and \gls{tatt} models, and variations of them. We do not explore more complex models given that DR1, while expecting a significant improvement over previous surveys, will still deliver only a fraction of the final mission’s constraining power.

\subsubsection{NLA}

The simplest model describing the \gls{ia} of galaxies involves a relation between the shape of the galaxies and the tidal field of their surrounding \gls{lss}. This leads to a linear response to the tidal field, as first proposed in the \gls{la} model \citep{Linear_alignment_1, Linear_alignment_2}. In this scenario, the intrinsic shear of a galaxy is described by
\begin{equation}\label{eq:intrinsic_shear}
    \gamma^{\rm I} = -\frac{\bar{C_{1}}}{4\pi G_{\rm{N}}}\left ( \Delta_{x}^{2}-\Delta_{y}^{2}, 2\Delta_{x}\Delta_{y} \right )S[\psi _{\rm P}]\,,
\end{equation}
where $\bar{C_{1}} = 5 \times 10^{-14}\, M_{\odot}^{-1} h^{-2} \, \mathrm{Mpc}^{3}$ is a normalisation factor calibrated by \citet{C1_value} using low-redshift observations from the SuperCOSMOS survey \citep{SuperCOSMOS}, $G_{\rm{N}}$ is the Newtonian gravitational constant, $x$ and $y$ are the Cartesian coordinates in the plane of the sky, $S$ is a smoothing function applied to the Newtonian potential $\psi_{\rm P}$ at the epoch of galaxy formation, and $\Delta$ corresponds to the comoving spatial derivative.

From Eq.~(\ref{eq:intrinsic_shear}), the intrinsic-intrinsic ($P_{\rm II}$) and matter-intrinsic ($P_{\rm \delta I}$) power spectra can respectively be defined as
\begin{equation}\label{eq:power_spectra_nla}
\begin{aligned}
    P_{\rm II}(z, k) = C_{1}^{2}(z)P_{\rm \delta\delta}(z, k)\,, \\
    P_{\rm \delta I}(z, k) = C_{1}(z)P_{\rm \delta\delta}(z, k)\,,
\end{aligned}
\end{equation}
where, assuming a power-law dependence with redshift, $C_{1}(z)$ can be written as
\begin{equation}\label{eq:c1_IA}
    C_{1}(z) = -A_{1}\bar{C_{1}}\frac{\Omega_{\rm m} \,\rho_{\rm crit}}{D(z)} \left ( \frac{1 + z}{ 1 + z_0} \right )^{\eta_{1}}\,,
\end{equation}
with $\rho_{\rm crit}$ the critical density today, $D(z)$ the growth factor, and $z_0$ the pivoting redshift. The free parameters $A_{1}$ and $\eta_{1}$ indicate the overall amplitude and the redshift dependence, respectively. In the \gls{la} model, $P_{\rm \delta\delta}$ in Eq.~(\ref{eq:power_spectra_nla}) corresponds to the linear matter power spectrum. However, the \gls{nla} model \citep{NLA_1, NLA_2} proposes the substitution of the linear by the non-linear matter power spectrum, which allows for a better description of the observed \gls{ia} signal. Although this is not a theoretically motivated approach, \gls{nla} constitutes one of the most frequently used \gls{ia} models. As such, we adopt it as one of the \gls{ia} models used in this work. In particular, we considered two scenarios within this model, the minimal \gls{nla} (which we simply refer to as \gls{nla}, if not specified otherwise) and the \nlaz models, satisfying $\eta_{1} = 0$ and $\eta_{1}  \neq  0$, respectively. 

\subsubsection{TATT}

A more complex \gls{ia} model can be derived by performing perturbative expansions of the intrinsic shear field, $\gamma^{\rm I}_{ij}$, in terms of the matter density field, $\delta_{\rm m}$, and the tidal field tensor, $s_{ij}$. By expanding up to the second order, we obtained the intrinsic shear field as described by the \gls{tatt} model \citep{TATT_blazek}:
\begin{equation}\label{eq:TATT_field_expansion}
    \gamma^{\rm I}_{ij} = C_{1} s_{ij} + C_{1 \delta} \delta_{\rm m} s_{ij} + C_{2} \left ( \sum_{k} s_{ik} s_{kj} - \frac{1}{3}\delta_{ij}s^{2} \right )+\ldots \,,
\end{equation}
where the first term reduces to the general \gls{nla} model, the second term can be related to the fact that we only observe \glspl{ia} at the positions of the galaxies, and the third term describes the tidal torquing effect, which is thought to explain the angular momentum-driven \glspl{ia} of spiral galaxies. The term $s^{2}\equiv s^{ij}s_{ij}$, where we have assumed Einstein notation.

Within the \gls{tatt} model, the intrinsic shear field naturally decomposes into $E$- and $B$-mode contributions. The linear tidal-alignment term generates only $E$-modes, as in the \gls{nla} model, while the density-weighted and quadratic tidal-torquing terms introduce additional $E$-mode power and generate $B$ modes. Evaluated at one-loop order, the expressions of the $E$- and $B$-mode intrinsic-intrinsic power spectra, are respectively given by
\begin{equation}\label{eq:P_II_TATT_Euclid}
\begin{aligned}
    P_{\rm II}^{EE}(z, k) &= C_{1}^{2}(z)P_{\delta\delta}(z, k)\\
            &+2C_{1}(z)C_{1\delta}(z)D^4(z)\left[A_{0|0E}(k)+ C_{0|0E}(k)\right]\\
            &+C_{1\delta}^{2}(z)D^{4}(z)
             A_{0E|0E}(k)
             +C_{2}^{2}(z)D^4(z)
            A_{E2|E2}(k)\\
            &+2C_{1}(z)C_{2}(z)D^4(z)
            \left[A_{0|E2}(k)+B_{0|E2}(k)\right]\\
            &+2C_{1\delta}(z)C_{2}(z)D^4(z)
            D_{0E|E2}(k)\,,
\end{aligned}
\end{equation}
and
\begin{equation}\label{eq:P_II_BB_TATT_Euclid}
\begin{aligned}
    P_{\mathrm{II}}^{BB}(z, k) &= C_{1 \delta}^2(z) A_{0 B | 0 B}(k)\\
    &+C_2^2(z) A_{B 2 \mid B 2}(k)+2 C_{1 \delta}(z) A_2(z) D_{0 B \mid B 2}(k)\,,
\end{aligned}
\end{equation}
while the matter-intrinsic power spectrum is
\begin{equation}\label{eq:P_mI_TATT_Euclid}
\begin{aligned}
    P_{\rm \delta I}(z, k) &= C_{1}(z)P_{\rm \delta\delta}(z, k)
            +C_{1\delta}(z)D^4(z)\left[A_{0|0E}(k)+ C_{0|0E}(k)\right]
            \\ &+C_{2}(z)D^4(z)\left[ A_{0|E2}(k)+B_{0|E2}(k)\right]\,,
\end{aligned}
\end{equation}
where $C_1(z)$ corresponds to the one introduced in Eq.~(\ref{eq:c1_IA}) and 
\begin{equation}
\begin{aligned}
    &C_2(z) = 5 A_2 \bar{C_1} \frac{\Omega_{\rm m} \,\rho_{\rm crit}}{D^{2}(z)} 
            \left ( \frac{1 + z}{1 + z_0} \right )^{\eta_{2}}\,, \\
    &C_{1\delta}(z) = b_{\rm{TA}} C_1(z)\,,
\end{aligned}
\end{equation}
with $A_1$, $A_2$, $b_{\rm{TA}}$, $\eta_1$, and $\eta_2$ as the free \gls{tatt} parameters. The one-loop order terms in Eqs.~(\ref{eq:P_II_TATT_Euclid}--\ref{eq:P_mI_TATT_Euclid}) are defined in \citet{TATT_blazek}.

Note that the $A_1$ and $\eta_1$ parameters are shared with the \nlaz model, which is recovered in the case where $A_2=\eta_2=b_{\rm{TA}}=0$. We considered four cases for the \gls{tatt} model in this work: the minimal \gls{tatt} model (which we simply refer to as \gls{tatt}, if not specified otherwise), with $\eta_{1} = \eta_{2} = 0$; the \tattz model, with $\eta_{1}  \neq  0$ and $\eta_{2}  \neq  0$; the \tattzbta model, with $\eta_{1}  \neq  0$, $\eta_{2}  \neq  0$, and $b_{\rm{TA}}=b_{1}=1$,  where $b_{1}$ is the linear galaxy bias of source galaxies, assumed to be one~\citep{DES_Troxel, DES_Y1_samuroff}; and the \nlak model \citep{KiDS_Wright_cosmic_shear}, with $A_2=\eta_2=0$, which can also be thought of as an extension of the \nlaz model. Table~\ref{tab:ia_models} summarises the different \gls{ia} models employed in this work, specifying their free parameters.

\begin{table}
\centering
\caption{\gls{ia} models considered in this work.}
\label{tab:ia_models}
\begin{tabular}{c c}
\hline
\hline
\gls{ia} model & Free parameters \\ \hline
\gls{nla} & $A_{1}$ \\ 
\nlaz & $A_{1}$, $\eta_{1}$ \\ 
\nlak & $A_{1}$, $\eta_{1}$, $b_{\rm{TA}}$ \\
\gls{tatt} & $A_{1}$, $A_{2}$, $b_{\rm{TA}}$ \\
\tattzbta & $A_{1}$, $A_{2}$, $\eta_{1}$, $\eta_{2}$ \\
\tattz & $A_{1}$, $A_{2}$, $b_{\rm{TA}}$, $\eta_{1}$, $\eta_{2}$ \\
\hline
\end{tabular}
\end{table}

\section{Forecast pipeline}\label{sec:forecast_pipeline}

In this section we present the pipeline used to forecast the impact of \glspl{ia} on the baseline $3\times 2$\,pt \Euclid analyses. We first describe how we build and sample the likelihood. Next, we introduce the set of parameters sampled in the analysis, justifying their fiducial values and priors. We later describe how the covariance matrix is built and, finally, we present the scale-cut choices that we employ.

The analysis was performed with the \texttt{CosmoSIS}\,\footnote{\url{https://github.com/joezuntz/cosmosis}} code \citep{CosmoSIS}, which allowed us to compute the theoretical predictions introduced in Sect.~\ref{sec:Modelling} in a module-like setup, where users can specify their modelling choices. \texttt{CosmoSIS} connects the theoretical predictions with samplers, allowing us to explore the parameter space when given a specific \gls{dv}, covariance matrix, and scale cuts. The linear matter power spectrum and other underlying background quantities are computed using the \texttt{CAMB} Boltzmann solver code \citep{CAMB_1, CAMB_2}. We employed two different recipes for the computation of the non-linear matter power spectrum. For the generation of synthetic \glspl{dv} in Sect.~\ref{sec:generation_synt_DVs}, with which we performed the core of the analysis, we used the \texttt{HMCode-2020} implementation with baryonic feedback \citep{mead_2020_feedback}, which covers the full range of the priors of the sampled cosmological parameters (see Table~\ref{tab:parameters_and_priors}). For the determination of the \gls{ia} parameters in the Flagship simulation (see Sect.~\ref{sec:IA_values_FS}), we used the \texttt{EuclidEmulator2}\footnote{\url{https://github.com/miknab/EuclidEmulator2}} \citep{EuclidEmulator2} since it more accurately describes the signal measured in Flagship. The one-loop terms in Eqs.~(\ref{eq:P_II_TATT_Euclid}--\ref{eq:P_mI_TATT_Euclid}) were computed using the \texttt{FAST-PT} code \citep{Fast_pt_1, Fast_pt_2}, which uses an integration based on \gls{fft} methods to accelerate the computation.

While in this work we used \texttt{CosmoSIS}, we note that the Euclid Consortium is developing its own inference pipeline (\texttt{cloelib} and \texttt{cloelike}) that will be publicly hosted prior to the first \Euclid cosmology analysis at the GitHub organisation cloe-org\footnote{\url{https://github.com/cloe-org}} \citep{CLOE_org}. This pipeline builds upon previous work published in \citet{CLOE_cardone}, \citet{CLOE_joudaki}, \citet{CLOE_Canas_herrera}, \citet{CLOE_martinelli}, \citet{CLOE_goh}, and \citet{CLOE_blot}.

\subsection{Likelihood}\label{sec:likelihood}

Given a \gls{dv}, $\vec{D}$, and a theoretical vector given by a model $M$, $\vec{T_M}$, we performed a likelihood analysis to constrain a set of parameters of interest, $\vec{p}$. 
We defined a Gaussian likelihood $L$ such that
\begin{equation}\label{eq:likelihood}
    \ln{L(\vec{D}|\vec{p}, M)} = -\frac{1}{2}\left [ (\vec{D}-\vec{T_M})^{T} \tens{C}^{-1} (\vec{D}-\vec{T_M}) \right] + \mathcal{C}\,,
\end{equation}
where $\tens{C}$ corresponds to the covariance matrix of the data, presented in Sect.~\ref{sec:covariance_matrix}, and $\mathcal{C}$ is a normalisation constant. The \gls{dv} in our analysis is given by
\begin{equation}\label{eq:datavector_likelihood}
    \vec{D} =  \left\{C_{ij}^{\rm LL}(\ell), C_{ij}^{\rm GL}(\ell), C_{ij}^{\rm GG}(\ell)\right\}\,,
\end{equation}
which represents the concatenation of the theoretical quantities defined in Eqs.~(\ref{eq:cl_wl}--\ref{eq:cl_gc_phot}) evaluated at the fiducial values in the parameter space $\vec{p}$ and is presented in Sect.~\ref{sec:synt_DVs}. Given that $\vec{D}$ is a synthetic \gls{dv}, the theoretical vector $\vec{T_M}$ has a similar definition as in Eq.~(\ref{eq:datavector_likelihood}), with the difference that it is not necessarily evaluated at the fiducial values of $\vec{p}$ but on any value in the parameter space:
\begin{equation}
    \vec{T_M}
    =  \left\{C_{ij}^{\rm LL}(\ell|\vec{p}
    ), C_{ij}^{\rm GL}(\ell|\vec{p}
    ), C_{ij}^{\rm GG}(\ell|\vec{p}
    )\right\}\,.
\end{equation}

The posterior probability distribution of the set of parameters $\vec{p}$, which indicates how likely different values of the parameters are when given a \gls{dv} and an assumed model, is provided by Bayes' theorem as
\begin{equation}\label{eq:posterior_probability}
    P(\vec{p}|\vec{D}, M) = \frac{L(\vec{D}|\vec{p}, M) P(\vec{p}|M)}{P(\vec{D}|M)}\,,
\end{equation}
where $P(\vec{p}|M)$ is the prior probability distribution and $P(\vec{D}|M)$ is the Bayesian evidence.

To sample the posterior distribution, we employed the \texttt{NAUTILUS} sampler \citep{NAUTILUS}, which is interfaced in \texttt{CosmoSIS}. \texttt{NAUTILUS} employs importance nested sampling techniques \citep{Importance_nested_sampling}, an extension of traditional nested sampling methods \citep{Nested_sampling}. The latter technique allowed us to compute the Bayesian evidence, and it samples the posterior distribution as a by-product, while the former improves the process by keeping all likelihood evaluations and, in the case of \texttt{NAUTILUS}, efficiently sampling the posterior distribution using neural networks. The \texttt{NAUTILUS} configuration we used discards the points drawn in the exploration phase. We relied on 6000 live points and required a minimum effective sample size of $N_{\mathrm{eff}}=15\,000$, with 1200 likelihood evaluations at each step. We used 16 neural networks in each network ensemble.

\subsection{Parameters and priors}\label{sec:parameters_and_priors}

We constrained the posterior probability distributions of the parameters of interest for two cosmological models: a flat $\Lambda$CDM and $w_{0}w_{a}$CDM. The latter is characterised by a redshift-dependent dark energy equation-of-state parameter $w(z)=w_{0} + w_{a}(1-a)$, where $w_{0}$ is the equation of state today and $w_{a}$ quantifies the dependence with redshift \citep{CPL_1, CPL_2}. The $\Lambda$CDM model corresponds to the special case where $w_{0}=-1$ and $w_{a}=0$. We varied three cosmological parameters for the flat $\Lambda$CDM case: the total matter density in units of physical density, $\Omega_{\rm m}h^{2}$; the Hubble parameter, $h$; and the amplitude of the primordial scalar density perturbations, $A_{\rm{s}}$. In the case of the $w_{0}w_{a}$CDM, in addition to these three cosmological parameters, we also varied $w_{0}$ and $w_{a}$. We fixed the values of the other cosmological parameters to reduce computation time since they are not that well constrained by $3\times 2$\,pt analyses. However, we note that in real data analysis, these parameters are usually sampled. Restricting the inference to this reduced parameter space allowed us to isolate modelling-induced biases without the additional impact of projection effects arising from poorly constrained parameter directions. If extra cosmological parameters were allowed to vary, degeneracies between constrained and unconstrained directions could partially project or amplify modelling biases, potentially modifying the parameter shifts obtained for a given scale cut \citep{DESY6_modelling}. The shifts reported in this work should therefore be interpreted as being obtained in the absence of such projection effects.

We fixed both the baryonic and the curvature density in physical units, $\Omega_{\rm b}h^{2}$ and $\Omega_{K}h^{2}$; the spectral index of the primordial power spectrum, $n_{\rm s}$; and the optical depth of reionisation, $\tau$. We set the number of massive neutrinos to one, with $N_{\rm {eff}} = 3.046$ and $\Sigma m_{\nu} = 0.06\,\rm{eV}/c^{2}$, where this minimum mass arises from that allowed by oscillation experiments \citep{Review_particle_physics}. \texttt{CosmoSIS} also allows some extra cosmological parameters to be derived, such as $\sigma_{8}$ and $S_{8} \equiv \sigma_8 \sqrt{\Omega_{\rm m}/{0.3}}$, where the former quantifies the amplitude of matter perturbations and the latter captures the $\Omega_{\rm m}$--$\sigma_8$ degeneracy that \gls{wl} measurements are most sensitive to.

In addition to the cosmological parameters, we also considered other systematic and astrophysical parameters. The number of parameters sampled to capture \glspl{ia} will depend on the \gls{ia} model we consider, as shown in Table~\ref{tab:ia_models}. We set the pivoting redshift to $z_0=0.62$, in accordance with P26. Baryon feedback was included based on the \gls{agn} feedback strength, modelled via the sub-grid heating parameter, $T_{\mathrm{AGN}}$. We assumed a linear galaxy bias approach, with $b_{i}$ the linear galaxy bias per lens bin $i$. This way, we focused on the effect of \glspl{ia} and avoided opening up the parameter space with more complex galaxy bias relations, such as a non-linear galaxy bias approach \citep{Nonlin_gal_bias}, which is left for future work. To take into account the uncertainty in the shear calibration procedure, we used a multiplicative shear bias factor per source tomographic redshift bin, $m_{i}$, relating the observed and the true shear such that
\begin{equation}\label{eq:shear_multiplicative_bias_shear}
    \gamma^{\mathrm{obs}}_{i} = (1+m_{i})\gamma^{\mathrm{true}}_i\,,
\end{equation}
which in turn modifies the angular power spectra \citep{Kitching_2019} described in Eqs.~(\ref{eq:cl_wl}) and (\ref{eq:cl_cross}) by
\begin{align}\label{eq:shear_multiplicative_bias_cl}
    C_{ij}^{\rm LL}(\ell) &\rightarrow 
    (1+m_i)(1+m_j) C_{ij}^{\rm LL}(\ell)\,, \\
    C_{ij}^{\rm GL}(\ell) &\rightarrow 
    (1+m_i) C_{ij}^{\rm GL}(\ell)\,.
\end{align}

We did not consider an additive bias term since this can be estimated and subtracted from the data \citep{Kitching_2019, additive_bias_ref}.

Any \gls{photo-z} biases were parametrised by considering a shift ($\Delta z$) in the mean of the $n(z)$ redshift distribution of both sources and lenses such that
\begin{equation}\label{eq:delta_z_photoz}
    n_{i}(z) \rightarrow n_{i}(z-\Delta z_{i})\,,
\end{equation}
where $\Delta z_{i}$ is set for each redshift bin. Finally, we also sampled the magnification bias parameter, $\alpha_{i}$, introduced in Eq.~(\ref{eq:kernel_magnification}). Considering that our DR1 setup comprises six lens and six source tomographic redshift bins (see Sect.~\ref{sec:Sample_definition}), this analysis leads to 34 sampled parameters (without considering $w_{0}$ and $w_{a}$) which, in combination with the \gls{ia} ones, amount to 35--39 in the case of $\Lambda$CDM and 37--41 in the case of $w_{0}w_{a}$CDM, depending on the assumed \gls{ia} model.

Table~\ref{tab:parameters_and_priors} shows the fiducial values and priors for all parameters considered in this analysis. In the case of the cosmological parameters, we followed the fiducial values from the Flagship simulation \citep{Flagship} and the priors derived in the \gls{des} Y3 $3\times 2$\,pt analysis \citep{DES_Y3_3x2pt}. The values of the \gls{ia} parameters were also derived from the Flagship simulation and are further justified in Sect.~\ref{sec:IA_values_FS}. The \gls{ia} priors were assumed to be broad, although in Sect.~\ref{sec:constraining_power_ia_priors} we show the negligible effect of narrowing the \gls{ia} priors in our analyses. For the halo model parameter, we assumed the prior range of the combined \texttt{BAHAMAS} \citep{BAHAMAS} and \texttt{COSMO-OWLS} \citep{OWLS} case, extracted from the halo model \texttt{HMCODE-2020}, and chose a fiducial value widely used in that work \citep{mead_2020_feedback}. We derived the galaxy bias parameters by fitting the angular correlation function, $w(\theta)$, to the sample used in this analysis (Sect.~\ref{sec:Sample_definition}), and set broad priors to recover the fiducial. The priors of the remaining parameters were defined as Gaussian, $\mathcal{N}(\mu, \sigma)$, with $\mu$ representing the mean and $\sigma$ the standard deviation. The priors of the multiplicative shear bias were based on the expected uncertainties for \Euclid DR1. A similar scenario occurs for the priors on the source and lens \gls{photo-z} errors, which were defined as $\Delta z_{\rm s}^{i}=0.01(1+z_{\rm s})$, so that they are of the same order as the final stage-III \gls{photo-z} priors (e.g. \citealt{photo_z_priors_kids}), and which are also expected to be similar to those of \Euclid DR1. Finally, in the case of the magnification bias, the fiducial values were computed from the slope of the cumulative number counts at the magnitude cut of the lenses defined in Sect.~\ref{sec:Sample_definition}, while the priors were defined as broad to recover the fiducial.

\begin{table}
\centering
\caption{Model parameters, fiducial values, and priors used for the $\Lambda$CDM and $w_{0}w_{a}$CDM analyses.}
\label{tab:parameters_and_priors}
\resizebox{0.5\textwidth}{!}{
\begin{tabular}{c c c}
\hline
\hline
Parameter & Fiducial & Prior \\ \hline
& \textbf{Cosmology} & \\
$\Omega_{\rm m}h^{2}$ & 0.143 & [0.045, 0.404] \\ 
$h$ & 0.67 & [0.55, 0.91] \\
$A_{\rm s} \times 10^9$ & 2.1 & [0.5, 5.0] \\
$\Omega_{\rm b}h^{2}$ & 0.022 & Fixed \\
$\Omega_{K}h^{2}$ & 0.0 & Fixed \\
$n_{\rm s}$ & 0.96 & Fixed \\
$w_{0}$ & $-$1.0 & Fixed or [$-$2.0, $-$0.333] \\
$w_{a}$ & 0.0 & Fixed or [$-$3.0, 3.0] \\
$\tau$ & 0.0697186 & Fixed \\
& \textbf{Intrinsic alignments} & \\
$A_{1}$ & 0.92 & [0.0, 3.0] \\
$A_{2}$ & 0.40 & [$-$2.0, 2.0] \\
$\eta_{1}$ & 2.26 & [0.0, 4.0] \\
$\eta_{2}$ & 2.69 & [$-$6.0, 6.0] \\
$b_{\rm{TA}}$ & $-$0.83 & [$-$5.0, 3.0] \\
$z_{0}$ & 0.62 & Fixed \\
& \textbf{Halo model parameters} & \\
$\logten (T_{\rm AGN}/\rm{K})$ & 7.8 & [7.6, 8.3] \\
& \textbf{Galaxy bias} & \\
$b_{i}$ & 1.14, 1.20, 1.32, 1.42, 1.52, 1.94 & [0.8, 3.5] \\
& \textbf{Multiplicative shear bias} & \\
$m_{i}$ & 0.0 & $\mathcal{N}(0.0, 0.01)$ \\
& \textbf{Source photo-$z$ errors} & \\
$\Delta z_{\rm s}^{1}$ & 0.0 & $\mathcal{N}(0.0, 0.014)$ \\
$\Delta z_{\rm s}^{2}$ & 0.0 & $\mathcal{N}(0.0, 0.016)$ \\
$\Delta z_{\rm s}^{3}$ & 0.0 & $\mathcal{N}(0.0, 0.018)$ \\
$\Delta z_{\rm s}^{4}$ & 0.0 & $\mathcal{N}(0.0, 0.020)$ \\
$\Delta z_{\rm s}^{5}$ & 0.0 & $\mathcal{N}(0.0, 0.023)$ \\
$\Delta z_{\rm s}^{6}$ & 0.0 & $\mathcal{N}(0.0, 0.028)$ \\
& \textbf{Lens photo-$z$ errors} & \\
$\Delta z_{\rm l}^{1}$ & 0.0 & $\mathcal{N}(0.0, 0.013)$ \\
$\Delta z_{\rm l}^{2}$ & 0.0 & $\mathcal{N}(0.0, 0.014)$ \\
$\Delta z_{\rm l}^{3}$ & 0.0 & $\mathcal{N}(0.0, 0.016)$ \\
$\Delta z_{\rm l}^{4}$ & 0.0 & $\mathcal{N}(0.0, 0.017)$ \\
$\Delta z_{\rm l}^{5}$ & 0.0 & $\mathcal{N}(0.0, 0.019)$ \\
$\Delta z_{\rm l}^{6}$ & 0.0 & $\mathcal{N}(0.0, 0.023)$ \\
& \textbf{Magnification bias} & \\
$\alpha_{1}$ & 0.516 & [0.1, 1.0] \\
$\alpha_{2}$ & 0.682 & [0.1, 1.5] \\
$\alpha_{3}$ & 0.652 & [0.1, 1.5] \\
$\alpha_{4}$ & 0.804 & [0.1, 2.0] \\
$\alpha_{5}$ & 1.178 & [0.1, 2.5] \\
$\alpha_{6}$ & 1.972 & [0.1, 4.0] \\
\hline
\end{tabular}
}
\tablefoot{Priors between brackets correspond to flat priors, while priors indicated with $\mathcal{N}(\mu, \sigma)$ are Gaussian priors, with mean $\mu$ and standard deviation $\sigma$.}
\end{table}

\subsection{Covariance matrix}\label{sec:covariance_matrix}

We computed the covariance matrix using the \texttt{Spaceborne} code\,\footnote{\url{https://github.com/davidesciotti/Spaceborne}} (Euclid Collaboration: Sciotti et al., in prep.), which has been specifically designed to produce covariance matrices for the \Euclid mission. Three terms contribute to the total covariance matrix of this work: Gaussian covariance, super-sample covariance, and connected non-Gaussian  covariance. The Gaussian covariance accounts for the case where the fields under study are Gaussian distributed, which is a good approximation for large angular scales. This is proportional to the squared amplitude of the observables. The super-sample covariance allows for the inclusion of super-survey modes, which couple with the in-survey modes observed by \Euclid. In particular, the super-sample covariance implemented in \texttt{Spaceborne} is computed taking into account the cross-covariance of the super-survey modes at different redshifts. Finally, the connected non-Gaussian covariance accounts for the coupling of in-survey modes. This term is most relevant at small scales, where the assumption of Gaussianity of the power spectra is not satisfied, due to non-linear gravitational interactions. The relevant equations employed for the covariance computation can be found in Euclid Collaboration: Sciotti et al. (in prep.). The shape noise, sky fraction, and number densities of sources and lenses -- necessary for the computation of the covariance matrix -- are specified in Sect.~\ref{sec:Sample_definition}.

\subsection{Scale cuts}\label{sec:scale_cuts}

Since current theories become unreliable on small, highly non-linear scales, we impose scale cuts to ensure we only use scales where predictions are robust. We set scale cuts based on \citet{Doux_2021, Doux_2022}, who define a small-scale physical mode, $k_{\mathrm{max}}$, that is converted to a multipole $\ell_{\mathrm{max}}$ using the relation $k=(\ell + 1/2)/\chi(z)$. Therefore, for a given cosmological probe, the same $k_{\mathrm{max}}$ is employed for all redshift bins. The idea behind this technique is that the matter power spectra that enter Eq.~(\ref{eq:general_cl_limber}) are valid for $k<k_{\mathrm{max}}$, so that we effectively remove multipoles $\ell$ receiving information from smaller scales. This was done by integrating the $C_{\ell}$ up to a given $k_{\mathrm{max}}$ such that the fraction over the total $C_{\ell}$ is $\alpha<1$:
\begin{equation}\label{eq:scale_cut}
    \int_{-\infty}^{\ln k_{\mathrm{max}}} {\rm d} \ln k \frac{{\rm d}C_{\ell}}{{\rm d} \ln k} = \alpha C_{\ell} \,.
\end{equation}

We set $\alpha=0.95$ so that the wavenumbers $k>k_{\mathrm{max}}$ only contribute 5\% to the total $C_{\ell}$ \citep{Doux_2021, Doux_2022}. We considered different $k_{\mathrm{max}}$ based on the cosmological probe we analysed. We set $k_{\mathrm{max}}=0.3\,\invhMpc$ for \gls{ggl} and \gls{gc} since we want to stay in a regime where linear galaxy bias is valid.\footnote{Euclid Collaboration: Gouyou Beauchamps et al. (in prep.) analyse the validity of the linear galaxy bias prescription in angular statistics for a \Euclid sample defined in Flagship. The authors recommend defining a maximum $k_{\rm{max}}=0.2\,\invhMpc$ to avoid significant deviations from the linear galaxy bias regime. However, given that this recommendation is made for the Flagship simulation, and the main analyses of this work are performed with synthetic \glspl{dv}, the slightly more optimistic scale cut of $k_{\rm{max}}=0.3\,\invhMpc$ that we employed does not yield any bias in our constraints, just a small increase in the constraining power for \gls{ggl} and \gls{gc}.} For \gls{wl}, we defined two scale cuts, $k_{\mathrm{max}}=1\,\invhMpc$ and $k_{\mathrm{max}}=3\,\invhMpc$, to test how the constraining power varies when including smaller scales. In this way, we captured the regime where systematic uncertainties are sub-dominant with respect to statical ones \citep{takahashi_2012, Doux_2021, Doux_2022, FLAMINGO_2023, EuclidEmulator2}.

The scale cuts were derived from Eqs.~(\ref{eq:cl_wl}--\ref{eq:cl_gc_phot}), which describe the \gls{wl}, \gls{ggl}, and \gls{gc} observables. We considered two scenarios: (i) deriving scale cuts for the determination of the \gls{ia} values from the Flagship simulation (Sect.~\ref{sec:IA_values_FS}) and (ii) deriving scale cuts for the main analysis of this work. In the first case, since the Flagship simulation allows positions without the effect of magnification to be obtained, Eqs.~(\ref{eq:cl_cross}--\ref{eq:cl_gc_phot}) do not include the magnification contributions. In the second case, all the terms in Eqs.~(\ref{eq:cl_wl}--\ref{eq:cl_gc_phot}) are included. We note that the scale cuts will depend on the assumed \gls{ia} model, so we computed different sets of scale cuts based on the model considered. However, we also note that the differences of scale cuts between \gls{ia} models are not very significant.

\begin{figure}
    \centering
    \includegraphics[width=0.48\textwidth]{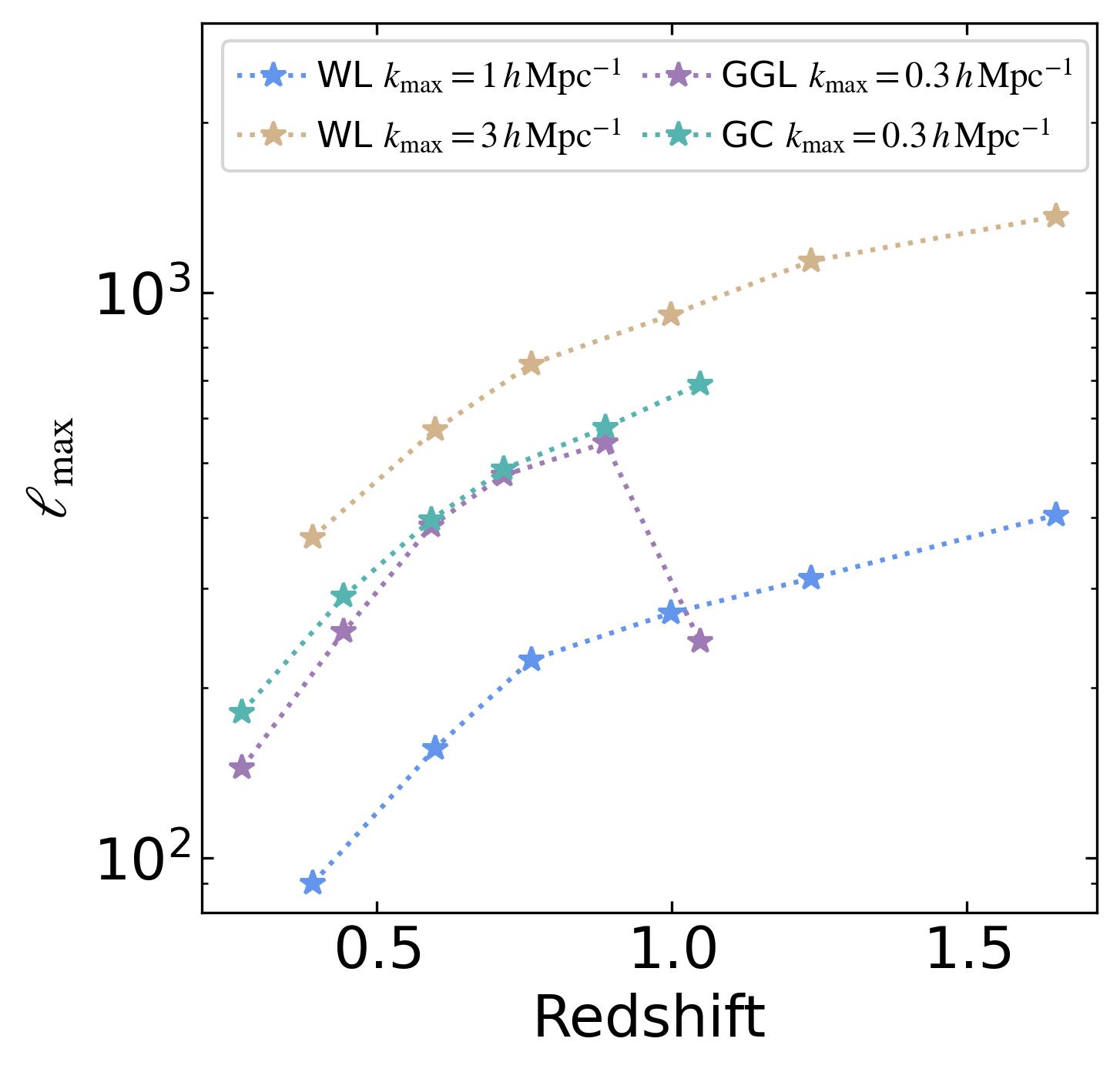}
    \caption{Scale cuts for \gls{wl}, \gls{ggl}, and \gls{gc} as a function of redshift for different $k_{\mathrm{max}}$. The assumed \gls{ia} model is \nlaz, and magnification contributions are included.}
    \label{fig:scale_cuts}
\end{figure}

Figure~\ref{fig:scale_cuts} shows the scale cuts in terms of $\ell_{\mathrm{max}}$ as a function of redshift for the \gls{wl} case with $k_{\mathrm{max}}=1\,\invhMpc$ and $k_{\mathrm{max}}=3\,\invhMpc$, and for \gls{ggl} and \gls{gc} at $k_{\mathrm{max}}=0.3\,\invhMpc$. For simplicity, we only show the scale cuts for the auto-correlation of tomographic redshift bins. In this case, we adopt the \nlaz model and consider the effect of magnification. Note how the derived $\ell_{\mathrm{max}}$ for the \gls{wl} probe increases when increasing the $k_{\mathrm{max}}$. We find that the $\ell_{\mathrm{max}}$ values derived for $k_{\mathrm{max}}=1\,\invhMpc$ and $k_{\mathrm{max}}=3\,\invhMpc$ are consistent with those reported in \citet{Doux_2021, Doux_2022}, although they are more conservative than those in other \Euclid forecasts \citep{scale_cuts_optimistic_1, scale_cuts_optimistic_2, CLOE_blot}. The drop in $\ell_{\mathrm{max}}$ in the last redshift bin for the \gls{ggl} case is due to a large contribution to the $C_{\ell}$ from the magnification term. We tested determining the scale cuts when removing the magnification contributions ($\rm{g I}$ and $\mu \rm I$) from Eq.~(\ref{eq:cl_cross}) and the drop in $\ell_{\mathrm{max}}$ at high redshift did not appear.

\section{Synthetic data vectors}\label{sec:synt_DVs}

Since the observed differences between the \nlaz and \tattz models are highly dependent on the assumed values on the \gls{ia} parameters, we first constrain these values from the Flagship simulation \citep{Flagship}. This is a galaxy mock catalogue based on an $N$-body simulation, developed to reproduce the expected galaxy samples that \Euclid will observe to help prepare its scientific analyses. Importantly for \glspl{ia}, Flagship was calibrated to match the observed correlation functions of \glspl{ia} \citep{Hoffmann_IA} from previous surveys. After measuring the \gls{ia} values from Flagship, we generate synthetic data vectors assuming a \Euclid DR1 setup.

\subsection{Flagship simulation}\label{sec:Flagship}

The Flagship simulation\footnote{accessible via \url{https://cosmohub.pic.es} \citep{CosmoHub_1, CosmoHub_2}.} is based on a state-of-the-art $N$-body simulation that evolved four trillion dark matter particles, generating a lightcone on-the-fly that covers one octant of the sky up to redshift $z=3$ \citep{Flagship}. From these dark matter particles, 16 billion dark matter haloes were identified with the \texttt{ROCKSTAR} halo finder \citep{ROCKSTAR_halo_finder}. These haloes were populated with mock galaxies using \gls{hod} and \gls{am} techniques. The parameters of the galaxy mock were calibrated to reproduce observed galaxy correlations and basic galaxy properties. \Gls{photo-z} estimates were computed from the expected observed fluxes and errors using the official \Euclid pipeline, the Nearest Neighbours Photometric Redshifts (\glsentryshort{nnpz}\glsunset{nnpz}; \citealt{NNPZ}). As a result, the Flagship simulation provides a complete magnitude-limited sample down to $\HE<26$, with 3.4 billion galaxies including several galaxy properties. The  simulation assumes a $\Lambda$CDM cosmology with $\Omega_{\rm m}=0.319$, $\Omega_{\rm b}=0.049$, $h=0.67$, $n_{\rm s}=0.96$, and $A_{\rm s}=2.1 \times10^{-9}$.

The \glspl{ia} are implemented in the Flagship simulation following \citet{Hoffmann_IA}. This implementation takes place in two steps: modelling galaxy shapes and modelling galaxy orientations. First, each galaxy is approximated as a 3D ellipsoid and its axis ratios are assigned based on the galaxy redshift, colour, and magnitude. Then, these 3D ellipsoids are projected along the observer's \gls{los} to obtain 2D galaxy shapes, and are calibrated to follow the galaxy shape distribution from the COSMOS survey \citep{Cosmos_2015, ACSGC}. Second, the galaxy orientations are assigned separately for central and satellite galaxies, where the former have their principal axes aligned with those from the dark matter haloes in which they reside, while the latter have their major axes aligned towards the centre of their haloes. Furthermore, to accurately match the \gls{ia} statistics in observations and simulations, one has to introduce a misalignment angle that reduces the alignment between the axes of galaxies and haloes. This misalignment angle depends on the galaxy type (central or satellite), colour, redshift, and magnitude.

\subsection{Sample definition}\label{sec:Sample_definition}

To match the approximate upper-limit area expected for DR1, we defined a 30-degree cone in the Flagship simulation, corresponding to an area of $\sim \! 2758 \,\rm{deg}^{2}$ and a sky fraction of $f_{\mathrm{sky}}=0.067$. We only selected objects with a \gls{snr} $\geq 5$ -- where \gls{snr} is defined as the ratio of the flux over the flux uncertainty in the $\IE$ band -- and with the \gls{nnpz} flag equal to zero, indicating no issues with the computed \glspl{photo-z}. The lens sample was further selected with a magnitude cut in the $\IE$ band of $\IE\leq23.5$ -- including the continuum flux, emission lines as described in Model 3 of \citet{Pozzetti_model_3}, internal attenuation, and Milky Way extinction. We did not apply a direct magnitude cut to the source sample, which resulted in a very large catalogue. Therefore, to speed up the computation time, we reduced the source catalogue by removing objects based on their shear weights. However, Flagship does not incorporate shear weights yet, which depend on the shape measurement algorithm and require image simulations. Instead, we approximated the shear weights as a function of the \gls{snr} by fitting a simple functional form that follows the \texttt{lensfit} \citep{lensfit_1, lensfit_2} shear weights from the public \gls{kids}-1000 catalogue \citep{KiDS-1000}:
\begin{equation}\label{eq:shear_weight}
w(\mathrm{S/N}) = \frac{1}{1 + \left(\frac{10}{\mathrm{S/N}}\right)^{2.5}} \,.
\end{equation}
Then, we only kept objects with weights larger than uniform random draws, which preferentially removes low \gls{snr} objects, preserves the $n(z)$, and minimises the shape noise.

\begin{figure}
    \centering
    \includegraphics[width=0.48\textwidth]{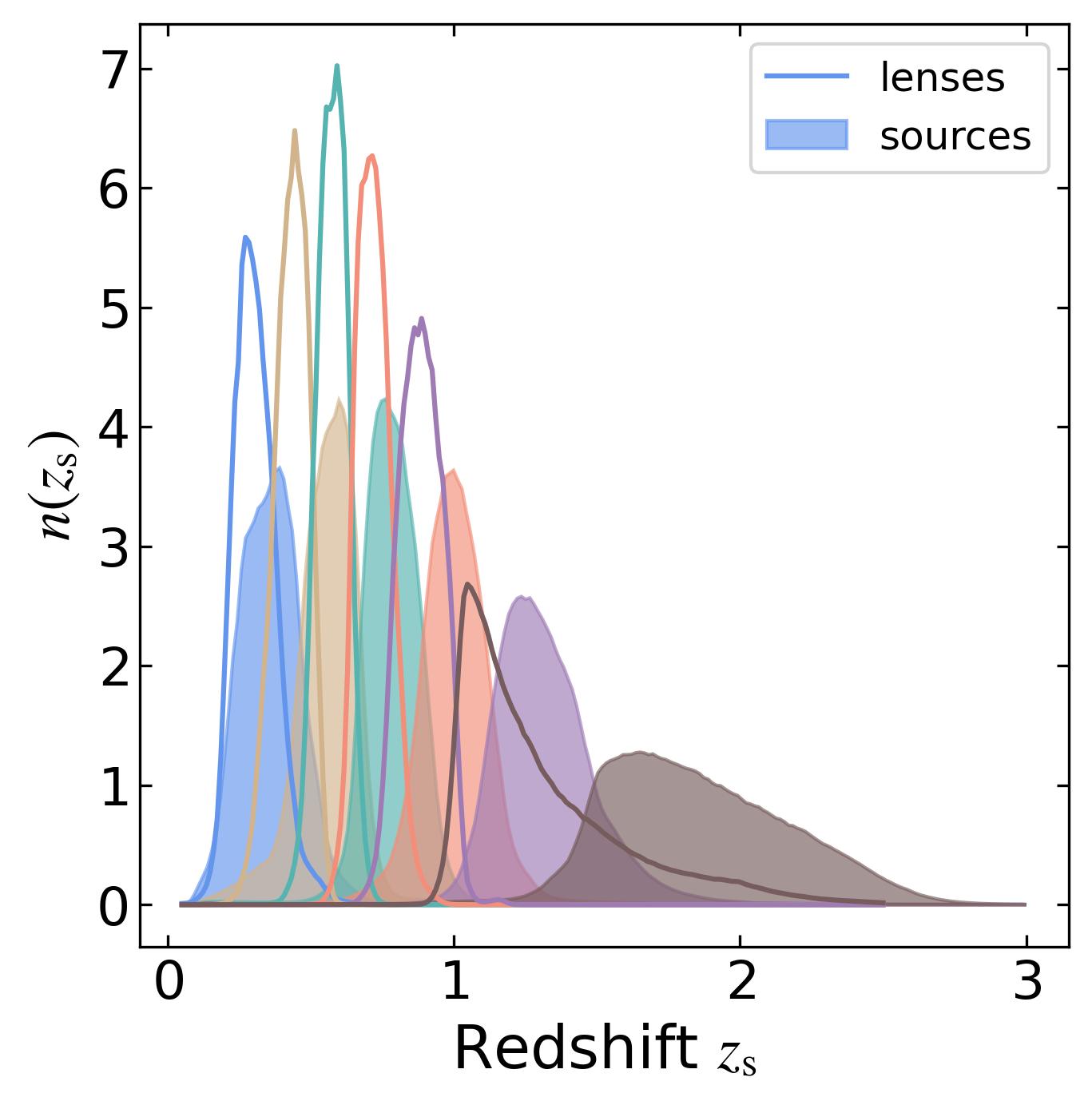}
    \caption{Normalised true redshift distribution, $z_{\mathrm{s}}$, of sources and lenses for the galaxy samples defined in Sect.~\ref{sec:Sample_definition}.}
    \label{fig:n_z}
\end{figure}

Both source and lens samples were further restricted with a cut in the \gls{photo-z} range $0.2<z_{\mathrm{b}}<2.5$, and each sample was divided into six equipopulated tomographic redshift bins. The resulting shape noise is $\sigma_{e} \! \sim \! 0.37$ (with $\sigma_{e}$ the total shape disperion), and the number densities of sources and lenses are $\sim \! 5.99$ and $\sim \! 1.69$ galaxies/arcmin$^{2}$, respectively, in each tomographic bin. Figure~\ref{fig:n_z} shows the resulting $n(z_{\mathrm{s}})$ distribution of both sources and lenses, with $z_{\mathrm{s}}$ the true redshift values. We note that, although we consider these samples as \Euclid DR1, the final samples that will be used in the cosmology analyses are still subject to change.

\subsection{Measurement of IA values in Flagship}\label{sec:IA_values_FS}

Here, we derive the \gls{ia} parameters that best describe the samples defined in the previous section. First, we measure the \gls{wl}, \gls{gc}, and \gls{ggl} $C_{\ell}$ using the \texttt{Heracles} \citep{Heracles} code.\footnote{\url{https://heracles.readthedocs.io/stable/index.html}} This code allows us to measure angular power spectra by employing discrete data sets, obtained at the galaxies' positions. The key ingredients that \texttt{Heracles} needs to compute the angular power spectra are the galaxy positions, the galaxy shears (and their intrinsic ellipticities and shear weights), the visibility map, and the $n(z)$ distribution of sources and lenses. The $C_{\ell}$ that \texttt{Heracles} outputs are convolved with the visibility map, so that they have an amplitude proportional to the area covered by the 30\,deg cone in our case, while also accounting for its mask. This complicates the computation of the covariance matrix (Sect.~\ref{sec:covariance_matrix}), because it also needs to be convolved with the mask. To avoid this, we chose to deconvolve the measured $C_{\ell}$ by dividing their associated correlation function by the correlation function of the mask, and transforming back to harmonic space (Ruiz-Zapatero et al. in prep.). This approach is similar to the \texttt{PolSpice} methodology \citep{PolSpice}, with the crucial difference that spin-2 fields are transformed with their associated Wigner matrices, as opposed to the opposite spin matrix (as described in \citealt{deconvolve_1, deconvolve_2, deconvolve_3}). To make the correlation function of the mask invertible, we set its inverse to zero in scales outside of the mask. We applied this regularisation using a smooth logistic function to avoid the ringing when mapping back to harmonic space. This approach is equivalent to regularizing small eigen values in singular value decomposition algorithms but using a smooth transition between the regularised and unregularised scales.

\begin{figure}
    \centering
    \includegraphics[width=0.48\textwidth]{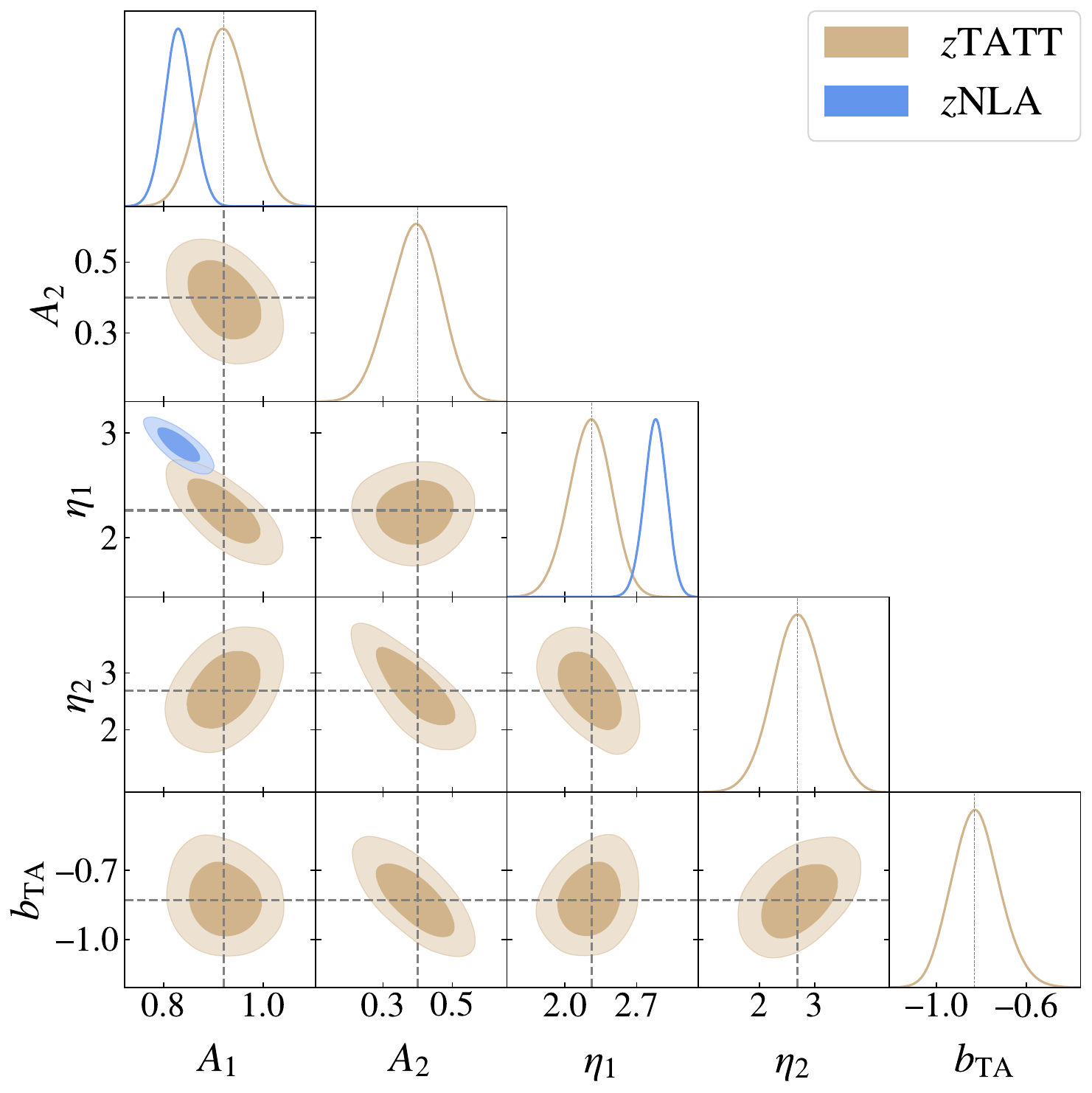}
    \caption{\gls{ia} constraints for the measured Flagship $C_{\ell}$ described in Sect.~\ref{sec:IA_values_FS} when using the \tattz (tan) and the \nlaz (blue) \gls{ia} models. The dashed lines correspond to the maximum of the posterior distributions of the \tattz parameters, which are used to generate the synthetic \glspl{dv} in Sect.~\ref{sec:generation_synt_DVs}.}
    \label{fig:IA_values_FS}
\end{figure}

Once the $C_{\ell}$ were computed, we determined the \gls{ia} values by sampling over $A_{1}$, $A_{2}$, $\eta_{1}$, $\eta_{2}$, and $b_{\rm{TA}}$. We considered the flat priors $A_1\in[0,3]$, $A_2\in[-2,2]$, $\eta_1\in[0,4]$, $\eta_2\in[-4, 4]$, and $b_{\rm{TA}}\in[-2,3]$ based on the results from P26. We fixed the cosmological and other nuisance parameters to their fiducial values (Table~\ref{tab:parameters_and_priors}). We performed this analysis with \texttt{CosmoSIS}, following the implementation described in Sect.~\ref{sec:forecast_pipeline}. In particular, for this analysis, we used the \texttt{EuclidEmulator2} to obtain the non-linear matter power spectrum since it accurately describes the measurements performed in Flagship. Figure~\ref{fig:IA_values_FS} shows the constraints on the \gls{ia} values for both the \nlaz and \tattz models when considering a scale cut of $k_{\mathrm{max}}=3\,\invhMpc$ in the \gls{wl} $C_{\ell}$ and of $k_{\mathrm{max}}=0.3\,\invhMpc$ in the \gls{ggl} and \gls{gc} $C_{\ell}$. We can see different posterior distributions for $A_{1}$ and $\eta_{1}$ between both \gls{ia} models, which are explained by the fact that the higher-order \tattz terms capture additional information, particularly at small scales, while the \nlaz model tries to describe the Flagship \glspl{dv} by shifting the $A_{1}$ and $\eta_{1}$ values. We show the comparison between the measured $C_{\ell}$ in Flagship and the \nlaz and \tattz theory predictions, based on the constraints of Fig.~\ref{fig:IA_values_FS}, in Appendix~\ref{sec:comparison_FS_Cls_synth_DVs_CosmoSIS}. There, we can see that \tattz describes the Flagship \gls{dv} better, with maximum differences of 2$\sigma$ between both. For this reason, and also because we need all \tattz terms to generate different modelling scenarios in our results (see Table~\ref{tab:ia_models}), we employed the maximum of the posterior distribution of the \tattz constraints from Fig.~\ref{fig:IA_values_FS} as \gls{ia} fiducial values, which are specified in Table~\ref{tab:parameters_and_priors}. 

Finally, we are interested in understanding how the \gls{ia} parameter constraints from P26 compare to those derived in this work. P26 also employed the Flagship simulation, although the authors used a slightly different galaxy sample selection and a different estimator to compute the \gls{ia} signal. We show this comparison in Appendix~\ref{sec:comparison_IA_values_romain} and confirm the agreement between both cases. Additionally, for our fiducial sample definition, we show the level of overlap in the \tattz constraints when removing the last redshift bins. This is an important test since P26 found that the \gls{ia} implementation in Flagship deviates from a power-law dependence at high redshift. Even though we find good agreement with the \tattz constraints for both scenarios, which indicates that the constraints in Fig.~\ref{fig:IA_values_FS} are valid, we should be cautious in future works with real data about the redshift dependence of \glspl{ia}.

\subsection{Generation of synthetic DVs}\label{sec:generation_synt_DVs}

We generated synthetic noiseless \glspl{dv} using the fiducial values described in Table~\ref{tab:parameters_and_priors} and the pipeline described in Sect.~\ref{sec:forecast_pipeline}. We used an $f_{\mathrm{sky}}=0.067$, as described in Sect.~\ref{sec:Sample_definition}, and the $n(z)$ for sources and lenses. As for the \gls{ia} parameters, we used the fiducial values computed with the \tattz model in Sect.~\ref{sec:IA_values_FS}. We generated two synthetic \glspl{dv} from the \tattz constraints: one generated with \nlaz (fixing $A_{2}=\eta_{2}=b_{\rm{TA}}=0$) and the other with \tattz.

\section{Results}\label{sec:results}
In this section we present the main results for the different \gls{ia} models considered in this analysis. First, we analyse how the constraining power changes as a function of different analysis choices. Second, we analyse the bias in both \gls{ia} and cosmological parameters that arise from mismodelling \glspl{ia}. Finally, we focus on analysing the degeneracies between \gls{ia} and \gls{photo-z} nuisance parameters.

\subsection{Constraining power}\label{sec:constraining_power}

Here we study how the constraining power on cosmological parameters depends on the \gls{ia} model (\nlaz vs. \tattz) used to generate the synthetic \glspl{dv} (Sect.~\ref{sec:IA_values_FS}), together with other analysis choices, such as scale cuts (defined in Sect.~\ref{sec:scale_cuts}), cosmological probe combinations ($3\times 2$\,pt, $2\times 2$\,pt, and \gls{wl}), the assumed cosmological model ($\Lambda$CDM vs. $w_{0}w_{a}$CDM), and \gls{ia} priors. A summary of the resulting constraining power for the different cases is presented in Table~\ref{tab:perc_err}, in terms of the 68\% percentiles in the \gls{ia} and cosmological parameters.

\begin{table*}
\centering
\caption{Sixty-eighth percentile of the \gls{ia} and cosmological parameters.}
\label{tab:perc_err}
\begin{tabular}{c c c c c c c c c c}
\hline
\hline
Case & $A_{1}$ & $A_{2}$ & $\eta_{1}$  & $\eta_{2}$ & $b_{\rm{TA}}$ & $\Omega_{\rm{m}}$ & $S_8$ & $w_{0}$ & $w_{a}$  \\ \hline
$\Lambda$CDM \nlaz $k_{\rm{max}}=1\,\invhMpc$ $3\times 2$\,pt & 0.063 &  - &  0.277 &  - &  - &  0.019 &  0.010 &  - &  -  \\
$\Lambda$CDM \tattz $k_{\rm{max}}=1\,\invhMpc$ $3\times 2$\,pt & 0.079 &  0.380 &  0.301 &  2.184 &  0.584 &  0.020 &  0.010 &  - &  -   \\
$\Lambda$CDM \nlaz $k_{\rm{max}}=3\,\invhMpc$ $3\times 2$\,pt &  0.054 &  - &  0.234 &  - &  - &  0.017 &  0.009 &  - &  - \\
$\Lambda$CDM \tattz $k_{\rm{max}}=3\,\invhMpc$ $3\times 2$\,pt & 0.067 &  0.111 &  0.268 &  0.673 &  0.197 &  0.017 &  0.009 &  - &  -  \\
$\Lambda$CDM \nlaz $k_{\rm{max}}=3\,\invhMpc$ $2\times 2$\,pt &  0.094 &  - &  0.546 &  - &  - &  0.022 &  0.014 &  - &  - \\
$\Lambda$CDM \tattz $k_{\rm{max}}=3\,\invhMpc$ $2\times 2$\,pt & 0.118 &  0.659 &  0.596 &  2.505 &  1.002 &  0.022 &  0.014 &  - &  -  \\
$\Lambda$CDM \nlaz $k_{\rm{max}}=3\,\invhMpc$ WL &  0.069 &  - &  0.280 &  - &  - &  0.025 &  0.010 &  - &  - \\
$\Lambda$CDM \tattz $k_{\rm{max}}=3\,\invhMpc$ WL & 0.118 &  0.202 &  0.337 &  0.939 &  0.357 &  0.025 &  0.010 &  - &  -  \\
$w_{0}w_{a}$CDM \nlaz $k_{\rm{max}}=3\,\invhMpc$ $3\times 2$\,pt &  0.116 &  - &  0.248 &  - &  - &  0.038 &  0.022 &  0.357 &  0.946  \\
$w_{0}w_{a}$CDM \tattz $k_{\rm{max}}=3\,\invhMpc$ $3\times 2$\,pt & 0.131 &  0.138 &  0.273 &  0.651 &  0.215 &  0.04 &  0.022 &  0.370 &  0.951  \\
$\Lambda$CDM \nlaz $k_{\rm{max}}=3\,\invhMpc$ narrow \gls{ia} prior 1 & 0.054 &  - &  0.226 &  - &  - &  0.017 &  0.009 &  - &  -  \\
$\Lambda$CDM \tattz $k_{\rm{max}}=3\,\invhMpc$ narrow \gls{ia} prior 1 &  0.066 &  0.110 &  0.265 &  0.670 &  0.194 &  0.017 &  0.009 &  - &  -  \\
$\Lambda$CDM \nlaz $k_{\rm{max}}=3\,\invhMpc$ narrow \gls{ia} prior 2 & 0.047 &  - &  0.152 &  - &  - &  0.016 &  0.009 &  - &  -  \\
$\Lambda$CDM \tattz $k_{\rm{max}}=3\,\invhMpc$ narrow \gls{ia} prior 2 & 0.055 &  0.069 &  0.151 &  0.261 &  0.156 &  0.017 &  0.009 &  - &  -   \\
\hline
\end{tabular}
\tablefoot{The cases shown depict different \gls{ia} models, scale cuts, cosmological probes, cosmological models ($\Lambda$CDM versus $w_{0}w_{a}$CDM), and \gls{ia} priors.}
\end{table*}

\subsubsection{\gls{ia} modelling and scale cuts}\label{sec:constraining_power_scale_cuts}

\begin{figure*}
    \centering
    \includegraphics[width=0.48\textwidth]{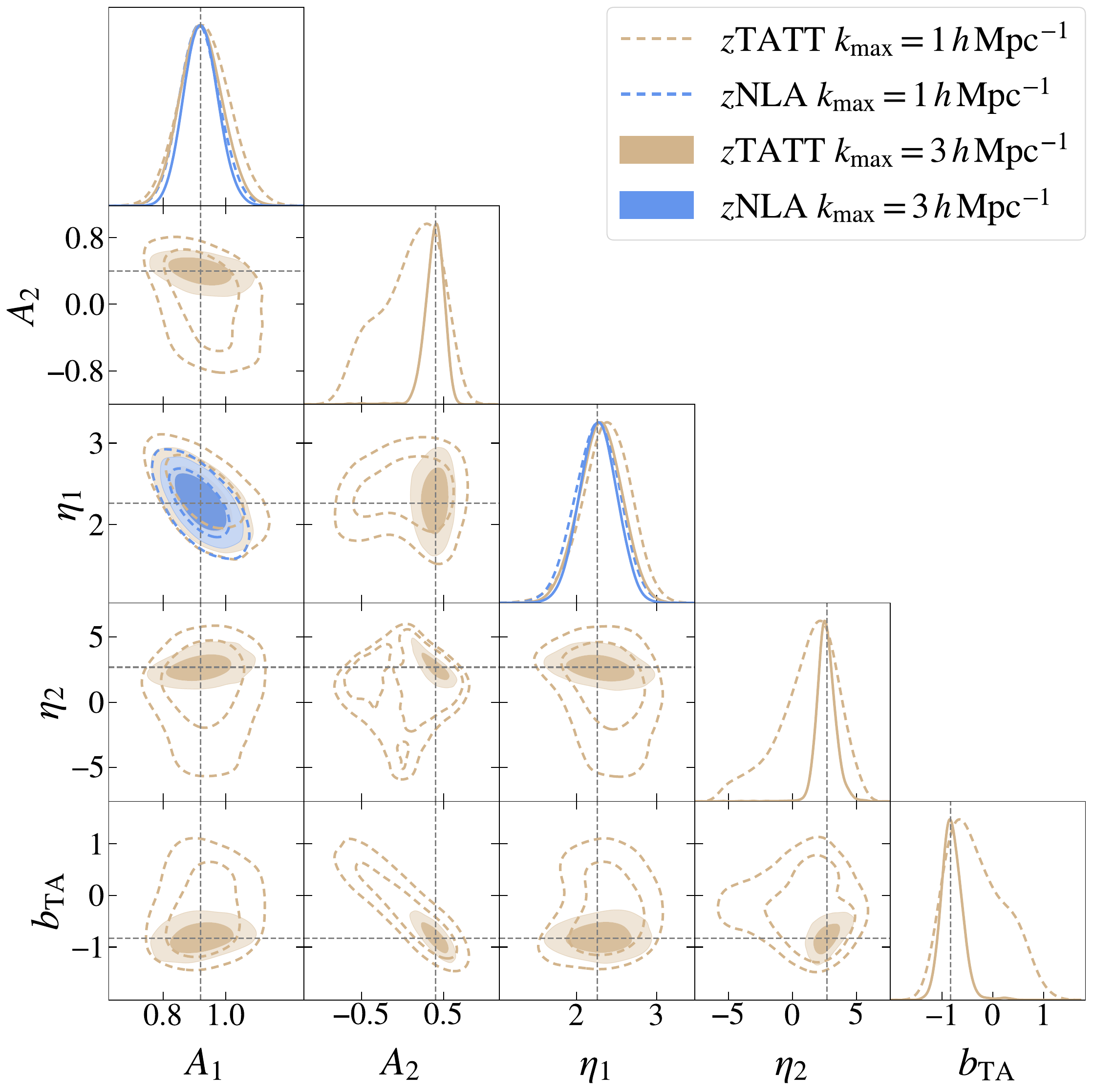}
    \includegraphics[width=0.48\textwidth]{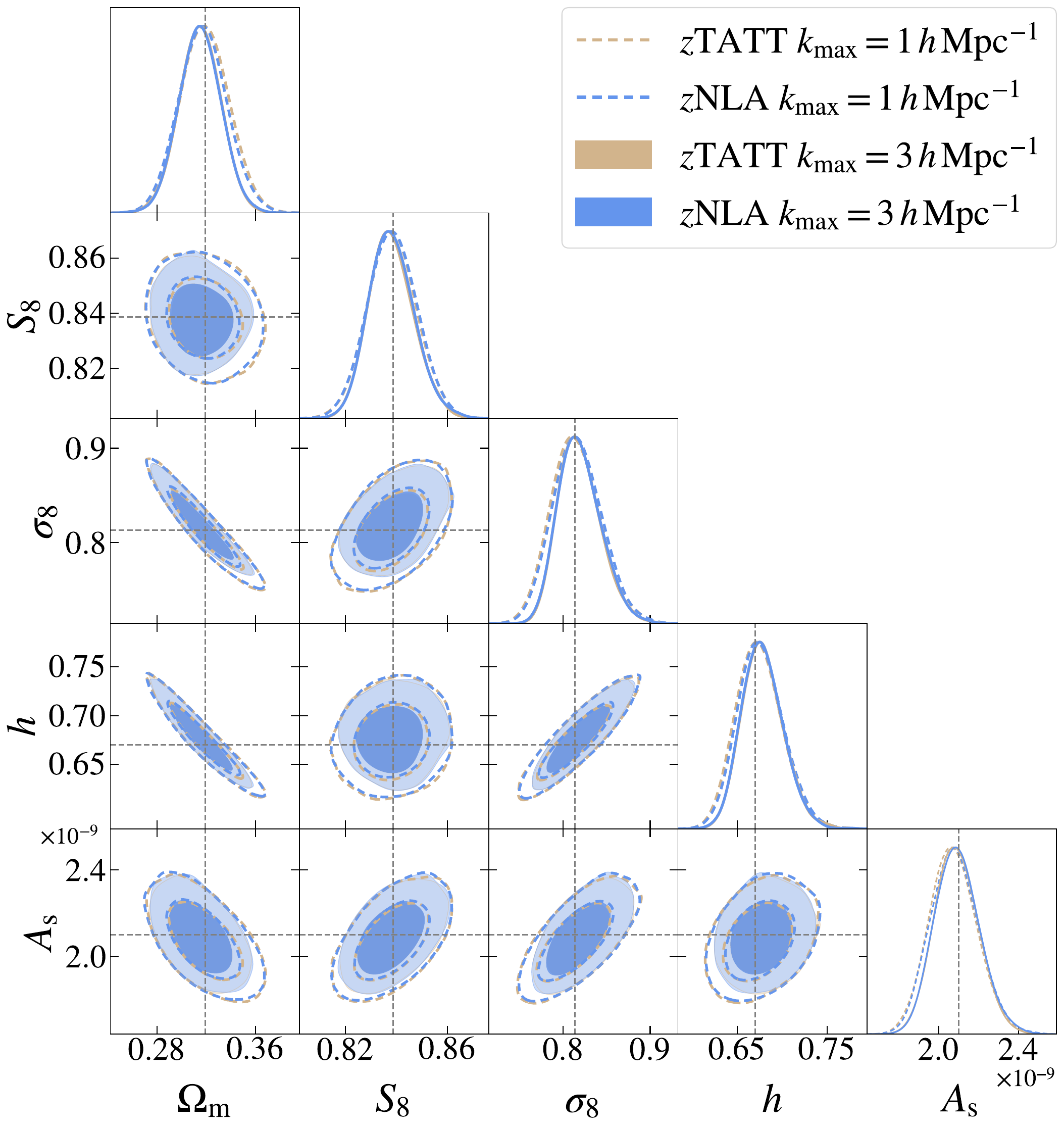}
    \caption{\gls{ia} \emph{(left)} and cosmological parameter \emph{(right)} constraints for \nlaz (blue) and \tattz (tan) at different scale cuts in the \gls{wl} $C_{\ell}$, $k_{\rm{max}}=1\,h\,\rm{Mpc}^{-1}$ (unfilled) and $k_{\rm{max}}=3\,h\,\rm{Mpc}^{-1}$ (filled). The dashed lines indicate the fiducial values of the parameters. The tan contours on the right plot are not well distinguished because they overlap with the blue ones.}
    \label{fig:TATT_NLA_scale_cuts}
\end{figure*}

Figure~\ref{fig:TATT_NLA_scale_cuts} shows the \gls{ia} (left) and cosmological (right) parameter constraints for four scenarios of $3\times 2$\,pt analyses: \nlaz (blue) and \tattz (tan), with a scale cut of $k_{\rm{max}}=1\,\invhMpc$ (unfilled) and $k_{\rm{max}}=3\,\invhMpc$ (filled) for the \gls{wl} $C_{\ell}$. We analyse the change in constraining power from two perspectives: (i) fixing the scale cuts and varying the \gls{ia} model and (ii) fixing the \gls{ia} model and varying the scale cuts. 

The first case shows that \nlaz yields slightly tighter constraints in $A_{1}$ and $\eta_{1}$ than \tattz (a factor of $\sim \!1.25$ in $A_{1}$ and $\sim \!1.1$ in $\eta_{1}$, based on Table~\ref{tab:perc_err}) since the former has fewer parameters. However, this does not translate into tighter constraints in the cosmological parameters, with \nlaz and \tattz yielding similar constraining power, as it was also found in a \gls{des} configuration-space analysis \citep{DESY6_modelling}. This allowed us to use a more flexible \gls{ia} model (with more parameters) without losing constraining power. 

\begin{figure*}
    \centering
    \includegraphics[width=0.98\textwidth]{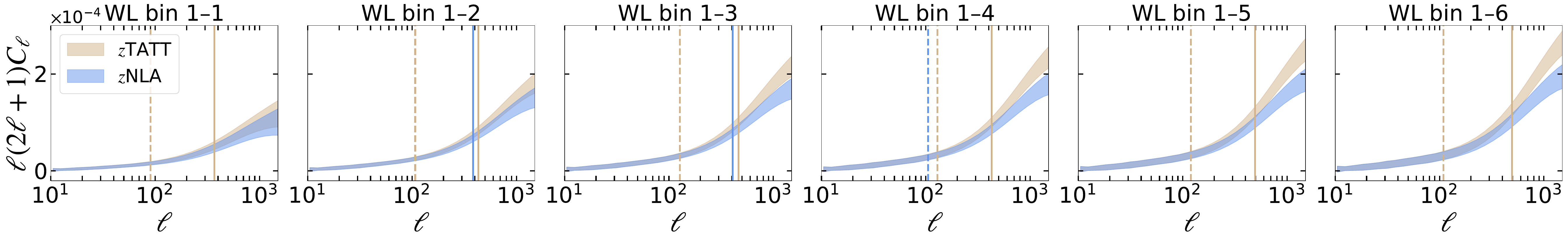}
    \caption{Comparison of the \nlaz (blue) and \tattz (tan) \gls{wl} $C_{\ell}$ for the cross-correlation of the source redshift bin 1 with the other bins. Vertical lines show the scale cuts applied to the analysis for $k_{\rm{max}}=1\,h\,\rm{Mpc}^{-1}$ (dashed line) and $k_{\rm{max}}=3\,h\,\rm{Mpc}^{-1}$ (solid line) in the \gls{wl} $C(\ell)$. The colour of the vertical lines indicate whether they were computed employing \nlaz (blue) or \tattz (tan).}
    \label{fig:comparison_nla_tatt_synth_dv_cl}
\end{figure*}

The second case indicates that the constraining power in the \gls{ia} parameters increases for both \nlaz and \tattz with the inclusion of smaller scales. This is especially true for \tattz, which presents bimodalities in the posterior distributions of its higher-order terms ($A_{2}$, $\eta_{2}$, and $b_{\rm{TA}}$) at $k_{\rm{max}}=1\,\invhMpc$, but not at $k_{\rm{max}}=3\,\invhMpc$, where there is a factor of roughly three increase in the constraining power. Such a result is expected because the higher-order terms are mainly constrained by small scales. This can be better seen in Fig.~\ref{fig:comparison_nla_tatt_synth_dv_cl}, where we show the cross-correlation of source redshift bin 1 with the rest of the redshift bins\footnote{We only include the source redshift bin 1 in this comparison since it is the bin that presents the largest difference between \nlaz and \tattz.} for the \gls{wl} $C_{\ell}$ -- which is the probe where the scale cuts are varied -- for both the \nlaz (blue) and \tattz (tan) \gls{ia} models. The vertical lines show the scale cuts, in terms of $\ell_{\rm max}$, which correspond to $k_{\rm{max}}=1\,\invhMpc$ (dashed line) and $k_{\rm{max}}=3\,\invhMpc$ (solid line).\footnote{Note that the scale cuts defined for \nlaz and \tattz may differ (blue for \nlaz and tan for \tattz), given that the \gls{ia} signal is also taken into account when computing the scale cuts (Eq.~\ref{eq:scale_cut}).} The poor constraints on $A_{2}$, $\eta_{2}$, and $b_{\rm{TA}}$ for \tattz at $k_{\rm{max}}=1\,\invhMpc$ arise because the \tattz signal is very similar to the \nlaz one up to that scale cut. As a result, the higher-order terms in \tattz are not well constrained unless smaller scales are included. Motivated by the greater impact of baryonic effects when including smaller scales, we also checked if there were degeneracies between the \gls{ia} parameters and $T_{\mathrm{AGN}}$, but did not find any. Regarding the constraining power of the cosmological parameters in Fig.~\ref{fig:TATT_NLA_scale_cuts}, it increases when changing the scale cuts, in a similar way to the constraints of $A_{1}$ and $\eta_{1}$.

\subsubsection{Combination of cosmological probes}\label{sec:constraining_power_cosmological_probes}

\begin{figure}
    \centering
    \includegraphics[width=0.48\textwidth]{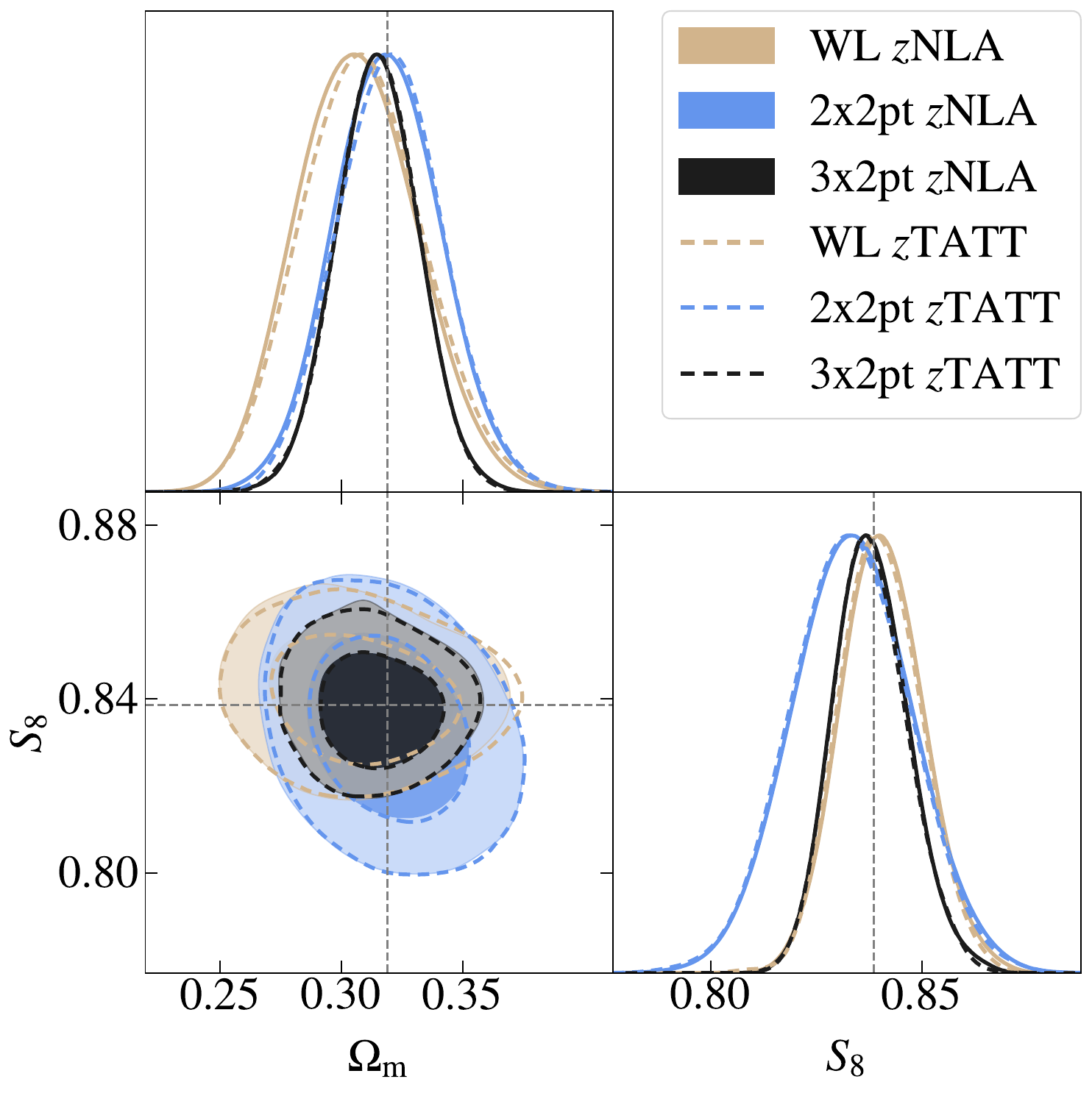}
    \caption{Constraints on $S_{8}$ and $\Omega_{\rm{m}}$ for \nlaz (filled contours) and \tattz (unfilled contours), and for different combinations of cosmological probes: WL (tan), $2\times 2$\,pt (blue), and $3\times 2$\,pt (black). The dashed lines indicate the fiducial values of the parameters.}
    \label{fig:constraining_power_different_probes}
\end{figure}

Figure~\ref{fig:constraining_power_different_probes} shows the constraining power on the cosmological parameters $S_{8}$ and $\Omega_{\rm{m}}$ for different combinations of cosmological probes: $3\times 2$\,pt, $2\times 2$\,pt (\gls{ggl} and \gls{gc}), and \gls{wl}. We show the constraints for \nlaz (filled contours) and \tattz (unfilled contours) for a fixed scale cut of $k_{\rm{max}}=3\,\invhMpc$ in the \gls{wl} $C_{\ell}$. As expected, the most constraining case is $3\times 2$\,pt since it combines all three probes. Similar to Fig.~\ref{fig:TATT_NLA_scale_cuts}, we observe virtually the same constraining power in $S_{8}$ and $\Omega_{\rm{m}}$ when employing \nlaz and \tattz, for both $2\times 2$\,pt and \gls{wl}-only cases.

Table~\ref{tab:perc_err} indicates that the \nlaz parameters are better constrained in the $3\times 2$\,pt and \gls{wl} cases, compared to $2\times 2$\,pt, since the main constraining probe for \glspl{ia} is \gls{wl}, due to the inclusion of smaller scales. That is also the case for \tattz, except from the $A_{1}$ constraints, which are similar for $2\times 2$\,pt and \gls{wl}.

We include in Appendix~\ref{sec:constraining_power_fixed_nuisance} a similar analysis to that performed in Figs.~\ref{fig:TATT_NLA_scale_cuts} and \ref{fig:constraining_power_different_probes}, but only sampling over the cosmological parameters ($\Omega_{\rm m}h^{2}$, $h$, and $A_{\rm s} \times 10^9$) and the \gls{ia} parameters, while fixing all other parameters. This is important because other nuisance parameters can interfere with \glspl{ia} when studying the constraining power. Thus, if we do not find any differences in the constraining power of \nlaz and \tattz in the best-case scenario, which would be to know the true value of all the other nuisance parameters, we will not find them in a less constraining scenario. To summarise, although fixing the remaining nuisance parameters improves the overall constraining power, Appendix~\ref{sec:constraining_power_fixed_nuisance} shows that the relative constraining power of \nlaz and \tattz remains essentially the same.

\subsubsection{Cosmological model}\label{sec:constraining_power_cosmological_model}

Table~\ref{tab:perc_err} includes constraints when considering a $w_{0}w_{a}$CDM model at $k_{\rm{max}}=3\,\invhMpc$. We study the change in constraining power from also two perspectives: (i) varying the \gls{ia} model in $w_{0}w_{a}$CDM and (ii) comparing the $w_{0}w_{a}$CDM and $\Lambda$CDM models for a given \gls{ia} model. The first case is very similar to what we saw for the $\Lambda$CDM model, with the \gls{ia} parameters being a $\sim \! 1.1$ factor better constrained in \nlaz than in \tattz and the cosmological parameters being virtually unaffected. The second case shows that the overall constraining power in the \gls{ia} and the cosmological parameters decreases for the $w_{0}w_{a}$CDM model, with $A_{1}$ and $A_{2}$ being a factor of $\sim \! 2.2$ and $\sim \! 1.2$ less constrained, respectively, and the cosmological parameters, $\Omega_{\rm{m}}$ and $S_{8}$, being a factor of $\sim \! 2.4$ less constrained.

\subsubsection{\gls{ia} priors}\label{sec:constraining_power_ia_priors}

We studied the case of considering more stringent priors on the \gls{ia} parameters to understand if we would lose constraining power with the \gls{ia} priors defined for the fiducial analysis. In this way, we tested the importance of knowing a priori the allowed range of the \gls{ia} parameters. We considered two scenarios in addition to the fiducial \gls{ia} priors, and we consequently reduced the allowed \gls{ia} prior range based on the constraints of Fig.~\ref{fig:TATT_NLA_scale_cuts}: 
\begin{itemize}
    \item Fiducial \gls{ia} prior: $A_{1}\in[0, 3]$, $A_{2}\in[-2, 2]$, $b_{\rm{TA}}\in[-5, 3]$, $\eta_{1}\in[0, 4]$, $\eta_{2}\in[-6, 6]$.
    \item Narrow \gls{ia} prior 1: $A_{1}\in[0.5, 1.5]$, $A_{2}\in[-0.5, 1]$, $b_{\rm{TA}}\in[-3, 1]$, $\eta_{1}\in[1, 3.5]$, $\eta_{2}\in[0, 6]$.
    \item Narrow \gls{ia} prior 2: $A_{1}\in[0.7, 1.1]$, $A_{2}\in[0.2, 0.6]$, $b_{\rm{TA}}\in[-1.2, -0.4]$, $\eta_{1}\in[2, 2.5]$, $\eta_{2}\in[2.2, 3]$.
\end{itemize}

Although Table~\ref{tab:perc_err} indicates that the constraining power in the \gls{ia} parameters slightly increases, especially for the second set of narrow \gls{ia} priors, where the prior boundaries of the \gls{ia} parameters are hit, the constraints on the cosmological parameters for both sets of narrow priors are virtually the same as the ones obtained from the fiducial analysis. This indicates that, for the setup considered in this work, the choice of the \gls{ia} priors is not that critical. However, studies of direct constraints on \glspl{ia} are still necessary, both because future \Euclid data releases will decrease statistical uncertainties, and to constrain \gls{ia} evolution with redshift.

\subsection{Mismodelling of \glspl{ia}}\label{sec:mismodelling}

Since the `true' model describing \glspl{ia} is unknown, we assess the level of cosmological parameter bias that can arise, for the adopted analysis choices, when the assumed \gls{ia} model differs from the one describing the data. For that, we employ the \glspl{dv} described in Sect.~\ref{sec:generation_synt_DVs}, one generated with \nlaz and the other with \tattz, and model them with the six \gls{ia} models described in Sect.~\ref{sec:IA_modelling}: \gls{nla}, \nlaz, \gls{tatt}, \tattz, \nlak, and \tattzbta. Here, we assume $k_{\rm{max}}=3\,\invhMpc$ for the \gls{wl} $C_{\ell}$, given that using $k_{\rm{max}}=1\,\invhMpc$ does not constrain well the higher-order \tattz parameters (as seen in Fig.~\ref{fig:TATT_NLA_scale_cuts}). As a result, analyses limited to $k_{\rm{max}}=1\,\invhMpc$ could suffer from exacerbated biases in the \gls{ia} and cosmological parameters.

Before performing any analysis, we can gain some intuition on which cases are more prone to biases. On the one hand, we expect a low bias when employing the \gls{dv} generated with \nlaz and model it with either \tattz or \nlak since \nlaz constitutes a subspace of these models. However, it is still interesting to study the cases where the modelling is done with redshift-independent models, such as \gls{nla} and \gls{tatt}, or by fixing $b_{\rm{TA}}=1$, as in the case of \tattzbta. On the other hand, when employing a synthetic \gls{dv} generated with \tattz, the use of \gls{nla}-like models (\gls{nla}, \nlaz, and \nlak) will miss the \tattz higher-order terms. Moreover, using a redshift-independent \gls{tatt} model or \tattzbta will also miss some of the dependencies of the \tattz synthetic \gls{dv}. As a consequence, the mismodelling of the synthetic \gls{dv} generated with \tattz is more prone to biases in both \gls{ia} and cosmological parameters.

\begin{figure*}
    \centering
    \includegraphics[width=0.48\textwidth]{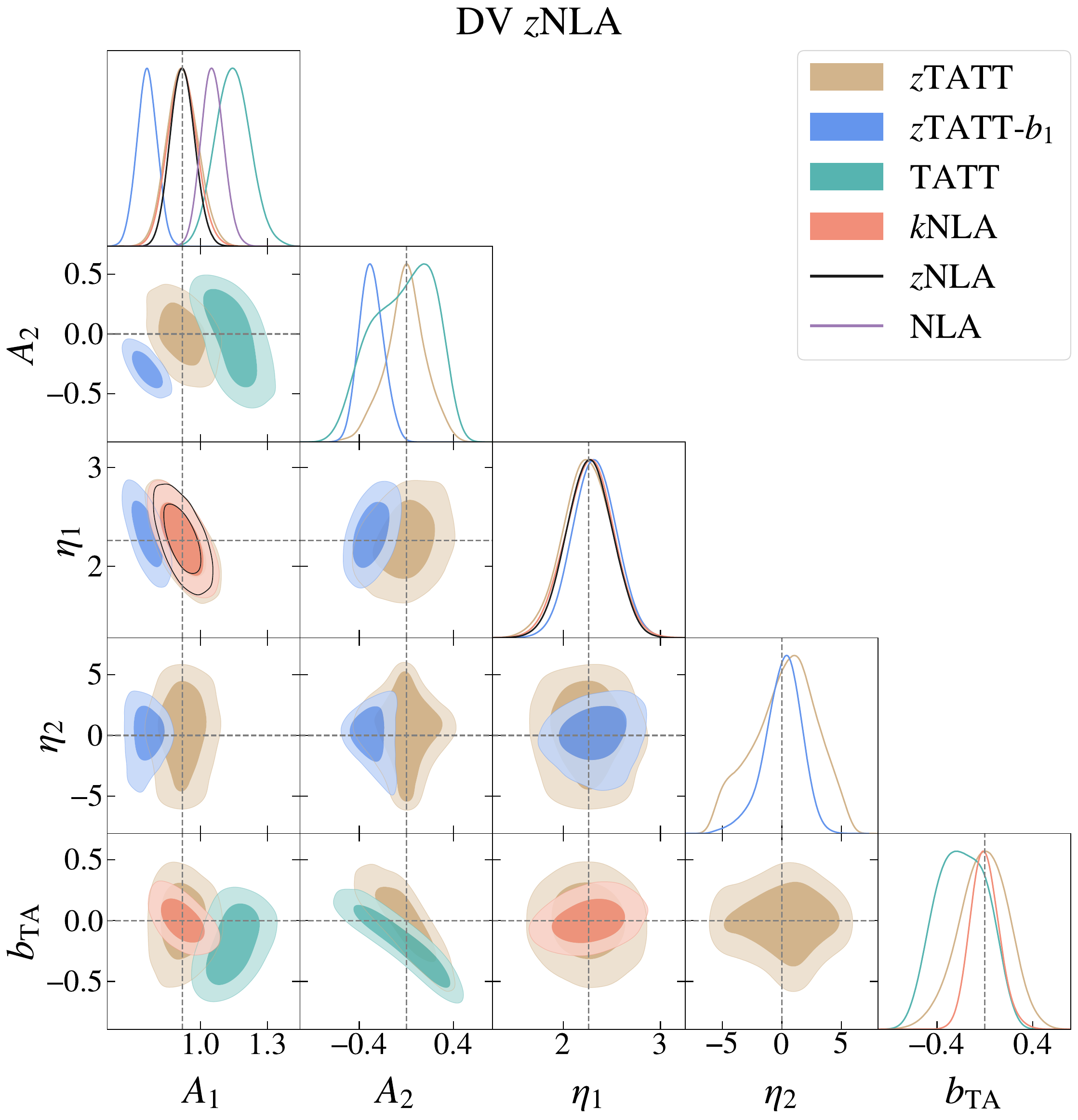}
    \includegraphics[width=0.48\textwidth]{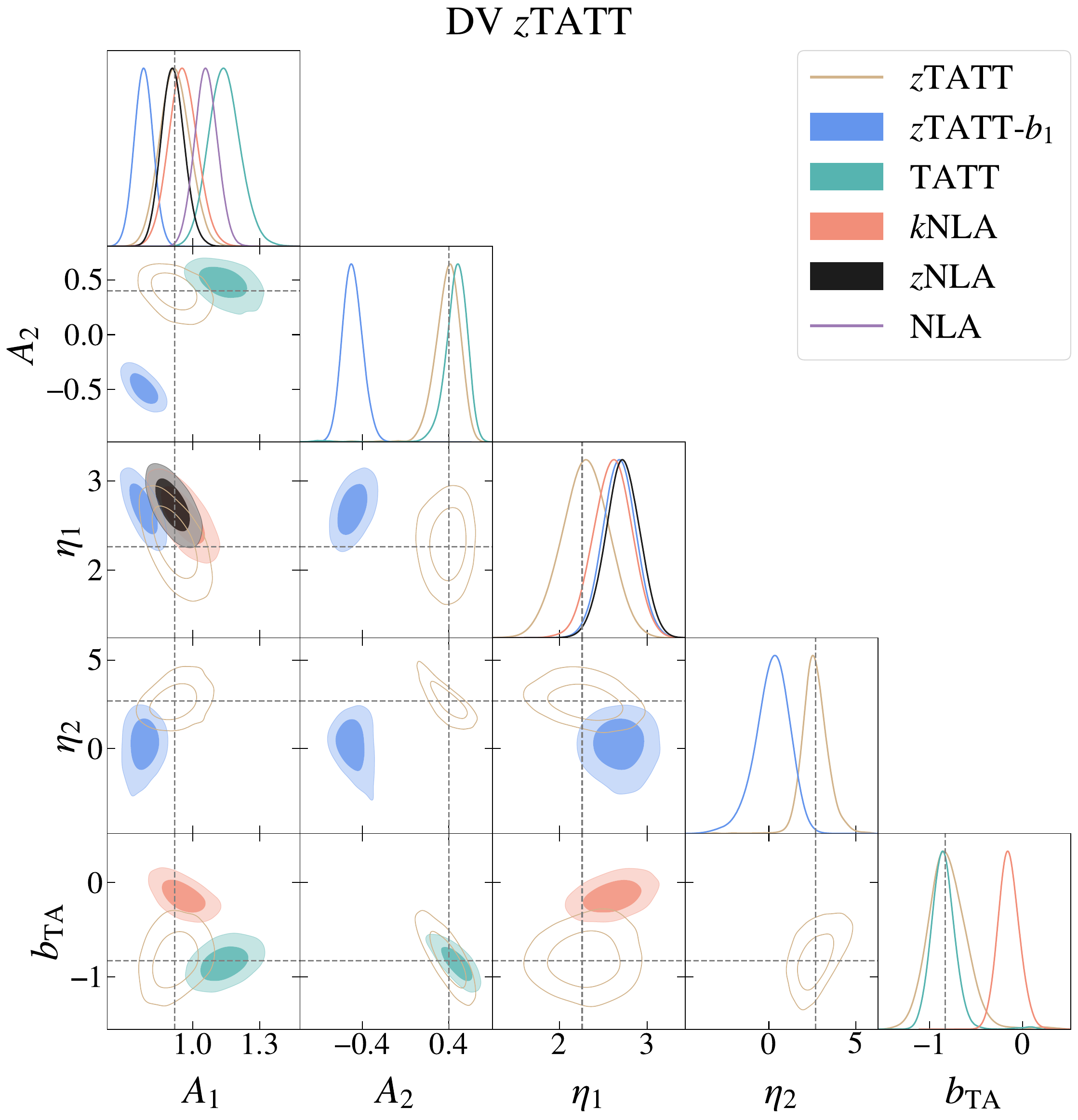}
    \caption{\gls{ia} parameter constraints when mismodelling synthetic \glspl{dv} generated with \nlaz \emph{(left)} or with \tattz \emph{(right)} with all possible \gls{ia} models presented in Sect.~\ref{sec:IA_modelling} and for the $\Lambda$CDM model. The case where the \gls{ia} model is the same as the one used to generate the synthetic \gls{dv} is depicted as black \emph{(left)} or tan \emph{(right)} unfilled contours. The dashed lines indicate the fiducial values of the parameters.}
    \label{fig:mismodelling_DV_IA}
\end{figure*}

Figure~\ref{fig:mismodelling_DV_IA} shows the constrained \gls{ia} parameters when mismodelling the \nlaz (left) and the \tattz (right) synthetic \glspl{dv}. As a reference, we show that the fiducial values are recovered when the model used to generate and analyse \glspl{ia} coincides, depicted as black (left) and tan (right) unfilled contours. The plot on the left shows that the fiducial values of the \nlaz \gls{dv} are also recovered with both \tattz and \nlak, as expected, with their higher-order terms centred on 0. However, the redshift-independent models (\gls{nla} and \gls{tatt}) show a significant bias in $A_{1}$, while \tattzbta exhibits non-negligible biases in both $A_{1}$ and $A_{2}$. The plot on the right shows that none of the \gls{ia} models, except for \tattz, are able to recover the fiducial \gls{ia} parameters describing the synthetic \gls{dv}. However, the biases are different depending on the assumed \gls{ia} model, with \nlaz and \nlak describing the \gls{dv} better since they include the two first-order \tattz terms, $A_{1}$ and $\eta_{1}$. As for the redshift-independent models, \gls{nla} and \gls{tatt}, they both have significant biases in $A_{1}$, although \gls{tatt} is able to recover the fiducial values of both $A_{2}$ and $b_{\rm{TA}}$. Finally, it is interesting to note that \tattzbta, even though it samples four/five parameters from \tattz, shows a large bias in all of its four parameters. This is because the assumption $b_{\rm{TA}}=1$ is not satisfied in the Flagship \glspl{dv}, as seen in Fig.~\ref{fig:IA_values_FS}. Interestingly, the assumption of $b_{\rm{TA}}>0$ is not satisfied in other studies either, such as the \texttt{MASSIVE-BLACK II} hydrodynamical simulation \citep{Massive_black_ii_simulation_1, Massive_black_ii_simulation_2} at high redshifts \citep{Samuroff_hydro}, the \gls{des}Y3 redMaGiC low-$z$ sample \citep{Samuroff_DESY3_eboss}, and the analysis from P26. Although we do not show the results for the $w_{0}w_{a}$CDM scenario in this work for conciseness, the biases are very similar, although with broader contours in the \gls{ia} parameters.

\begin{figure*}
    \centering
    \includegraphics[width=0.48\textwidth]{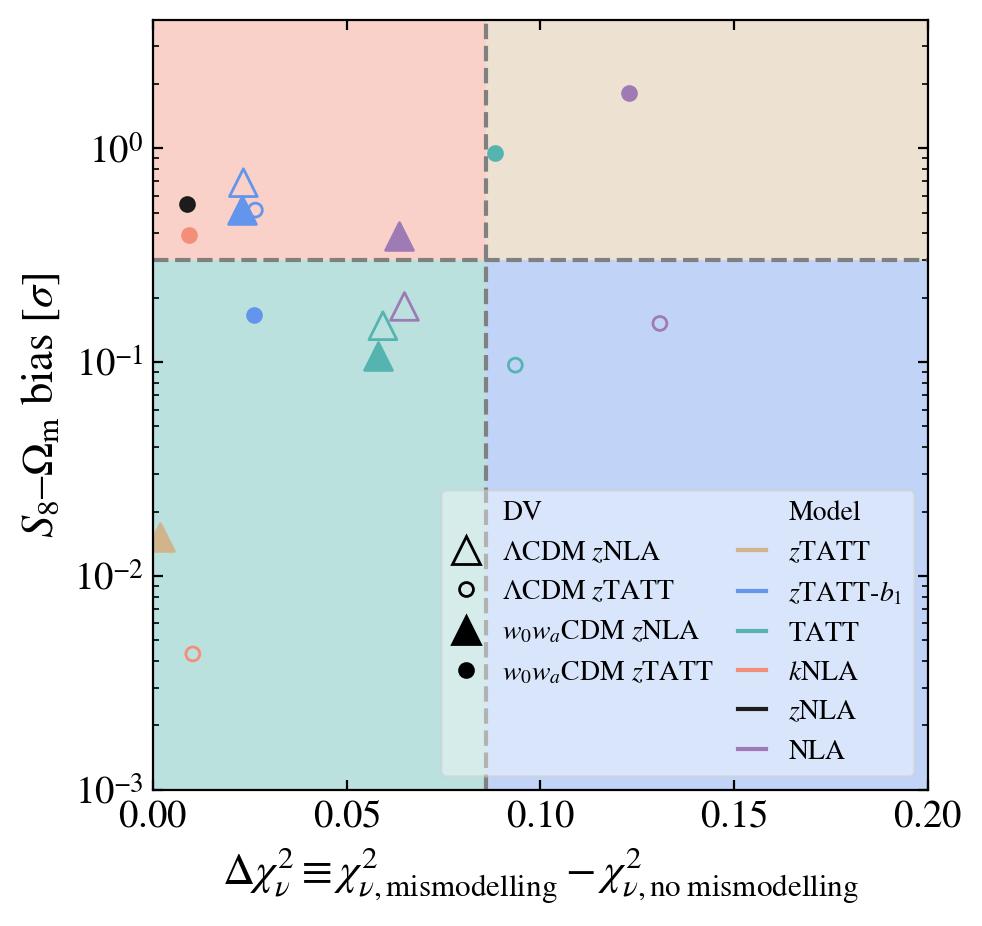}
    \includegraphics[width=0.48\textwidth]{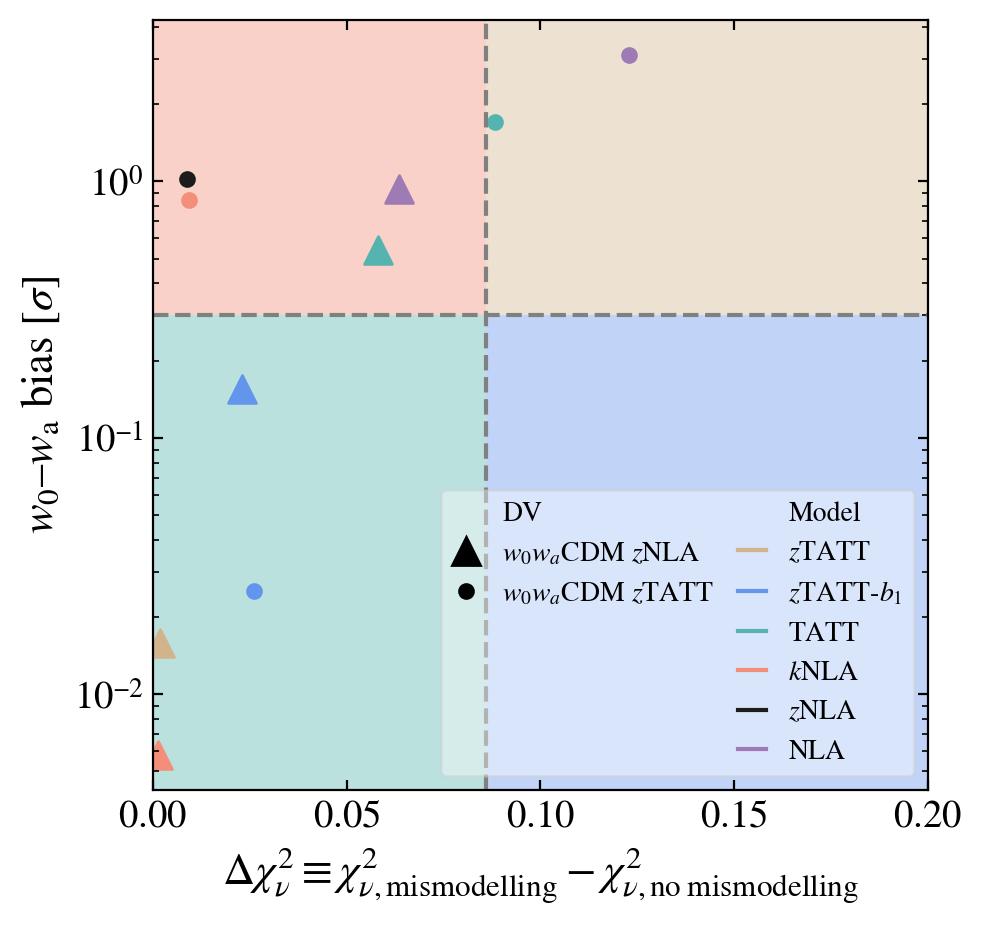}
    \caption{Bias in the $S_{8}$--$\Omega_{\rm{m}}$ plane \emph{(left)} and the $w_{0}$--$w_{a}$ plane \emph{(right)} versus the $\Delta\chi^{2}_{\nu}$, defined as the difference between the $\chi^{2}_{\nu}$ when performing mismodelling and when employing the same \gls{ia} model that was used to generate the \gls{dv}. The horizontal line depicts a 0.3\,$\sigma$ bias, while the vertical line depicts the $\Delta\chi^{2}_{\nu}$ value that corresponds to a probability of exceeding 2\%. The shaded colour areas indicate the different regimes in relation with the level of cosmological bias and $\Delta\chi^{2}_{\nu}$. The cases not shown have biases below the y-axis limit.}
    \label{fig:mismodelling_DV_cosmo}
\end{figure*}

In Fig.~\ref{fig:mismodelling_DV_cosmo}, we show the effect of mismodelling \glspl{ia} in the cosmological parameters $S_{8}$, $\Omega_{\rm{m}}$, $w_{0}$, and $w_{a}$, to which $3\times 2$\,pt analyses are most sensitive. We assessed the mismodelling effect by defining two metrics: (i) bias in the 2D parameter planes of $S_{8}$--$\Omega_{\rm{m}}$ and $w_{0}$--$w_{a}$, computed as the distance between the peaks of the marginalised posteriors, in terms of $n\sigma$; and (ii) $\Delta \chi^{2}_{\nu}$, defined as the difference between the $\chi^{2}_{\nu}:=\chi^{2}/\nu$ (with $\nu$ the degrees of freedom) of the mismodelled and correctly modelled analyses, where each $\chi^{2}$ is evaluated from the maximum $\chi^{2}$ posterior distribution of each chain. Moreover, we compute the $\Delta\chi^{2}_{\nu}$ value that corresponds to a \gls{pte} of 2\%, which is $\Delta\chi^{2}_{\nu}\!\sim\!0.08$. This quantity captures how unlikely it is to find a $\Delta\chi^{2}_{\nu}$ larger than that given by a \gls{pte} value. That is, it marks a region of very unlikely large $\Delta\chi^{2}_{\nu}$, which may be produced by mismodelling effects. Following previous works \citep{Krause_multiprobe_modelling, Campos_model_selection, Leonard_IA_photoz_forecast}, we define $0.3\,\sigma$ as the limit at which an \gls{ia} model is introducing a significant bias. 

The cosmological bias can be computed in two different ways: bias with respect to the fiducial values or relative to the results of an analysis without \gls{ia} mismodelling. Given that the correctly modelled chain can still deviate from the fiducial values, due to factors such as projection effects or constraining power, we compute the bias with respect to the correctly modelled chain. This ensures that the estimated $n\sigma$ differences only reflect the impact of the different \gls{ia} models.

The plots in Fig.~\ref{fig:mismodelling_DV_cosmo}, which show the bias versus the $\Delta\chi^{2}_{\nu}$, are useful to understand which cases can be problematic for a successful analysis. If the $\Delta\chi^{2}_{\nu}$ and the bias increase at the same time (tan shaded area), this indicates that a difference between the \gls{dv} and the model implies a bias in cosmology, and this can be detected from a large $\Delta\chi^{2}_{\nu}$ in the analysis. A low bias and a high $\Delta\chi^{2}_{\nu}$ (blue shaded area) implies that the residuals introduced by \gls{ia} mismodelling project only weakly onto the directions spanned by the cosmological parameters, producing a large mismatch between the \gls{dv} and the theoretical description, but only a small cosmological bias. Conversely, low values of both quantities (green shaded area), in the case where the fiducial \gls{ia} parameters in Fig.~\ref{fig:mismodelling_DV_IA} were not recovered, indicate that the model is flexible enough to absorb the differences in their parameter space. The problematic scenario occurs when the $\Delta\chi^{2}_{\nu}$ is low, but the bias is high (red shaded area). In these cases, the residual \gls{ia} differences are absorbed by a shift in cosmology, and the issue cannot be detected by studying the $\Delta\chi^{2}_{\nu}$ alone in real-data analyses.

The plot on the left of Fig.~\ref{fig:mismodelling_DV_cosmo} shows the bias in the $S_{8}$--$\Omega_{\rm{m}}$ plane as a function of $\Delta\chi^{2}_{\nu}$ for the cases where the \gls{dv} is generated with \nlaz (triangle) or with \tattz (circles) and for the $\Lambda$CDM (empty symbols) and $w_{0}w_{a}$ (filled symbols) cases. The different colours depict the \gls{ia} model used to analyse the \gls{dv}. Most of the points with a bias larger than $0.3\,\sigma$ are from $w_{0}w_{a}$ cases and correspond to points generated with \tattz. Only two points, which correspond to the \tattz $w_{0}w_{a}$CDM case modelled with \gls{tatt} and \gls{nla}, have a high $\Delta\chi^{2}_{\nu}$ and a large bias in $S_{8}$--$\Omega_{\rm{m}}$. However, there are six cases that present a low $\Delta\chi^{2}_{\nu}$, but a large bias, three of them modelled with \tattzbta. In contrast, two other cases have a large $\Delta\chi^{2}_{\nu}$ but a low bias, and correspond to the $\Lambda$CDM cases of the points with large $\Delta\chi^{2}_{\nu}$ and large bias. The plot on the right shows the bias in the $w_{0}$--$w_{a}$ plane versus $\Delta\chi^{2}_{\nu}$. The two previous cases of \tattz $w_{0}w_{a}$CDM case modelled with \gls{tatt} and \gls{nla} also show correlation with $\Delta\chi^{2}_{\nu}$ and bias in the $w_{0}w_{a}$ plane. Furthermore, there are four cases with low $\Delta\chi^{2}_{\nu}$ and high bias. As a summary, all the models, except from \tattz, which always covers the truth in the \nlaz and \tattz \glspl{dv}, can potentially bias cosmological parameter estimation.

\subsection{Degeneracies of IA with photometric redshifts}\label{sec:degeneracy_IA_photo_z}

The degeneracies between \gls{photo-z} and \gls{ia} parameters \citep{IA_degeneracies_shun_sheng, IA_degeneracies_fischbacher, Leonard_IA_photoz_forecast, IA_degeneracies_mcdonald}, if not properly considered, can lead to misinterpretations of the effects that \glspl{ia} can have on cosmological analyses. In particular, a poor estimation of \glspl{photo-z} might lead to biased \gls{ia} constraints and, consequently, to biased cosmological constraints.

\begin{figure}
    \centering
    \includegraphics[width=0.48\textwidth]{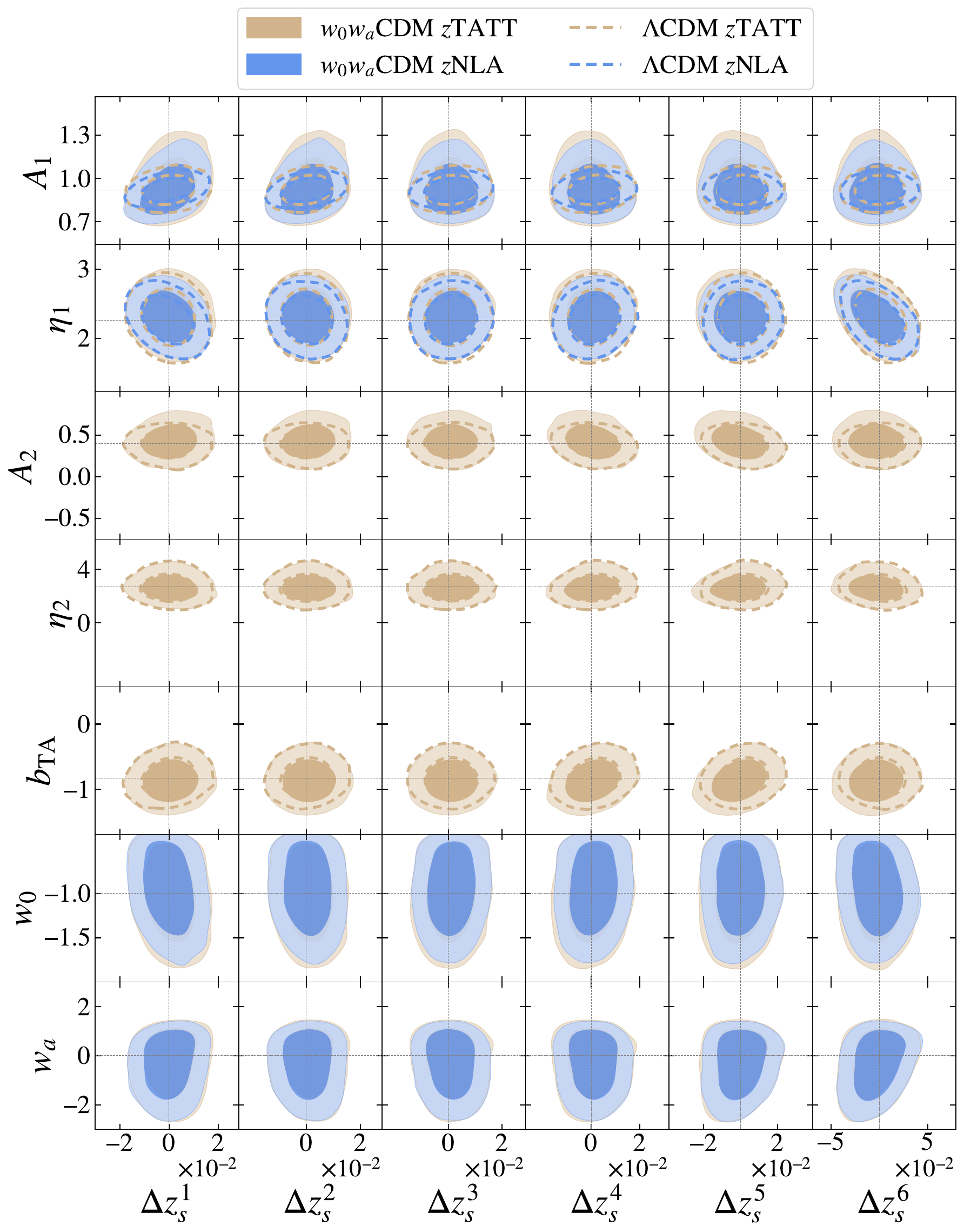}
    \caption{Degeneracies between \gls{ia}, $w_{0}$, $w_{a}$, and $\Delta z_{\rm s}$ parameters for \tattz (tan) and \nlaz (blue) and for $w_{0}w_{a}$CDM (filled contours) and $\Lambda$CDM (unfilled contours). The dashed lines indicate the fiducial values of the parameters.}
    \label{fig:degeneracies_photo_z}
\end{figure}

We inspect these degeneracies in Fig.~\ref{fig:degeneracies_photo_z}, which shows the constraints of the \gls{ia} parameters (and of $w_{0}$ and $w_{a}$ for some cases) versus the source \gls{photo-z} parameters for \nlaz (blue) and \tattz (tan) at $k_{\rm{max}}=3\,\invhMpc$ for the \gls{wl} $C_{\ell}$. We consider the case of a $\Lambda$CDM (unfilled contours) and a $w_{0}w_{a}$CDM (filled contours) scenario. Generally, we did not find significant degeneracies between \gls{ia} and \gls{photo-z} parameters for either the $\Lambda$CDM or $w_{0}w_{a}$CDM model. However, a mild degeneracy appears between the $\Delta z_{\rm s}$ parameter of bin 6 and the $\eta_{1}$ \gls{ia} parameter, especially for $w_{0}w_{a}$CDM. For the higher-order \tattz terms, we did not find degeneracies with $\Delta z_{\rm s}$. In order to understand why this analysis does not show the degeneracies from previous literature, we tested wider priors in $\Delta z_{\rm s}$, defined as flat priors with a 3\,$\sigma$ width of those defined in Table~\ref{tab:parameters_and_priors}. Although not shown here for conciseness, we reran all the cases of Fig.~\ref{fig:degeneracies_photo_z} with these wider priors, and observed strong degeneracies between $A_{1}$ and all redshift bins $\Delta z_{\rm s}$, $\eta_{1}$ and redshift bin 6, and $b_{\rm{TA}}$ and the first five redshift bins for $\Lambda$CDM. This implies that if the \gls{photo-z} Gaussian priors are not correctly centred, the inferred \gls{ia} parameters from \gls{wl} analyses will be biased. These effects can be revealed through, for example, colour-based split consistency tests, as shown in \citet{IA_degeneracies_shun_sheng}. As in the case with the fiducial Gaussian priors, the higher-order \tattz terms do not present degeneracies with $\Delta z_{\rm s}$. Thus, if the \glspl{photo-z} have at least the level of precision of stage-III surveys and are correctly centred, we will not be affected by strong degeneracies between \glspl{ia} and \glspl{photo-z}.

Finally, we also tested if there were other degeneracies between \glspl{ia} and other nuisance parameters -- such as galaxy bias, magnification, and multiplicative shear bias. We did not find any.

\section{Conclusions}\label{sec:Conclusions}

\Euclid will deliver a powerful survey that will observe billions of galaxies and constrain cosmological parameters with unprecedented precision. However, to achieve this goal, all systematic effects also need to be quantified and corrected at an unprecedented level. Amongst the sources of systematic effects that will affect \Euclid, we focused on the \glspl{ia} of galaxies since they are a major contaminant to the \gls{wl} and \gls{ggl} observables and can bias the estimation of cosmological parameters. 

Our main goal was to assess which of the \gls{ia} models currently used in stage-III survey analyses, \gls{nla} or \gls{tatt}, is preferred to describe \glspl{ia} for \Euclid DR1 analyses. We defined six tomographic redshift bins for sources and lenses in an area of $\sim \! 2500 \, \rm{deg}^{2}$ and included systematic effects from \glspl{ia}, magnification, galaxy bias, multiplicative shear bias, and \gls{photo-z} uncertainties. Given that a comparison between the \gls{nla} and \gls{tatt} models (and variations of them) depends on the assumed \gls{ia} central values, we employed the Flagship simulation to estimate these values for our \Euclid DR1 sample. These \gls{ia} values were used to generate synthetic \glspl{dv}, with which we performed $3\times 2$\,pt analyses to constrain the cosmological and \gls{ia} parameters.

We determined the best \gls{ia} modelling choice for \Euclid DR1 by considering three approaches:
\begin{itemize}
    \item Constraining power: We compared the constraining power when using \gls{nla} and \gls{tatt} as a function of different analysis choices, such as scale cuts, combination of cosmological probes, cosmological model ($\Lambda$CDM versus $w_{0}w_{a}$CDM), and \gls{ia} priors. Since \gls{nla} is a subspace of \gls{tatt}, we expected to have more constraining power with the former due to its fewer parameters. This will be true if no extra information comes from the higher-order \gls{tatt} terms at small scales. As a function of scale cuts, when setting $k_{\rm{max}}=1\,\invhMpc$ for the \gls{wl} $C_{\ell}$, the higher-order \gls{tatt} terms are unconstrained. However, if we set $k_{\rm{max}}=3\,\invhMpc$, these terms are constrained, and the loss of constraining power with respect to \gls{nla} is virtually null. As a result, if we include scales small enough, we can expect \gls{nla} and \gls{tatt} to yield very similar constraining powers for a $3\times 2$\,pt $\Lambda$CDM analysis. This is also true for a $2\times 2$\,pt and a \gls{wl}-only scenario as well as for the $w_{0}w_{a}$CDM case. Finally, we tested that the constraining power does not change when reducing the width of the \gls{ia} priors.
    \item \gls{ia} mismodelling: We analysed the cosmological bias caused by mismodelling \glspl{ia}, that is, modelling \glspl{ia} with a different model than that of the \gls{dv}. We studied the mismodelling effect for two \glspl{dv}: one generated with \gls{nla} and one generated with \gls{tatt}. We find that the only \gls{ia} model able to avoid cosmological bias is the \gls{tatt} model with redshift dependence (\tattz), as it is the more general model covering the synthetic \glspl{dv}. Importantly, we find that modelling \glspl{ia} without redshift dependence, or fixing the $b_{\rm{TA}}=1$, can induce significant cosmological biases. 
    \item Degeneracy between \gls{ia} and \gls{photo-z} parameters: We did not find significant degeneracies between \gls{ia} and \gls{photo-z} parameters when assuming a Gaussian prior, with a similar width of stage-III surveys, on the \gls{photo-z} nuisance parameters. However, stronger degeneracies appear when employing a flat prior with a width three times the standard deviation of the Gaussian priors. Thus, if the \gls{photo-z} uncertainties for \Euclid DR1 are kept at least at the same level of precision as in stage-III surveys, we will not suffer from degeneracies between \glspl{ia} and \glspl{photo-z}. It is important to note that we did not find degeneracies with the higher-order \gls{tatt} terms, $A_{2}$ and $\eta_{2}$, for any of the prior choices, which indicates that \gls{tatt} is as reliable as \gls{nla} in this regard.
\end{itemize}

The combination of the lessons learnt from these approaches suggests that the preferred \gls{ia} model to analyse \Euclid DR1 data is \tattz, assuming the redshift-dependence is correctly captured with a power-law, as it does not reduce the constraining power, it is the most robust model regarding the mismodelling of \glspl{ia}, and it does not present degeneracies with \glspl{photo-z}. We leave for future work a full analysis that employs the Flagship simulation measurements as \glspl{dv}, instead of synthetic \glspl{dv}, together with other \gls{ia} models not considered in this work.

\begin{acknowledgements}
The authors would like to thank Aniruddh Herle and Casper Vedder for useful comments on this work. DNG and H. Hoekstra acknowledge support from the European Research Council (ERC) under the European Union's Horizon 2020 research and innovation program with Grant agreement No. 101053992. IT~has been supported by the Ramon y Cajal fellowship (RYC2023-045531-I) funded by the State Research Agency of the Spanish Ministerio de Ciencia, Innovaci\'on y Universidades, MICIU/AEI/10.13039/501100011033/, and Social European Funds plus (FSE+).  IT, DNG, MC, and SGB~also acknowledge support from the same ministry, via projects PID2021-128989NB, PID2019-11317GB, PID2022-141079NB, PID2022-138896NB, PID2024-156844NB; and the European Research Executive Agency HORIZON-MSCA-2021-SE-01 Research and Innovation programme under the Marie Sk\l odowska-Curie grant agreement number 101086388 (LACEGAL) and the programme Unidad de Excelencia Mar\'{\i}a de Maeztu, project CEX2020-001058-M. We acknowledge the use of Spanish Supercomputing Network (RES) resources provided by the Barcelona Supercomputing Center (BSC) in MareNostrum 5 under allocations AECT-2024-3-0020, 2025-1-0045, and 2025-2-0046. \AckEC\AckCosmoHub
\end{acknowledgements}

%
%

\bibliography{references}

\begin{appendix}

\section{Comparison of Flagship \texorpdfstring{$C_{\ell}$}{cls} with theoretical descriptions from \texttt{CosmoSIS}}\label{sec:comparison_FS_Cls_synth_DVs_CosmoSIS}

\begin{figure*}
    \centering
    \includegraphics[width=0.98\textwidth]{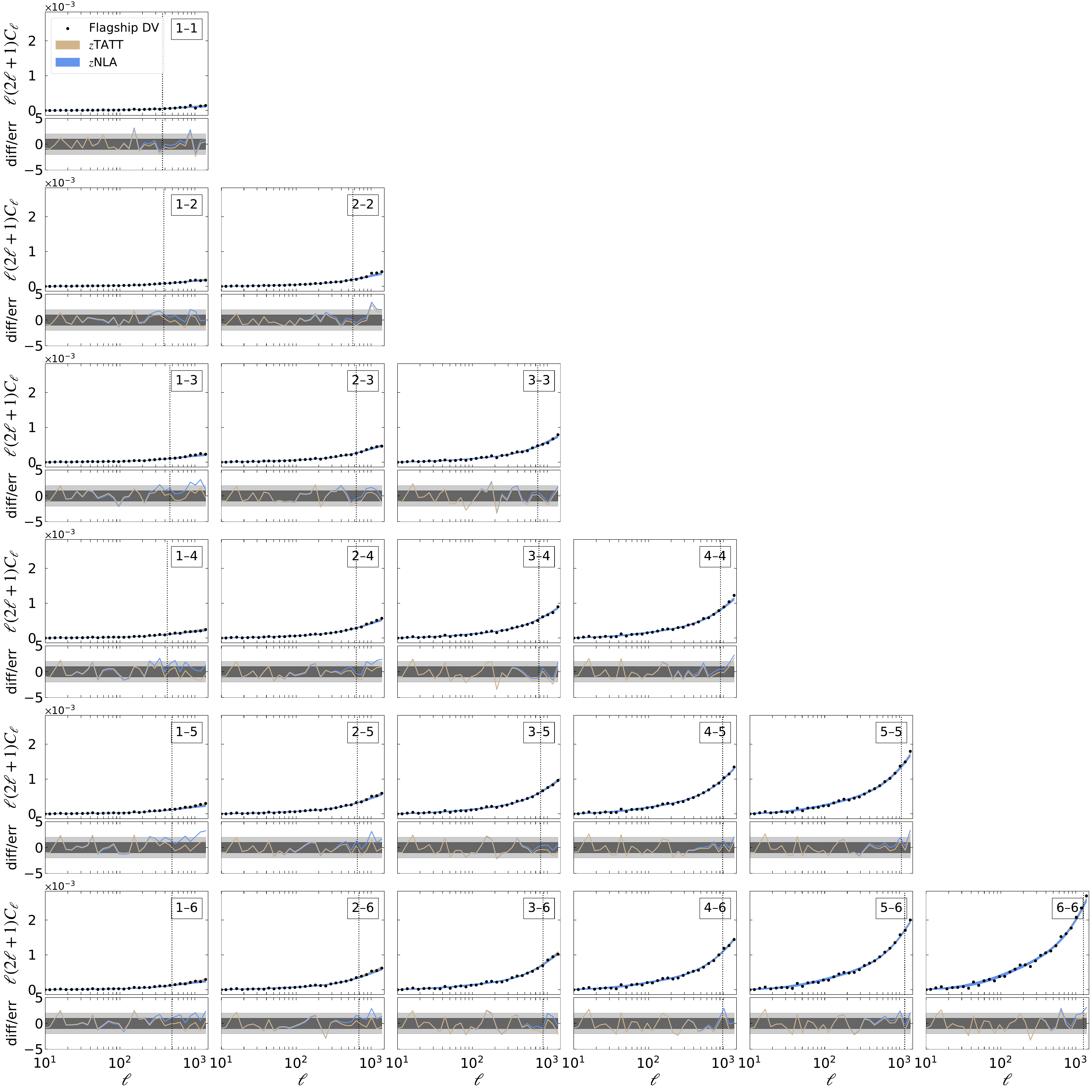}
    \caption{\textit{Top}: Comparison of the measured \gls{wl} $C_{\ell}$ from Flagship (black) against \tattz (tan) and \nlaz (blue) theoretical predictions for the different combinations of source tomographic redshift bins. Theoretical uncertainties assume a Gaussian covariance matrix with a DR1 area. Scale cuts of $k_{\rm max}=3\,h\,\rm{Mpc}^{-1}$ are shown as dotted vertical lines. \textit{Bottom}: Ratio of the difference between the measured $C_{\ell}$ and the theoretical predictions over the uncertainty.}
    \label{fig:Shear_cl_FS}
\end{figure*}

\begin{figure*}
    \centering
    \includegraphics[width=0.98\textwidth]{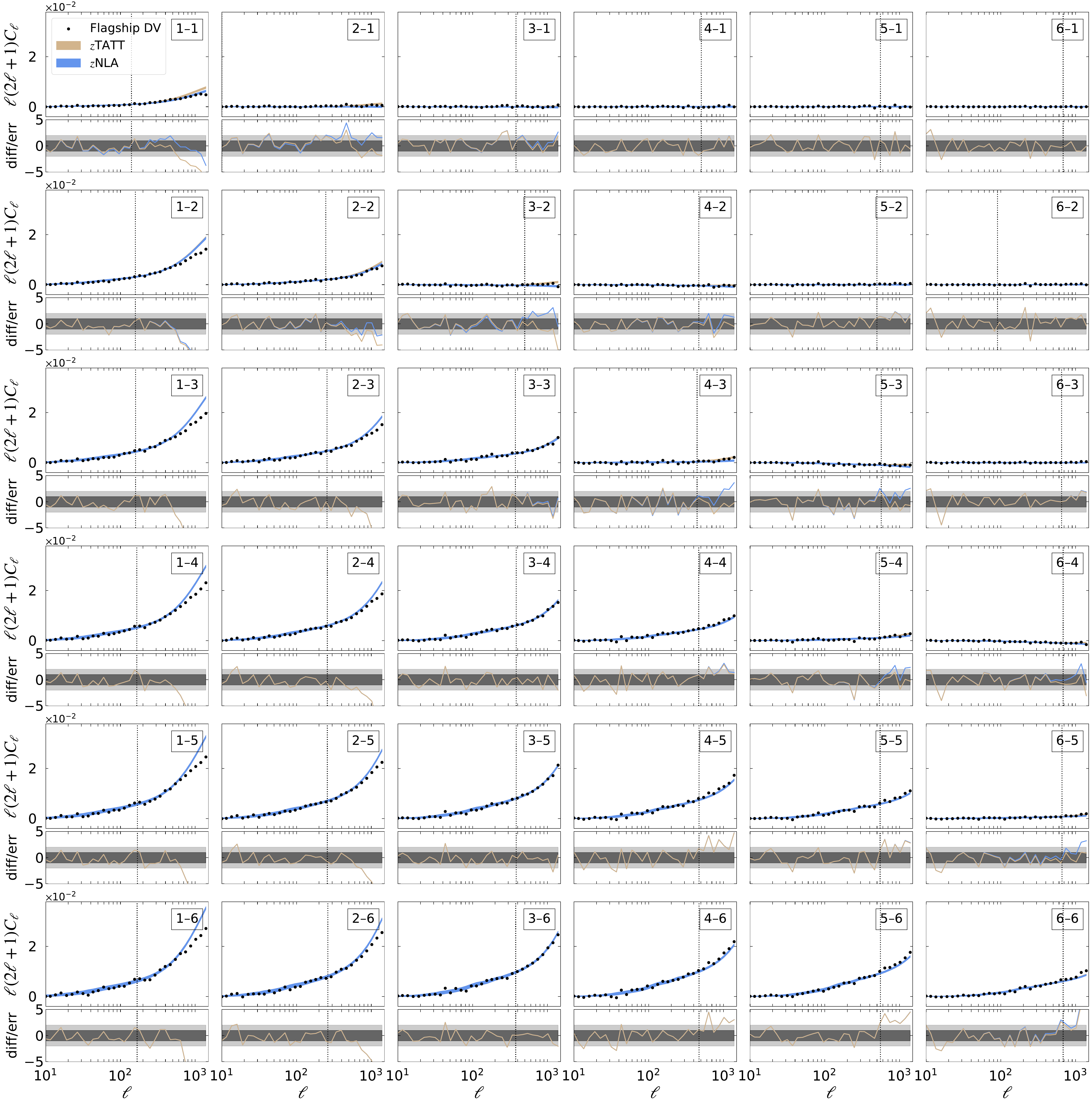}
    \caption{Same as Fig.~\ref{fig:Shear_cl_FS} but for the \gls{ggl} $C_{\ell}$ from Flagship, where the $i$--$j$ tomographic redshift bin combinations correspond to the $i$ lens and the $j$ source bins.}
    \label{fig:GGL_cl_FS}
\end{figure*}

\begin{figure*}
    \centering
    \includegraphics[width=0.98\textwidth]{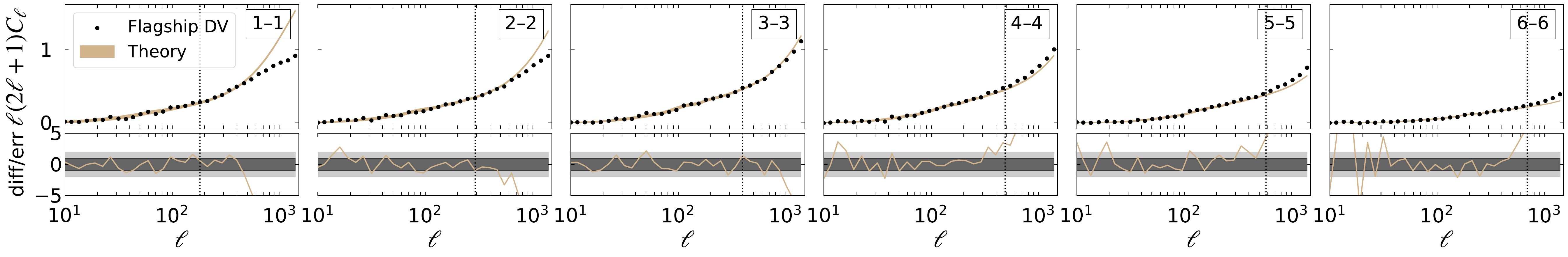}
    \caption{Same as Fig.~\ref{fig:Shear_cl_FS} but for the \gls{gc} $C_{\ell}$ from Flagship, where we only include the auto-correlation of the lens bins.}
    \label{fig:GC_cl_FS}
\end{figure*}

Here, we show how the measured $C_{\ell}$ in Flagship (Sect.~\ref{sec:IA_values_FS}) compare to the \nlaz and \tattz theoretical predictions, based on the constrained values from Fig.~\ref{fig:IA_values_FS}. Figure~\ref{fig:Shear_cl_FS} shows this comparison for the case of the \gls{wl} $C_{\ell}$ for all the $i$--$j$ combinations of source tomographic redshift bins. For each panel, in the upper plot we include the measured Flagship $C_{\ell}$ (black points), the \tattz (tan shaded area), and the \nlaz (blue shaded area) predictions, where the uncertainties from the theoretical predictions assume a Gaussian covariance for the \Euclid DR1 setup (Sect.~\ref{sec:Sample_definition}). In this case, we did not compute the super-sample and the connected non-Gaussian covariance terms because they are sub-dominant and are not expected to strongly affect the recovered \gls{ia} values. The bottom panels show the differences between the measured Flagship $C_{\ell}$ and the theoretical predictions divided by the uncertainty, where the tan and blue lines correspond to the \tattz and \nlaz cases, respectively. The black and grey shaded horizontal lines in the bottom plots show the 1\,$\sigma$ and 2\,$\sigma$ limits. The dotted black vertical lines show the scale cuts that were used to constrain the \gls{ia} parameters in Fig.~\ref{fig:IA_values_FS}, corresponding to $k_{\rm max}=3\,\invhMpc$. Both \nlaz and \tattz are able to describe the \gls{wl} $C_{\ell}$ measured in Flagship, generally with maximum differences of 2\,$\sigma$. However, \tattz allows us to better describe the Flagsip \gls{dv}, especially when cross-correlating low and high redshift bins. Figure~\ref{fig:GGL_cl_FS} shows the comparison for the case of the \gls{ggl} $C_{\ell}$, where both \nlaz and \tattz theory predictions are able to describe the measured $C_{\ell}$, up to the corresponding scale cuts, with a similar level of accuracy. Finally, Fig.~\ref{fig:GC_cl_FS} shows the comparison of the \gls{gc} $C_{\ell}$ for the auto-correlation of lens redshift bins, in general showing a good agreement. However, the high-redshift bin shows some discrepancy at large scales, which potentially originates from the evolution of the galaxy bias parameter in the redshift range covered by this wide bin (see Fig.\,\ref{fig:n_z}). As explained in Sect.~\ref{sec:parameters_and_priors}, the galaxy bias parameters are obtained by fitting the angular correlation functions to the samples of this analysis. However, if the galaxy bias evolves within the redshift bin, the estimated galaxy bias will only be an approximation, which can lead to the differences we observe in Fig.~\ref{fig:GC_cl_FS}. Nevertheless, this is not an issue for our study since we are able to accurately describe the Flagship \gls{wl} and \gls{ggl} $C_{\ell}$, which encode the \gls{ia} information, and the galaxy bias values we employ for our fiducial analyses are used for synthetic \glspl{dv}.

\section{IA values derived from Euclid Collaboration: Paper of Paviot et al.}\label{sec:comparison_IA_values_romain}

To assess a robust estimation of the \gls{ia} values in this work, we compare the \gls{ia} constraints obtained following our methodology with those from P26, which also used the Flagship simulation for their analysis on the \gls{ia} evolution with galaxy properties. However, there are some differences between the two analyses, both in terms of \gls{ia} estimators and in the galaxy sample definition, that need to be understood before performing this comparison. Regarding the \gls{ia} estimators, this work derives the \gls{ia} values by fitting \gls{wl}, \gls{ggl}, and \gls{gc} $C_{\ell}$, where we define a source and a lens catalogue; while P26 employ the so-called $w_{gg}$, $w_{g+}$, and $w_{++}$ correlation functions, which are commonly used in \gls{ia} studies. These correlation functions define density and shape samples, which trace the galaxy positions and the shapes of the galaxies, respectively. Moreover, they only correlate objects that are close to each other, in terms of angular and radial distance. In this way, they do not cross-correlate samples at different redshift bins, as in the case of our study. For the galaxy sample definition, we did not apply a direct magnitude cut in our source sample (which is the one mainly driving \glspl{ia}), while P26 performed an apparent magnitude cut such that $\IE<24.5$. Furthermore, P26 generated ten redshift bins of size $\Delta z = 0.2$ between $0.1<z<2.1$, and subsample the catalogue down to a galaxy number density of 1.5 arcmin$^{-2}$ to reduce computation time. Instead, we perform a cut in the \gls{photo-z} range of $0.2<z_{\mathrm{b}}<2.5$, and generate six equipopulated tomographic bins for sources and for lenses, where the number density of sources is $\sim \! 5.99$ galaxies/arcmin$^{2}$. Finally, the scale cuts employed by both analyses are applied in a different space. P26 define scale cuts based on configuration space such that $r_{\rm{p}}>5\,\hMpc$ (with $r_{\rm{p}}$ the transverse distance between objects), while this work applies cuts in harmonic space, leading to $k_{\mathrm{max}}<3\,\invhMpc$ for \gls{wl} $C_{\ell}$, and $k_{\mathrm{max}}<0.3\,\invhMpc$ for \gls{ggl} and \gls{gc} $C_{\ell}$.

\begin{figure}
    \centering
    \includegraphics[width=0.48\textwidth]{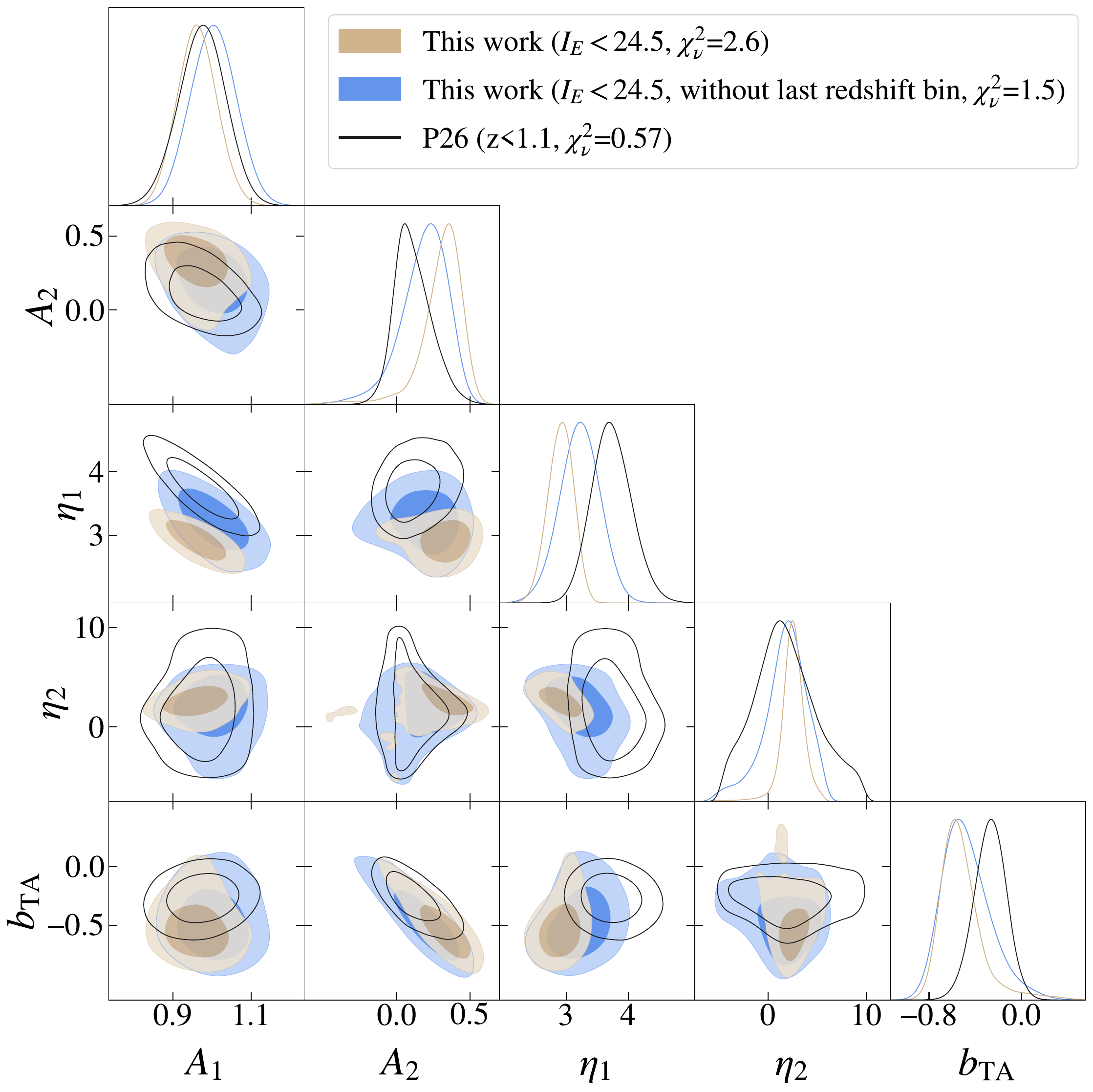}
    \caption{Comparison of the \tattz \gls{ia} parameter constraints and the $\chi^{2}_{\nu}$ estimates derived from Flagship for three cases: (i) applying an apparent magnitude cut $\IE<24.5$ to the source catalogue (tan), (ii) applying the magnitude cut and removing the last lens and source redshift bin (blue), and (iii) the constraints derived in P26 (black).}
    \label{fig:comparison_IA_values_romain}
\end{figure}

All these reasons make it unfeasible to directly compare the constraints we obtained in Fig.~\ref{fig:IA_values_FS} with those obtained in P26. Thus, we modify our sample definition to more closely resemble that of P26. In particular, we generate a new source catalogue by applying the same magnitude cut as in P26 ($\IE<24.5$). We also tested the effect of removing the last source and lens redshift bins since P26 are unable to fit the \gls{ia} parameters properly at $z>1.1$. This could be related to the fact that the \gls{ia} implementation in Flagship is not accurate enough at high redshifts, considering the lack of \gls{ia} observations in this regime, and the fact that Flagship was calibrated at $z=1$ with the hydro-dynamical simulation Horizon-AGN \citep{Horizon_AGN_1, Horizon_AGN_2} and extrapolated to higher redshifts \citep{Hoffmann_IA}. Figure~\ref{fig:comparison_IA_values_romain} shows the \tattz constraints derived by P26 (black unfilled contours), together with the constraints derived in this analysis by applying an $\IE<24.5$ magnitude cut (tan contours) and  by also removing the last source and lens tomographic redshift bin (blue contours). For each case, we include the $\chi^{2}_{\nu}$. We can see that, in general, the posterior distributions of the \tattz parameters from P26 overlap with ours with tensions below $2\sigma$, with the largest differences in the $\eta_{1}$ and $b_{\rm{TA}}$ parameters. We also note that removing the last redshift bins increases the overlap with P26, especially for the $A_{2}$ and $\eta_{1}$ parameters. As in the case of P26, we find a large $\chi^{2}_{\nu}$ when including the last redshift bins, which is much reduced when removing them. Although there are still some differences between both studies, which come from the different analysis choices described above, the level of overlap with the results from P26 allowed us to confirm the robust measurement of \gls{ia} values in this work.

\begin{figure}
    \centering
    \includegraphics[width=0.48\textwidth]{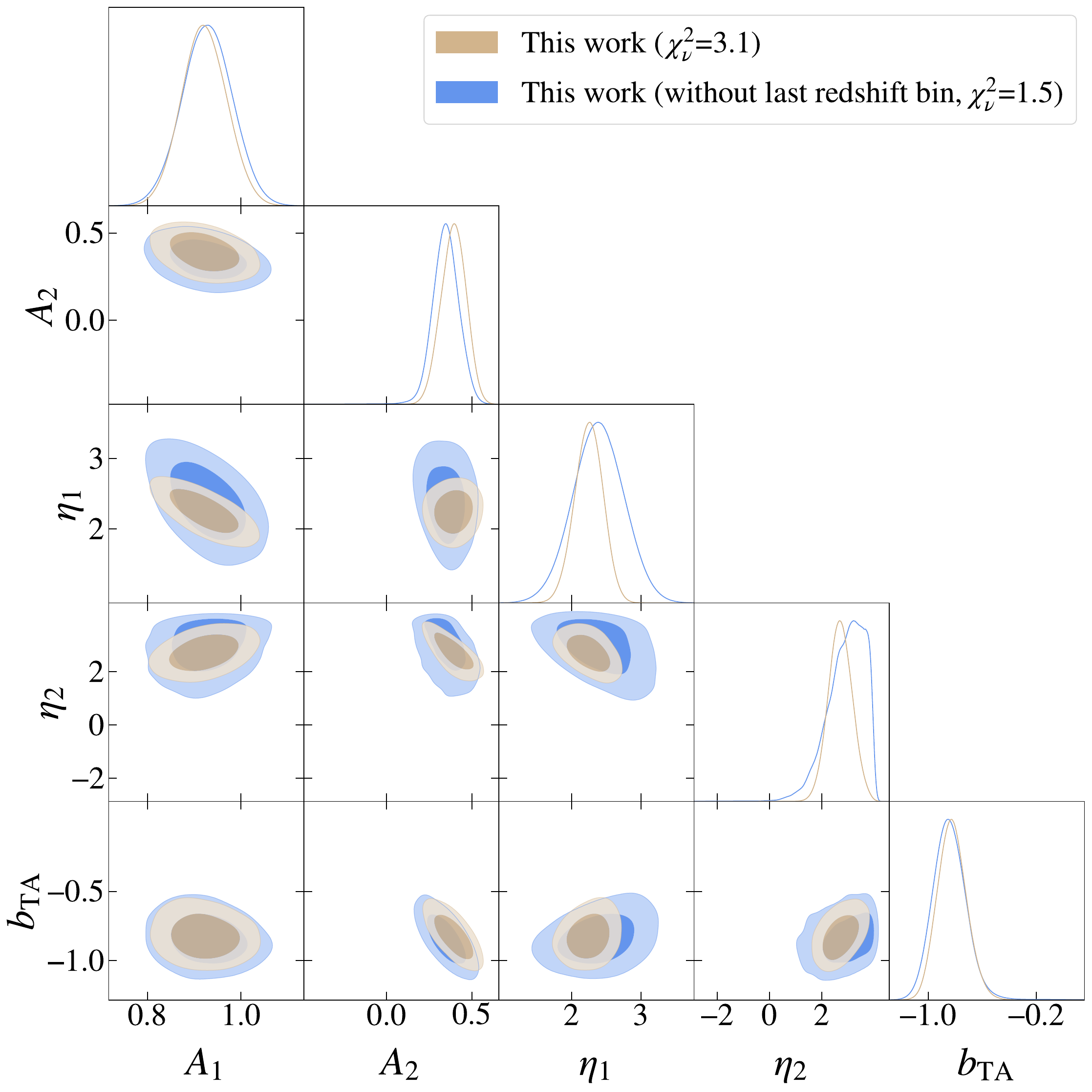}
    \caption{Comparison of the \tattz \gls{ia} parameter constraints and the $\chi^{2}_{\nu}$ estimates derived from Flagship for two cases: (i) fiducial sample selection (tan; Sect.~\ref{sec:Sample_definition}), and (ii) removing the last lens and source redshift bin from the fiducial sample selection (blue).}
    \label{fig:comparison_sample_definitions}
\end{figure}

Since we obtain a worse fit when including all the redshift bins in our modified sample selection (source catalogue with $\IE<24.5$), we check if this also happens when using the fiducial sample selection defined in Sect.~\ref{sec:Sample_definition}. Figure~\ref{fig:comparison_sample_definitions} shows the \tattz constraints derived in the fiducial analysis when considering all the redshift bins (tan, same \tattz constraints as in Fig.~\ref{fig:IA_values_FS}) and when removing the last lens and source redshift bins (blue), together with their $\chi^{2}_{\nu}$. As in the modified sample definition, we obtain a large $\chi^{2}_{\nu}$ when including all the redshift bins, while it behaves well if we remove the last ones. However, we note the large level of overlap in the \tattz constraints in both cases. This indicates that the \gls{ia} fiducial values that were used to generate the synthetic \glspl{dv} (Sect.~\ref{sec:synt_DVs}) are valid, as it was also seen in Fig.~\ref{fig:IA_values_FS} and Figs.~\ref{fig:Shear_cl_FS}--\ref{fig:GC_cl_FS}.

\section{Constraining power with fixed nuisance parameters}\label{sec:constraining_power_fixed_nuisance}

Here, we study the constraining power when only sampling over the cosmological and \gls{ia} parameters. In this way, we can assess whether the lack of difference in constraining power between \nlaz and \tattz in Sect.~\ref{sec:constraining_power} is preserved in a more optimal scenario, where all other nuisance parameters are fixed.

\begin{figure*}
    \centering
    \includegraphics[width=0.48\textwidth]{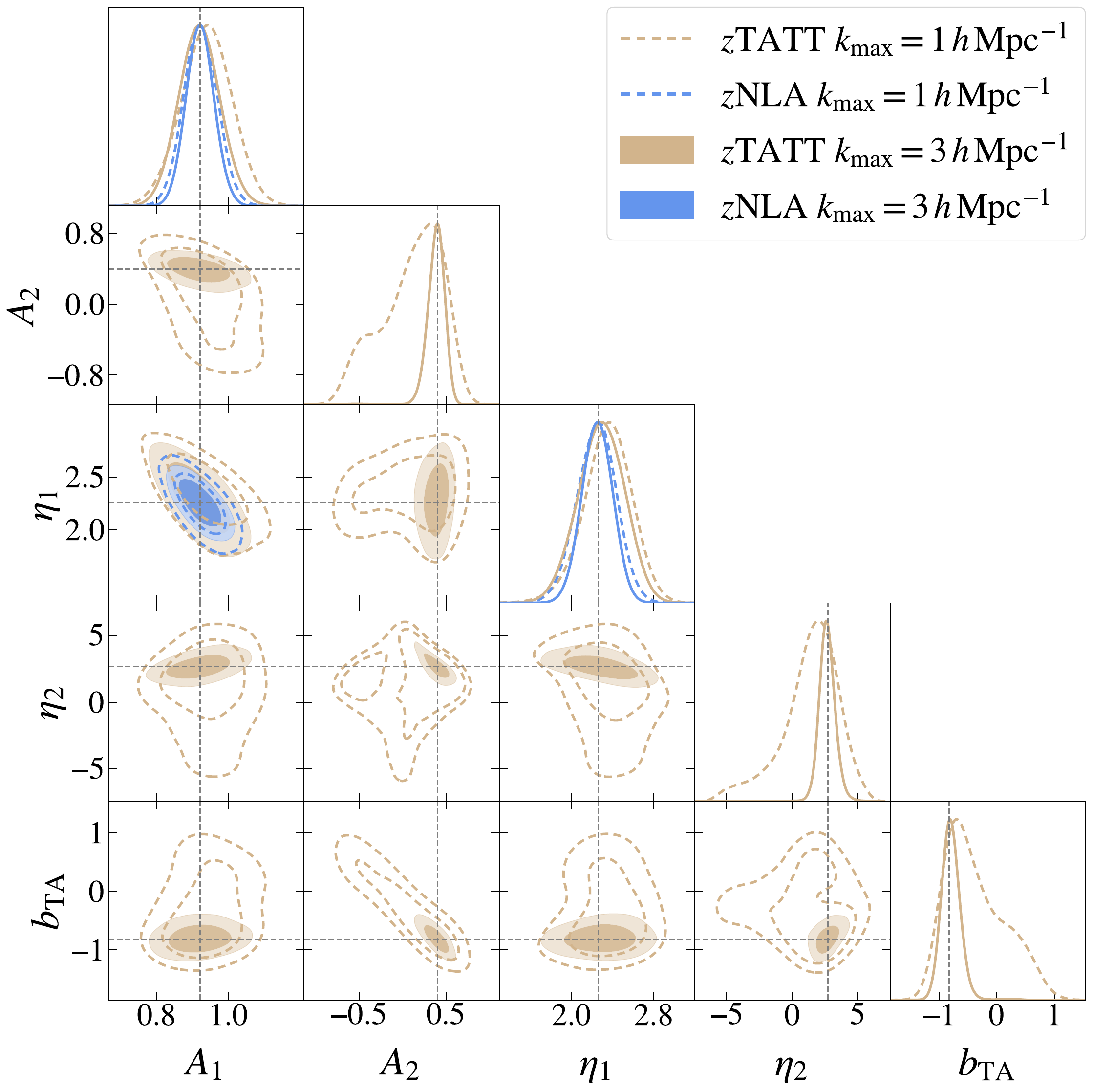}
    \includegraphics[width=0.48\textwidth]{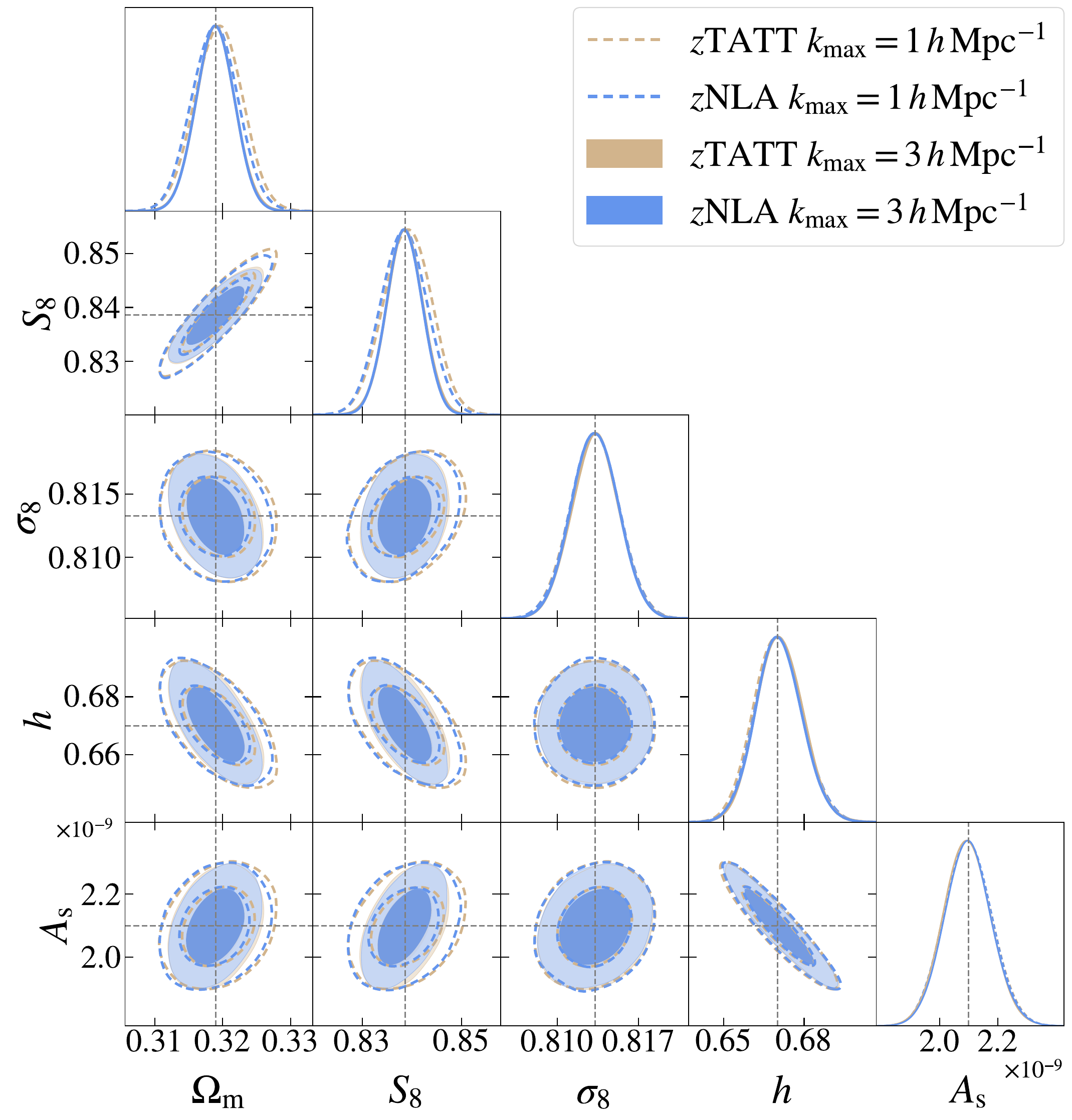}
    \caption{Same as Fig.~\ref{fig:TATT_NLA_scale_cuts} but only sampling over the \gls{ia} and cosmological parameters. Hence, the nuisance parameters are fixed to their fiducial values. The tan contours on the right plot are not well distinguished because they overlap with the blue ones.}
    \label{fig:TATT_NLA_scale_cuts_fixed_params}
\end{figure*}

Figure~\ref{fig:TATT_NLA_scale_cuts_fixed_params} shows the posterior distributions for the \gls{ia} (left) and the cosmological (right) parameters when modelling the \gls{dv} with \tattz (tan) or \nlaz (blue) as a function of the scale cuts applied to the \gls{wl} $C_{\ell}$: $k_{\mathrm{max}}=1\,\invhMpc$ (dashed unfilled contours) and $k_{\mathrm{max}}=3\,\invhMpc$ (filled contours). This is the same analysis performed in Fig.~\ref{fig:TATT_NLA_scale_cuts}, but with the nuisance parameters fixed. Even though both \gls{ia} and cosmological parameters are more constrained than in Fig.~\ref{fig:TATT_NLA_scale_cuts}, as expected when reducing the parameter space by fixing the nuisance parameters to their fiducial values, the same conclusion as in Fig.~\ref{fig:TATT_NLA_scale_cuts} arises: \nlaz and \tattz yield very similar constraining power on the cosmological parameters at $k_{\mathrm{max}}=3\,\invhMpc$, while the \tattz higher-order terms are unconstrained at $k_{\mathrm{max}}=1\,\invhMpc$. This indicates that we will not lose constraining power by choosing a more flexible model, such as \tattz, if we include scales small enough, even if we are able to reduce the uncertainties on the nuisance parameters with respect to our fiducial analysis, as defined in Table~\ref{tab:parameters_and_priors}.

\begin{figure}
    \centering
    \includegraphics[width=0.48\textwidth]{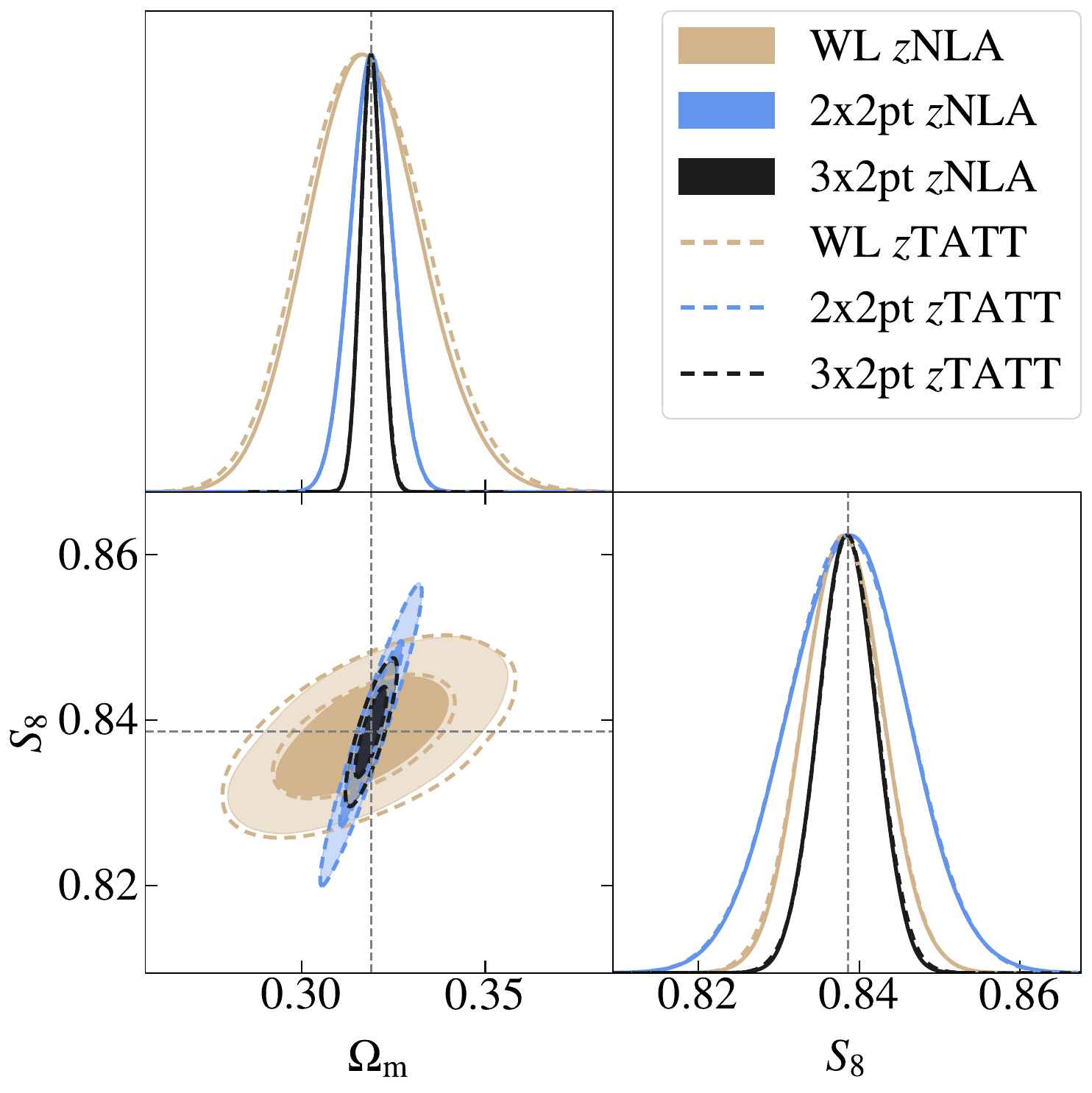}
    \caption{Same as Fig.~\ref{fig:constraining_power_different_probes} but only sampling over the \gls{ia} and cosmological parameters. The nuisance parameters have been fixed to their fiducial values.}
    \label{fig:constraining_power_different_probes_fixed_params}
\end{figure}

Figure~\ref{fig:constraining_power_different_probes_fixed_params} shows the $S_{8}$ and $\Omega_{\rm{m}}$ constraints for different combinations of cosmological probes ($3\times 2$\,pt, $2\times 2$\,pt, and \gls{wl}) for \tattz and \nlaz. This is the same configuration as Fig.~\ref{fig:constraining_power_different_probes}, where we saw that there is no difference in constraining power between \tattz and \nlaz for the $2\times 2$\,pt and \gls{wl} cases. This is also true when fixing all nuisance parameters (except from \glspl{ia}), as can be seen in Fig.~\ref{fig:constraining_power_different_probes_fixed_params}. However, in this case, we can see the typical degeneracy between $S_{8}$ and $\Omega_{\rm{m}}$ in the \gls{wl}-only case, which was diluted in Fig.~\ref{fig:constraining_power_different_probes} by the lower constraining power of that scenario. The tight degeneracy between $S_{8}$ and $\Omega_{\rm{m}}$ for $3\times 2$\,pt and $2\times 2$\,pt arises mainly from fixing the galaxy bias terms.

\end{appendix}

\end{document}